\documentstyle[12pt,aps,epsf,epsfig,floats]{revtex}
\begin{document}

\def\half{{1\over 2}}
\def\be{\begin{equation}}
\def\ee{\end{equation}}
\def\3s1{$^3$S$_1$}
\def\1s0{$^1$S$_0$}
\def\3p2{$^3$P$_2$}
\def\3p1{$^3$P$_1$}
\def\3p0{$^3$P$_0$}
\def\1p1{$^1$P$_1$}
\def\J{{\rm J}}

\title{\small \rm \begin{flushright} 
\small{MC-TH-96/21} \ 
\small{ORNL-CTP-96-09} \ 
\small{RAL-96-039} \\ 
\end{flushright} 
\Large \bf Higher Quarkonia \\
\vspace{0.8cm} }

\author{
T.Barnes,$^1$\thanks{$^1$barnes@orph01.phy.ornl.gov~$^2$fec@v2.rl.ac.uk~
$^3$prp@a13.ph.man.ac.uk~$^4$swanson@unity.ncsu.edu} 
F.E.Close,$^2$ 
P.R.Page,$^3$
E.S.Swanson$^4$
}

\address{
$^1$Theoretical and Computational Physics Section, 
Oak Ridge National Laboratory, \\
Oak Ridge, TN 37831-6373, USA  \\  
Department of Physics and Astronomy,
University of Tennessee, \\
Knoxville, TN 37996-1501, USA \\
$^2$Particle Theory, Rutherford-Appleton Laboratory, Chilton,
Didcot OX11 0QX, UK \\
$^3$Department of Physics and Astronomy, University of Manchester, 
Manchester M13 9PL,  UK \\
$^4$Department of Physics,
North Carolina State University,
Raleigh, NC 27695-8202, USA \\
}

\date{September 1996}
\vspace{1.5cm}

\maketitle

\begin{center}
{\bf Abstract}
\end{center}

\begin{abstract}
We discriminate gluonic hadrons from conventional $q\bar q$ states by
surveying radial and orbital
excitations of all 
I=0 and I=1 $n\bar n$ systems anticipated up to 2.1 GeV.
We give detailed predictions of their quasi-two-body branching fractions
and identify
characteristic decay modes that can isolate quarkonia. 
Several of the ``missing mesons'' with L$_{q\bar q}=2$ and L$_{q\bar q}=3$
are predicted to decay dominantly into certain S+P and S+D modes, and should 
appear in 
experimental searches for hybrids in the same mass region.
We also consider the topical issues
of whether some of the recently discovered or controversial
meson resonances, including
glueball and hybrid candidates, can be accommodated as
quarkonia. 

\end{abstract}
\newpage
\section{Introduction.}

Theoretical studies of light hadron spectroscopy have led to the widespread
belief that gluonic excitations are present in the spectrum of
hadrons, so more resonances should be observed than are predicted by 
the conventional $q\bar q$ and
$qqq$ quark model. 
The two general categories of gluonic mesons expected
are glueballs (dominated by pure glue basis states) and hybrids (dominated
by basis states in which a $q\bar q$ is combined with a gluonic excitation).

Some of these novel states, notably the light hybrids, 
are predicted to have exotic quantum
numbers (forbidden to $q\bar q$), such as J$^{PC}=1^{-+}$. The confirmation
of such a resonance would be proof of the existence of exotic non-$q\bar q$
states, and would be a crucial step towards establishing the spectrum
of gluonic states. There are detailed theoretical predictions for the
decays of these exotic hybrids \cite{ikp,cp95}, 
which have motivated several experimental
studies of purportedly favored hybrid channels such as $b_1\pi$ and $f_1\pi$.

Although one would prefer to find these unambiguously non-$q\bar q$
J$^{PC}$-exotics, glueballs and hybrids with
non-exotic quantum numbers are also expected. 
For example, in the flux tube model the lowest hybrid multiplet, 
expected at
$\approx 1.8$-$1.9$ GeV \cite{paton85,bcs}, contains the non-exotics 
J$^{PC} = 0^{-+}, 1^{\pm\pm}, 1^{+-}$ and 
$2^{-+}$ in addition to the exotics
$0^{+-}$, 
$1^{-+}$ and 
$2^{+-}$. 
To identify these non-exotic states one needs to distinguish
them from the ``background'' of radial and orbital $q\bar q$ excitations
in the mass region $\approx 1.5$-$2.5$ GeV, 
where the first few gluonic levels are anticipated\cite{glueball,wein}.
                  
Our point of departure is to calculate the 
two-body decay modes of all radial and orbital excitations of 
$n\bar n$ states ($n=u,d$) anticipated up to 2.1 GeV. This includes 2S, 3S,
2P, 1D and 1F multiplets, a total of 32 resonances in the
$n\bar n$ sector. 
We also summarize the 
experimental status and important decays 
of candidate members of these multiplets, and compare the 
predictions for decay rates with experiment.

We start by briefly reviewing 
the established 1S and 1P states that confirm that
\3p0 pair creation dominates most hadronic decays. SHO wavefunctions
are employed 
for convenience; these lead to analytic results for decay amplitudes and
are known to give reasonable empirical approximations. This
is sufficient for our main purpose, which is to emphasize selection rules
and to isolate major modes to aid in the identification of states. 
In addition to the 1S and 1P states we also find reasonable agreement
between the model and decays of 1D, 2P and 1F states where data exist; this
confirms the extended utility of the model and adds confidence to
its applications to unknown states. 

Examples of new results include the following.

$\bullet$ 
The radial 2$^3$P$_1$ $a_{1R} \to \rho \pi$ is strongly suppressed in S-wave,
and dominant in D-wave. This contrasts with the expectation for a hybrid $a_1$.
The model's prediction of a
dominant D-wave has been dramatically confirmed for the 
$a_1(1700)$\cite{ves951,suchung} and thereby establishes 1.7~GeV as the 
approximate mass of the $n\bar n$ members of the 2P nonets. This
includes the $0^{++}$ nonet whose I=0 members share the quantum
numbers of the scalar glueball.

$\bullet$ In the scalar glueball sector, we find that the decays of the
$f_0(1500)$
and the $f_{\J}(1710)$ are inconsistent with radially excited quarkonia. 
 
$\bullet$ 
We identify the 2S $0^{-+}$ nonet. The $\eta$ members are predicted to have
narrow widths relative to the $\pi$ counterpart. This is consistent with
the broad $\pi(1300)$ and the narrower candidates $\eta(1295)$ and $\eta(1440)$.

$\bullet$ 
The vector states $\rho(1465)$ and $\omega(1419)$ are interesting in that
the decay branching fractions appear
to show anomalous features requiring a hybrid component. We identify
the experimental signatures needed to settle this question. 

$\bullet$ The $\pi(1800)$ has been cited as a likely hybrid 
candidate\cite{cp95,fec94,ves}
on the strength of its decay fractions.
The 3S $0^{-+}$ $q\bar q$ $\pi$ is also anticipated in this region. We
find that the decays of the hybrid and  3S $0^{-+}$ 
have characteristic differences
which enable them to be distinguished. We identify modes that may enable
the separation of these two configurations. 

Our other results for the many $n\bar n$ states predicted up to 2.1 GeV 
should be useful in the identification of these higher quarkonia,
and in confirming that non-exotic gluonic or molecular states are indeed
inconsistent with quarkonium assignments. 

The order of discussion is 1S and 1P (section 2); 2S and $^3D_1$ (section 3);
3S (section 4); 2P (section 5); 1D (section 6); 1F (section 7). A summary and
an  outline for experimental strategy is in section 8.

\section {1S and 1P Testbed} 

First we will use the well known decays of light 1S and 1P $n\bar n$ states to
motivate and constrain the \3p0 decay model. 
Ackleh, Barnes and Swanson\cite{abs} have carried out a systematic study of
$q\bar{q}$ 
decays in the \3p0 and related pair creation decay models: 
in that work a \3p0-type amplitude was established as 
dominant in most light $n\bar{n}$ decays. (For other discussions of
$q \bar q$ decays in the \3p0 model see Ref.\cite{3p0}).  
Fig.1,  from Ref.\cite{abs}, 
shows \3p0 model predictions for the
decay widths.
Large widths are
indeed predicted to be large and smaller widths are found to be 
correspondingly small. If we choose the pair creation strength
$\gamma=0.5$ (Eq. A3) to set an approximately correct
overall width scale, then
$\Gamma (h_1 \rightarrow \rho \pi)$ and 
$ \Gamma (
a_1 \rightarrow 
\rho \pi)$ are both $ \approx  0.4$-$0.5$ GeV; 
$\Gamma(f_2 \rightarrow \pi\pi)$,
$\Gamma
(\rho \rightarrow \pi\pi)$ and $\Gamma(b_1 \rightarrow \omega
 \pi)$ are all $\approx 0.1$-$0.2$ GeV, and 
$\Gamma (a_2 \rightarrow \rho \pi)$ is smallest, 
$\approx 0.05$ GeV; all are reasonably
close to the observed widths. 

\begin{figure}
$$\epsfxsize=5truein\epsffile{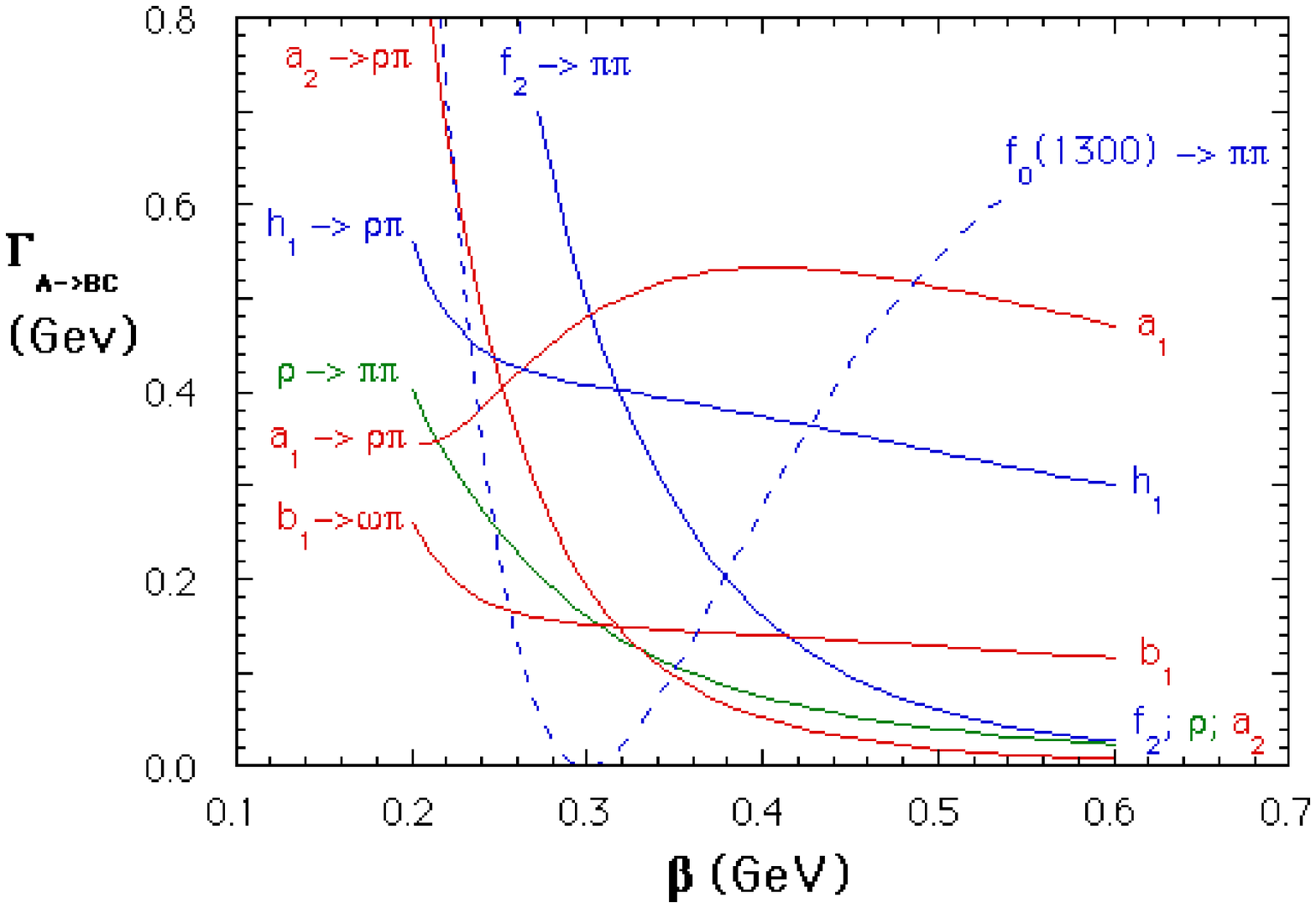}$$
{Figure~1.
Partial widths of light 1S and 1P $q\bar q$ mesons in the \3p0 model. 
The model parameters shown are $\beta=0.2$-$0.6$ GeV 
(with $\beta\approx 0.4$ GeV
preferred) and $\gamma=0.5$.
}
\end{figure}

The optimum parameter values found in a fit to the 
partial widths of Fig.1\cite{abs}
are $\beta=0.40$~GeV (which is 
actually the length scale most commonly
used in light $q\bar q$ decays)
and $\gamma=0.51$; with these values the
rms relative error for these six decays is
$\Delta\Gamma / \Gamma_{expt} = 29 \% $. 
In this work we have actually found that the pair production amplitude 
$\gamma=0.5$ is somewhat large 
for higher-L $q\bar q$ states, 
so in our discussions of higher quarkonia
we will instead use $\gamma=0.4$. 
In constrained-$\gamma$ fits we find that using $\gamma=0.4$ 
only moderately decreases the accuracy of the
fit to the light 1S and 1P decays, to 
$\Delta\Gamma / \Gamma_{expt} = 43 \% $, with an optimum
$\beta=0.36$~GeV.

A more sensitive test of the \3p0 model involves amplitude ratios
in the decays $b_1\to\omega\pi$ and $a_1\to\rho\pi$. In these decays 
both S- and D-wave final states are allowed, and the ratio of these decay
amplitudes is known to be D/S = $+0.260(35)$ for the $b_1$ and $-0.09(2)$ for the
$a_1$ \cite{pdg96}. This ratio is quite sensitive to the quantum numbers
of the produced pair; with \3p0 quantum numbers and the usual $\beta$
we find reasonable agreement in sign and magnitude, whereas a OGE pair
production mechanism gives the wrong sign for D/S \cite{abs}. This ratio
test for $b_1\to\omega\pi$ was historically very important in establishing the
\3p0 decay model \cite{early3p0}.

These successes of the \3p0 model 
motivate its use in predicting decays
of the less familiar radial and orbital excitations of light quarkonia.

\section{2S States}

We first consider the decays of the low-lying radially-excited pseudoscalar
and vector states. Our general approach will be to review recent data on
the state in question and compare these data to predictions for candidate
$q\bar q$ and (where appropriate)
hybrid states. In each case we will attempt to identify decay 
modes that distinguish between competing assignments 
most clearly.

\subsection{$0^{-+}$ $2^1$S$_0$: $ \pi$ and $\eta$}

$\bullet$ $\pi(1300)$

The $\pi(1300)$ 
was first reported by Bellini {\it et al.}\cite{bellini}
in 1982                                                                
but remains rather poorly known.  It is seen 
in $\pi\rho$, 
$\pi(\pi\pi)_S$ and $\pi f_0(1300)$, with a width of 
200-600 MeV; there is however no accurate
measurement of the 
branching fractions \cite{pdg94}. 
Recently higher statistics have been obtained for the $\pi(1300)$ 
by VES\cite{ves951,ves} and by E852 at BNL\cite{suchung}. 
The VES data shows a clear $\pi(1300)$ 
peak in $3\pi$, with a 
width of $\Gamma \approx 400$-$500$ MeV in both $\pi (\pi\pi)_S$ and
$\rho \pi$; the latter is particularly strong and dominates this channel
below 2 GeV. 

It should be noted, however, that the size of the Deck background in 
$\pi (\pi \pi)_S$ is uncertain,
and it is not clear whether the $\pi(1300)$ 
reported in  $\pi (\pi \pi)_S$ is actually due to 
the resonance. 
Fig.1c of Ref.\cite{ves951}
suggests that the Deck mechanism could cause {\it all} of the $\pi(1300)
\rightarrow \pi (\pi \pi)_S$ enhancement in Fig.4a of that reference.
We will assume that
this is essentially correct, and that 
the $\pi(1300)$ resonance decays dominantly to $\rho\pi$.

In the \3p0 decay model we expect $\rho\pi$ to be the dominant mode
of a 2S $q\bar q$ $\pi(1300)$, since this is the only open two-body channel.
(We assume that the $f_0(980)$ and $a_0(980)$ are
dominantly K$\bar {\rm K}$, so the mode
$\pi(1300)\to f_0(980)\pi $ is a more complicated three body or
virtual two-body decay.) 
With our parameter set $\gamma=0.4$ and $\beta=0.4$~GeV we predict a partial
width of

\begin{equation}
\Gamma(\pi(1300) \rightarrow \pi \rho) = 209 \; {\rm MeV} \ .
\end{equation}
This rate is given in Table B2 of Appendix B. (App.B is a tabulation of all
our numerical results for partial widths in the \3p0 model.)
In Fig.2 we show the dependence of this prediction on the
wavefunction length scale $\beta$. Evidently the prediction of a large
width, comparable to observation, follows from any plausible choice for
$\beta$. Thus the observed $\pi(1300)$ is consistent with expectations for
a 2$^1$S$_0$ $q\bar q$ state.

\begin{figure}
$$\epsfxsize=5truein\epsffile{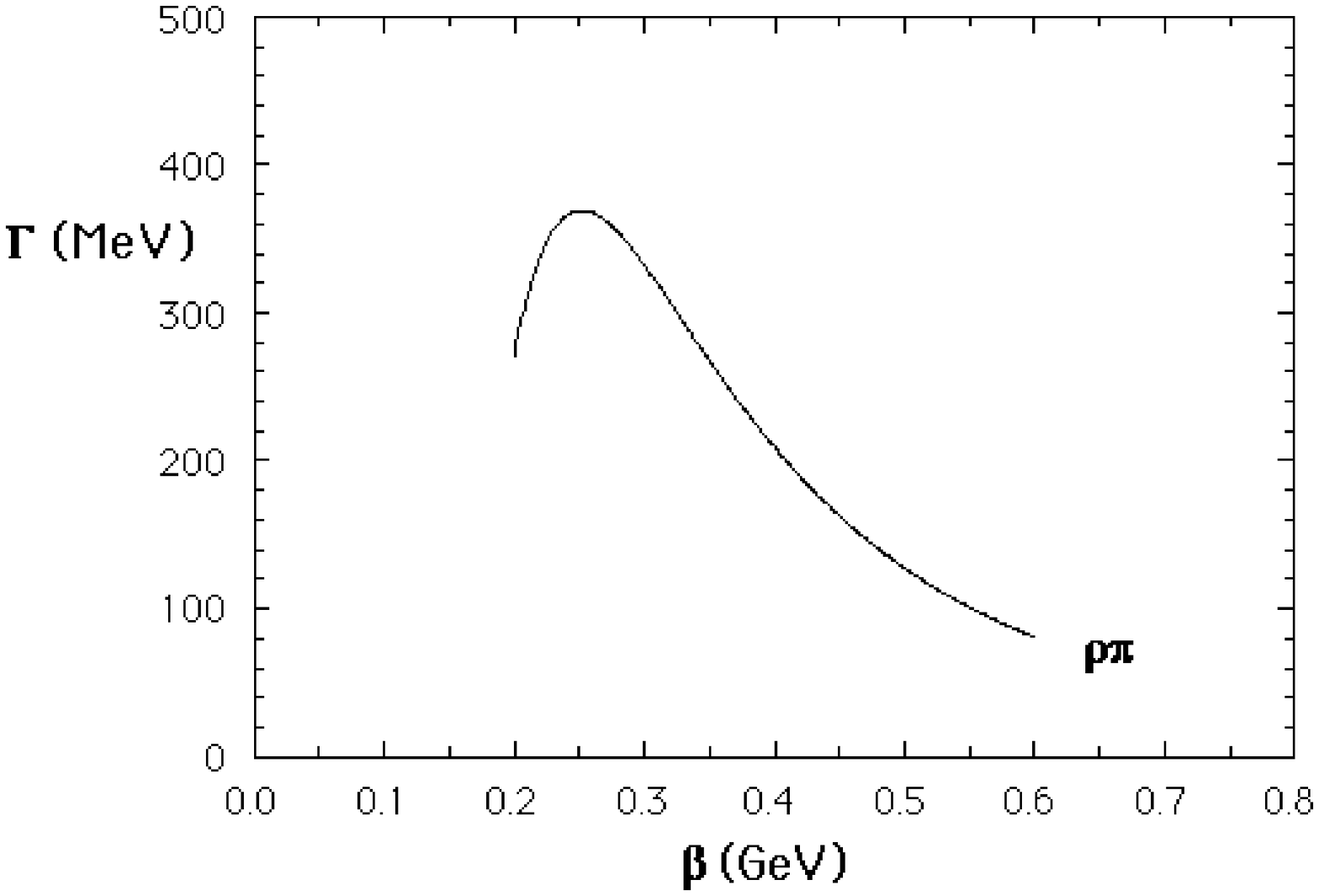}$$
{Figure~2.
The $\rho\pi$ partial width of a 2S $\pi(1300)$, with \3p0 model parameters
$\beta=0.2$-$0.6$ GeV and $\gamma=0.4$. 
}
\end{figure}

Although the mode $f_0^{q\bar q}(1300)\pi$ is nominally closed by phase space,
the $f_0(1300)$ is a very broad state, so one might anticipate a significant
$(\pi\pi)_S\pi$ mode through the
low-mass tail of the $f_0(1300)$. This possibility may be tested by 
varying $M(f_0^{q\bar q})$; the 
resulting $\Gamma(\pi(1300)\to f_0^{q\bar q}\pi )$ 
does not exceed $10$ MeV over the range
$M(f_0^{q\bar q}) = 400$-$1000$ MeV. Thus, the 
population of a $\pi (\pi\pi)_S$
mode by $\pi(1300)$ decays through an intermediate 
$f_0^{q\bar q}\pi $ state is predicted to be a small 
effect. If there actually is a large $\pi(1300)\to \pi (\pi\pi)_S$ mode,
rather than a nonresonant Deck effect, this 
would be in disagreement
with the \3p0 model. Thus it would be very interesting to establish the
branching fraction for $\pi(1300)\to \pi (\pi\pi)_S$ accurately
in future work.

$\bullet$ $\eta(1295)$

This state has a width of $\Gamma = 53(6)$ MeV\cite{pdg94},
much narrower
than its I=1 2$^1$S$_0$ partner $\pi(1300)$. It has
been reported in 
$a_0(980)\pi $ and
$\eta\pi\pi $.
This small width is natural if the  
$\pi(1300)$ does indeed decay
dominantly to $\rho\pi $, since G-parity forbids the analogous
processes 
$\eta_{n\bar{n}} \to \rho\pi$ and
$\eta_{n\bar{n}}\to \omega\eta$; to the extent that the $a_0(980)$ and
$f_0(980)$ are dominantly $K \bar{K}$ there are no quasi-two-body 
$q \bar q$ modes open to the
$\eta(1295)$. Consequently the decays must
proceed through the weaker direct three-body and virtual two-body channels
such as $a_0^{q\bar q}\pi$ and $f_0^{q\bar q}\eta$.

It is interesting to note the r\^ole that the 2S initial wavefunction
has played in our discussion.
Suppose for illustration that we had instead used 1S 
wavefunctions for the $\pi(1300)$ and $\eta(1295)$;
we would then have predicted
partial widths of several hundred
MeV into the low-energy tails of the modes 
$f_0^{q\bar q}\pi $ and 
$a_0^{q\bar q}\pi $, 
with consequent broad widths for the $\pi(1300)$
{\it and} the $\eta(1295)$, in contradiction with experiment.

$\bullet$ $\eta(1440)$

These successes raise provocative 
questions regarding the $\eta(1440)$ state(s). 
This is a purportedly complicated
region which may contain more than one resonance\cite{pdg94}.
The PDG width of the $\eta(1440)$ 
is only $\Gamma = 60(30)$ MeV, with signals 
reported in 
${\rm K}^* {\rm K} $, $a_0(980)\pi $,
$\eta(\pi\pi)_S$ and $\rho\gamma$. 

Except for $\rho\gamma$ 
these modes are not inconsistent with a dominantly $s\bar s$
state. The only two-body strong channel open for a 2$^1$S$_0$ $s\bar s$ 
$\eta(1440)$ 
is
${\rm K}^* {\rm K} $, but this could rescatter from
KK$\pi$ into the other reported modes 
$a_0(980)\pi$ and $\eta\pi\pi$.
The \3p0 model prediction for the 
partial width $\eta(1440) \to {\rm K}^* {\rm K}$ 
versus the wavefunction length scale $\beta$ is shown in Fig.3.
\begin{figure}
$$\epsfxsize=5truein\epsffile{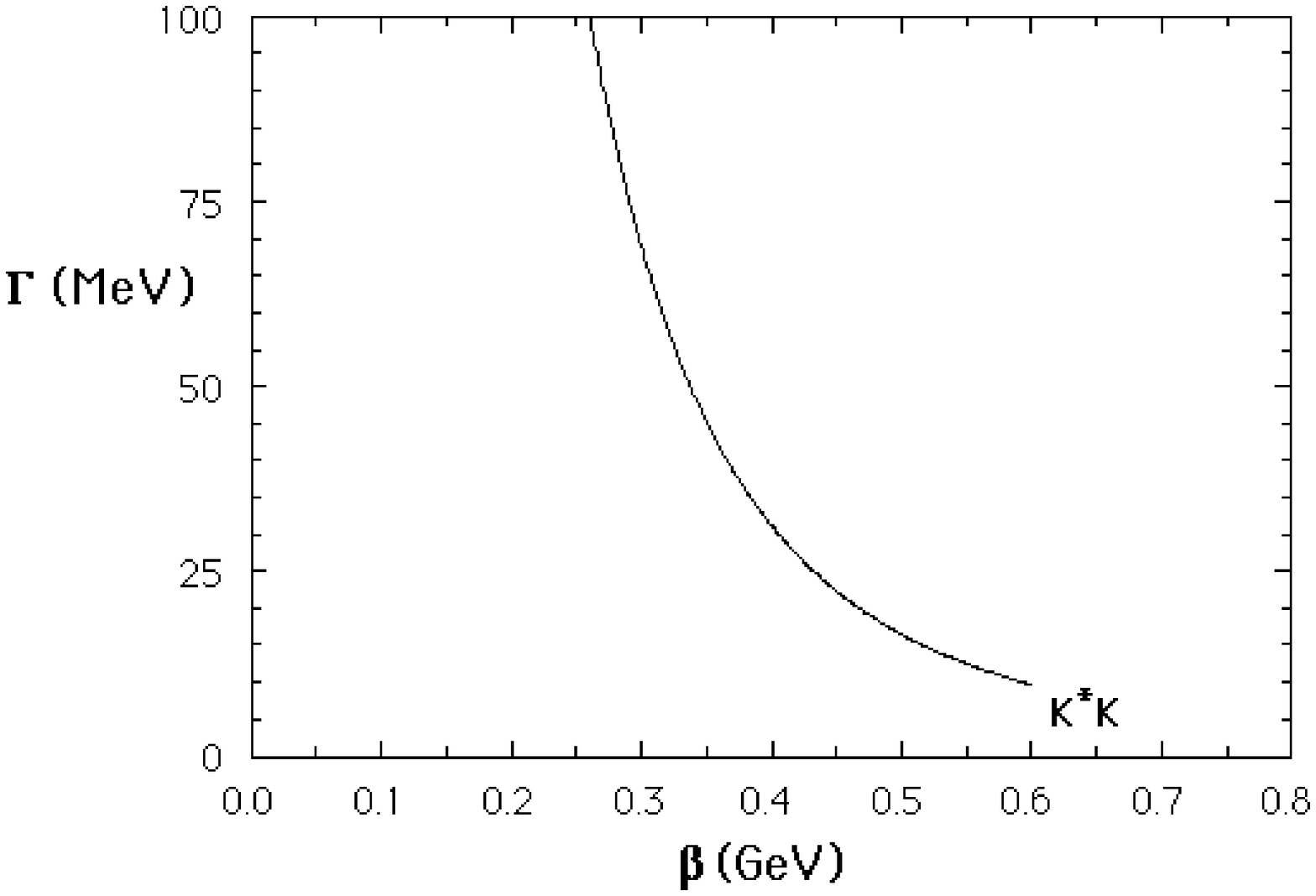}$$
{Figure~3.
The K$^*\bar {\rm K}$ + h.c. partial width of a 2$^1$S$_0$ $s\bar s$ 
$\eta(1440)$ in the \3p0 model.
Other two-body modes are excluded by phase
space.
}
\end{figure}
Evidently the predicted 
${\rm K}^* {\rm K} $ partial width is comparable to the observed
width, so a 2$^1$S$_0$ $s\bar s$ assignment appears possible for this
state. 

Of course the $\rho\gamma$ mode is not expected from $s\bar s$, and if confirmed
may imply large $n\bar n \leftrightarrow s\bar s$ mixing in this sector
as is observed in the 1S I=0 pseudoscalars. This
can be parameterized as
\begin{eqnarray}
&|\eta(1295)\rangle &= +\cos(\theta)|{n\bar{n}}\rangle  + 
\sin(\theta) |{s\bar{s}}\rangle  \\  
&|\eta(1440)\rangle &= -\sin(\theta)|{n\bar{n}}\rangle  + 
\cos(\theta) |{s\bar{s}}\rangle  \ .  \\   
\end{eqnarray}
A remeasurement of
$\eta(1440) \to \rho \gamma$, which should be possible at BEPC and TCF in
$\psi \to \gamma\gamma\rho $, would be 
very useful in clarifying the nature of this state. Ideally we would 
like to know the invariant mass distributions of $\rho\gamma$, $\omega\gamma$ 
and
$\phi\gamma$ final states, since these are flavor-tagging modes that allow
investigation of possible flavor mixing in the parent resonances. Similarly,
an accurate measurement of the branching fractions in the flavor-tagging 
$\psi\to V\eta(1440)$ and $V\eta(1295)$
hadronic decays, with $V=\omega, \phi$, would be
useful for the determination of the $n\bar n $-$s\bar s$ mixing angle.

In summary, from the total widths alone it is possible
to describe the $\eta(1295)$ and $\eta(1440)$ as unmixed $n\bar n$ and 
$s\bar s$ 2$^1$S$_0$ radial excitations. The report 
of a large $\eta(1440)\to\rho\gamma$ radiative mode however suggests 
flavor mixing between these states, and should be remeasured 
with
greater sensitivity together
with other $V\gamma$ modes. This mixing could also account for the 
large $\eta(1440)$ signal seen in $\eta (\pi\pi) $ by GAMS \cite{yup}. 

\subsection{$1^{--}$: 2$^3$S$_1$ and $^3$D$_1$ $\rho$ and $\omega$}

$\bullet${\bf $\rho(1465)$, $\rho(1700)$}

If one accepts that the $\pi(1300)$ and $\eta(1295)$ belong to a $2^1$S$_0$ 
$q\bar q$ nonet, it is then natural to assign the
$\rho(1465)$ and the $\omega(1419)$ 
\cite{pdg94,cd} to $2^3$S$_1$
states. Indeed, one expects the contact hyperfine interaction
to raise the mass of the vector nonet with respect to the pseudoscalar nonet
by approximately this amount \cite{god}.
It is unlikely that the vectors near 1.4-1.5 GeV are dominantly 
D-waves, 
since the $^3$D$_1$ $n\bar n$ states should 
lie close to the other 1D candidates such as the 
$\pi_2(1670)$,
$\rho_3(1691)$
and 
$\omega_3(1667)$.
In the Godfrey-Isgur potential model 
a mass of 1660 MeV was predicted for the $^3$D$_1$ state,
whereas they expect the
2$^3$S$_1$ radial excitation at 1450 MeV \cite{god}.
The $\rho(1465)$ also lies
well below flux-tube model expectations of
M$_H (1^{--}) \approx 1.8$-$1.9$ GeV\cite{paton85,bcs}
for vector hybrids, so although the possibility of
light vector hybrids has been discussed \cite{cp95,kalash}, these do not appear
likely unless the flux tube model for hybrids is misleading. 

The experimental branching fractions of these $1^{--}$ states 
are somewhat obscure, because there are at least two broad,
overlapping resonances in each flavor sector in this mass region.
The status of these vector states as seen in $e^+e^-$ annihilation 
was reviewed recently by Clegg and
Donnachie \cite{cd}.
In the $\rho$ sector they find that 
at least two states are present. The lighter state is
assigned a mass of M $ = 1.463(25)$ GeV and a width of 
$\Gamma = 0.311(62)$ GeV;
it couples strongly
to $4\pi$ states (including $a_1\pi$ but not $h_1\pi$) and $\omega\pi$,
and less strongly to $\pi\pi$. The higher state has 
M $ = 1.73(3)$ GeV,
$\Gamma = 0.40(10)$ GeV, 
couples most strongly to $4\pi$ ($a_1\pi$ and
$h_1\pi$ are not separated) and perhaps $6\pi$; $\pi\pi$ is also
important, but the $\omega\pi$ width is found to be small.

These states have also been reported recently by Crystal Barrel
\cite{cbrhor} in $\pi^-\pi^o$ states in $\bar p d\to
\pi^-\pi^o\pi^op$; both vectors appear in $\pi^-\pi^o$, with masses and
widths of 
M $ = 1.411(10)(10)$ GeV,
$\Gamma = 0.343(18)(8)$ GeV, 
and
M $ = 1.780{+34\atop -25}(14)$ GeV,
$\Gamma = 0.275(42)(17)$ GeV, quite similar to the $e^+e^-$ results.

The \3p0 model predictions for pure 2$^3$S$_1$ and $^3$D$_1$ $\rho$ states
at 1.465 GeV and 1.700~GeV are given in Table I (see also
Tables B1, B8), together 
with flux tube model predictions for a hypothetical 1.5~GeV vector hybrid. 
Very characteristic differences between the states are evident in
their couplings to $4\pi$ final states;
2S couples very weakly to these,  1D couples strongly to
both $a_1\pi$ and $h_1\pi$, and the hybrid couples strongly to $a_1\pi$
but not to $h_1\pi$. 
Both quarkonium states have moderately large couplings 
to $\pi\pi$ and $\omega\pi$,
whereas the hybrid couples strongly only to 
$a_1\pi$. 

\begin{table}
\caption{
Partial widths of 2S, 1D and hybrid $\rho$ states.} 
\label{tabrho}
\begin{tabular}{lccccccccc}
 & $\pi\pi$ & $\omega\pi$  & $\rho\eta$ & $\rho\rho$ & KK & K$^*$K
& $h_1\pi$ & $a_1\pi$  & total\\
\hline
$\rho_{2S}(1465)$ & 74. & 122. & 25. & - & 35. & 19. & 1. & 3. & 279.  \\
$\rho_{1D}(1700)$ & 48. & 35. & 16. & 14. & 36. & 26. & 124. & 134. & 435.  \\
$\rho_H(1500)$ & 0 & 5 & 1 & 0 & 0 & 0 & 0 & 140 & $\approx 150$  
\end{tabular}
\end{table}

Note that the $|q\bar q\rangle$ components are spin {\it triplet} whereas
the hybrid is spin {\it singlet}. This difference in spin underlies 
the characteristic pattern of branching fractions in Tables I and II.

Although there are many similarities between theory and experiment, 
there are problems in detail.
The important couplings of the lighter state to $\pi\pi$ and $\omega\pi$
found by Clegg and Donnachie are consistent with a 2S quarkonium, 
but we do not expect a significant coupling of a 2$^3$S$_1$ $\rho$ to
$4\pi$ final states. The dominant coupling of the heavier state to
$4\pi$ is as predicted for the D-wave quarkonium, but the 
reported absence of
$\omega\pi$ is not expected. 
The presence of two states (2$^3$S$_1$ and $^3$D$_1$)
in $\pi\pi$ with comparable strengths, 
reported by Crystal Barrel \cite{cbrhor}, is expected. 

Of course
it is difficult 
to distinguish the contributions from two broad states
with similar masses,
and the $4\pi$ final states themselves have not yet been
completely characterized. 
(The $a_1\pi$ and $h_1\pi$ modes of the $\rho(1700)$ in $e^+e^-$ for example 
have not been separated.)
It appears likely that the states and their branching fractions
are still inadequately resolved experimentally in this mass region, 
so it is not yet appropriate
to attempt a detailed fit, using for example 
linear combinations of the 2S and 1D basis states.

It is clear from our \3p0 results that in future 
it will be important to separate
the $a_1\pi$ and $h_1\pi$ contributions
(which tag 1D and H \cite{cp95,kalash} states), 
and that the $\pi\pi$ and $\omega\pi$ distributions
should also be studied carefully, since these are expected
to arise mainly from quarkonia rather than hybrids. 

$\bullet$ {\bf $\omega(1419)$ and $\omega(1649)$}

We anticipate similar problems with at least two broad overlapping resonances
in the I=0 sector. Clegg and Donnachie \cite{cd} discuss both one- and two-resonance
fits to the $\omega$ sector in the reactions $e^+e^- \to \rho\pi$ and
$\omega\pi\pi$. In their two-resonance fit they find a lower state with a
mass and width of
M$ = 1.44(7)$ GeV,
$\Gamma = 0.24(7)$ GeV,  
and a higher, quite narrow state with
M$ = 1.606(9)$ GeV,
$\Gamma = 0.113(20)$ GeV. 
The PDG quote masses and widths of
M$ = 1.419(31)$ GeV,
$\Gamma = 0.174(59)$ GeV,  
M$ = 1.649(24)$ GeV,
$\Gamma = 0.220(35)$ GeV; the parameters for the lighter state are
consistent but the width of the higher-mass $\omega$ state is broader than
Clegg and Donnachie estimate. 

Clegg and Donnachie find that both $\omega$
states couple strongly to $\rho\pi$. Only the second is found to
couple to $\omega\pi\pi$, and that coupling is rather weak. 
A fit with a single resonance finds instead that the $\omega\pi\pi$
branching fraction exceeds $\rho\pi$, so these should be
regarded as tentative conclusions.

\begin{table}
\caption{
Partial widths of 2S, 1D and hybrid $\omega$ states.} 
\label{tabomega}
\begin{tabular}{lcccccc}
 & $\rho\pi$ & $\omega\eta$  &  KK & K$^*$K & $b_1\pi$ & total \\
\hline
$\omega_{2S}(1419)$ & 328. & 12. & 31.  & 5.  & 1.   & 378.   \\
$\omega_{1D}(1649)$    & 101. & 13. & 35.  & 21. & 371. & 542.   \\
$\omega_H(1500)$    & 20     &  1   & 0   & 0    & 0     &  $\approx 20$
\end{tabular}
\end{table}

For comparison we again show the numerical predictions of the \3p0 model
for pure 2S, 1D and H states. The masses assumed are 1996 PDG 
values (see Tables B1 and B9). 
The large $\rho\pi$ couplings reported for the vector
states are evidently consistent with expectations for both 2S and 1D
quarkonia. Again the S+S modes are predicted to be small for
a hybrid, so they can be used to tag quarkonia or the $q\bar q$ components
of mixed states. 
Since none of the favored S+P modes is open to an I=0 hybrid at
1.5 GeV, such a state would be quite narrow, as shown in Table II. 
(The decay
$\omega_H\to b_1\pi$ 
is excluded by the 
``singlet selection rule'' 
\cite{cp95,abs}, which states 
that $({\rm S}_{q\bar q}=0) \not\to ({\rm S}_{q\bar q}=0) 
+ ({\rm S}_{q\bar q}=0) $ in the \3p0 model;
the $\omega_H$ hybrid has
${\rm S}_{q\bar q}=0$ in the flux tube model.
Interestingly, the singlet selection rule holds for 
both \3p0 and OGE quarkonium decay amplitudes
\cite{abs}.) 

A hybrid in this mass region should be visible as a narrow bump in the
$\rho\pi$ invariant mass distribution. (This channel is not favored for a
hybrid, but it is allowed at a reduced rate due to different $\rho$ and $\pi$ 
spatial wavefunctions.) 
Thus it may be useful to search $\rho\pi$ final states for narrow resonances
with improved statistics, although the signal would of course be
broadened by
the $\rho$ width.

The very large $b_1\pi$ mode predicted for the 1D quarkonium
is very interesting, because neither 2S nor hybrid vector states are expected
to couple significantly to $b_1\pi$. This two-body mode 
will appear as $\omega\pi\pi$;
Clegg and Donnachie do report an $\omega\pi\pi$ mode for their higher
$\omega$ state, but the coupling is not as strong as we predict. The
total width of their higher-mass state is also 
much smaller than expected. Since the 1D state
is predicted to have a very large width, $\approx 500$ MeV (Table B9),
this discrepancy may be due to a distortion of the
shape by threshold effects, with resulting inaccuracies in the
reported couplings.
Assuming that the \3p0 model
predictions are approximately correct, a study of the 
$1^{--}$ $\omega\pi\pi$ mass distribution should reveal the $^3$D$_1$ 
$\omega$ basis state in 
isolation. (It may be distributed over several resonances.) 
If the quasi-two-body approximation is correct, the
mass distribution of $\omega\pi$ pairs in the resonance contribution to
$\omega\pi\pi$  
should be consistent
with a $b_1(1231)$. 

\subsection{Mixing in the $1^{--}$ sector.}

Although we have considered the decay modes of pure 2S, 1D and H vector
states, the physical resonances are certainly linear combinations
of these and other basis states. Since the known resonances have
similar masses, we should consider the possibility that there is
significant mixing and introduce the linear combination

\begin{equation}
|{\rm V}\rangle =
\cos(\theta) \bigg( \cos(\phi) |2^3{\rm S}_1 \rangle +
\sin(\phi) |{}^3{\rm D}_1\rangle \bigg) + \sin(\theta) |{\rm H}\rangle \ .
\end{equation}

The mixing angles for each resonance can be determined from the
branching fractions to certain states. The S+S modes identify the
$q\bar q$ components of the state (see Tables~I and II). In the I=1 states
the $4\pi$ modes $a_1\pi$ and $h_1\pi$ are similarly characteristic;
the $h_1\pi$ mode is produced
only by the 1D basis state, and $a_1\pi$ comes from both 1D and hybrid
states. Similarly in I=0 the mode $b_1\pi$ tags the 1D quarkonium basis
state and 2S and 1D states both lead to strong $\rho\pi$ couplings.
Determination of the mixing angles in the physical states will be
possible given accurate measurements of the branching fractions to these 
characteristic modes. 

We have not carried out a fit to determine the mixing angles because the 
experimental results do not yet appear definitive. However we note that the
partial widths reported by Clegg and Donnachie for the
$\rho(1465)$, which include a large
$\Gamma_{a_1\pi}$  and a small $\Gamma_{h_1\pi}$, are inconsistent 
with 2S or 1D alone. These widths imply
a large H component in this state with the
possibility of considerable H-2S mixing.

Future experimental work could concentrate on an accurate determination
of the $\pi\pi$, $\omega\pi$, $h_1\pi$ and $a_1\pi$ branching fractions 
of the $\rho$ states. The $h_1\pi$ and  $a_1\pi$ modes are 
especially sensitive to the
nature of the initial state. Similarly the  
$\rho\pi$ and $b_1\pi$ branching fractions
of the $\omega$ states are the most interesting experimentally.

\section{3S States}

\subsection{$0^{-+}: \; 3^1$S$_0\ \pi(1800)$}

The same experiments\cite{ves951,ves,bellini,veseta}
 that see the $\pi(1300)$ in $\rho\pi $ and a possible
broad enhancement in $\pi (\pi \pi)_S$ also report
a prominent $\pi(1800)$ in $f_0(980)\pi $,
$f_0(1300)\pi $, $f_0(1500)\pi $ and
K$({\rm K} \pi)_S$. None of these experiments see the  $\pi(1800)$ 
in $\rho\pi$. This is striking, as also is the fact that the total width 
of $\approx 150$-$200$ MeV is considerably smaller than that of the $\pi(1300)$.
Furthermore, the presence of clear signals
in both $f_0(1300)\pi $ and $f_0(980)\pi $ is remarkable and was commented 
upon with some surprise\cite{ves}.

The decays into $\pi \rho$ and KK$^*$ are both suppressed; VES quote 
the limits \cite{ves} 

\begin{equation}
\frac{\pi(1800) \rightarrow \pi^- \rho^0}{\pi(1800) \rightarrow\pi^- f_0(980)_{|\to
\pi^+\pi^-}}
\;  < \; 0.14 \; \ \ (90 \% \ {\rm c.l.})
\end{equation}
and

\begin{equation}
\frac{\pi(1800) \to {\rm K}^- {\rm K}^*}{\pi(1800) 
\to {\rm K}^-{\rm K}^+ \pi ({\rm S-wave})} \;
 < \; 0.1 \; \ \ (95 \% \ {\rm c.l.})  \ . 
\end{equation}

A prominent KK$^*_0$ signal is present (observed as ${\rm K}({\rm K}\pi)_S$), so the
virtual transition
$\pi(1800) \to {\rm KK}^*_0 \to {\rm K}{\rm K}\pi \to f_0(980)\pi$ is probably
responsible for the coupling to $f_0(980)\pi$; this mode appears to be stronger than
$f_0(1300)\pi$.
The mass of this state makes it a candidate for either the radial $3^1$S$_0$ or 
the ground
state hybrid $\pi_H$. 
The predicted branching fractions for  $3^1$S$_0$ (Table B4) and 
$\pi_H$ hybrid states (from Ref.\cite{cp95}) near this mass are shown in 
Table III.

\begin{table}
\caption{
Partial widths of 3S and hybrid $\pi(1800)$ states.} 
\label{tabpi1800}
\begin{tabular}{lccccccc}
 & $\rho\pi$ & $\rho\omega$  &  $\rho(1465)\pi$ &  $f_0(1300)\pi $ & $f_2\pi$ 
& K$^*$K & total \\
\hline
$\pi_{3S}(1800)$ & 30. & 74. & 56.  &  6.   & 29.  & 36. & 231.  \\
$\pi_H(1800)$    &  30     &  0    &  30  &  170 & 6 & 5  
& $\approx 240$   
\end{tabular}
\end{table}

The decay amplitude for 
$3^1$S$_0 \to {}^3$S$_1 + {}^1$S$_0$ is actually close to a node
with these masses, so the
weak coupling to $\rho\pi$ is expected for both a 3S quarkonium and a hybrid.
The most important differences are in the $\rho\omega$ 
and $f_0(1300)\pi$ modes: $\rho\omega$ is predicted to be the largest
mode of a 3S $\pi(1800)$ state, whereas for a hybrid 
$\pi_H(1800)\to\rho\omega$ should be very weak 
(this is the usual selection rule against S+S final states). 
Conversely, $f_0(1300)\pi$ is predicted to be 
weak for 3S quarkonium but is expected
to be the dominant decay mode of a $\pi_H(1800)$ hybrid.
The observation
of a large $f_0(1300)\pi$ mode argues in favor of a hybrid assignment for
this state. One should note however that the \3p0 model also predicts
a small branching fraction for $\pi(1300)\to \pi (\pi\pi)_S$; if the observed
$\pi (\pi\pi)_S$ signal is really due to the $\pi(1300)$ rather than the 
Deck effect, the decay model may simply be inaccurate for 
N$^1$S$_0 \to ^1$S$_0 + ^3$P$_0$
transitions. 
There may for example be large OGE decay amplitudes in these
channels, as was found in the related
transition $^3$P$_0\to ^1$S$_0 + ^1$S$_0$ \cite{abs}; this can be checked in
a straightforward calculation \cite{bs}.
Thus the presence of a strong $\pi(1800)\to f_0(1300)\pi$
mode is indicative of a hybrid {\it assuming} that the \3p0 model
is accurate.

Although the strong $f_0(1300)\pi $ signal in the VES data may 
well have isolated the
$\pi_H(1800)$ hybrid, VES also finds evidence for a large $\rho\omega$ 
signal at a similar mass\cite{khok}. We expect $\rho\omega$
to arise from the 3S $\pi(1800)$ quarkonium state rather than from a
hybrid. 
These signals may be due to two different resonances;
the $\rho \omega$ signal 
is evident well below 1800 MeV, and persists to higher mass than the 
$f_0(1300)\pi $ distribution. Similarly the mode $f_2\pi $ is observed 
(Fig.4d of Ref.\cite{ves951}),
but at a mass of $\approx 1700$ MeV, well below the $\pi(1800)$ seen in
$f_0(1300)\pi $. This may also indicate a 3S state somewhat below a
hybrid $\pi(1800)$.
If two  
$0^{-+}$ $\pi$ 
resonances were to be isolated in this region, this would be strong
evidence through overpopulation 
for both a hybrid and a 3S $q\bar q$ excitation. 

Further
investigation of the modes $\rho\pi$, $\rho(1465) \pi$, 
$\rho\omega$, $f_0(1300)\pi$ and $f_2\pi$ could be useful
to clarify the resonances in the region of the
$\pi(1800)$; establishing the
branching fractions to these states is
especially important. The most characteristic are
$\rho\omega$ and $f_0(1300)\pi$, 
since the hybrid and 3S quarkonium predictions differ greatly
for these modes. 
Theoretical studies
of the stability of the decay amplitudes
under variation of parameters and wavefunctions and the
assumed decay mechanism \cite{abs} would also be interesting.

Searches for the multiplet partners of this state may be useful, since they
too have characteristic decay modes. A 3S $n\bar n$ 
$\eta(1800)$ quarkonium for example 
(Table B4) is predicted to have
large $\rho\rho$ and $\omega\omega$ modes, which should be zero for a hybrid.
An $\eta(1760)$ which couples to $\rho\rho$ and $\omega\omega$ was 
reported by 
MarkIII \cite{MIIIeta} 
and by 
DM2 \cite{DM2eta}.
The
conclusions regarding the presence of this pseudoscalar signal
in the MarkIII $4\pi$ data have since been disputed \cite{bsz}.

\subsection{$1^{--}: \; 3^3$S$_1 $}

If the $\pi(1800)$ is a 3S quarkonium we should expect to find 3S vector states
near 1.9~GeV. No candidates for these states are known at present below 2.1
 GeV, however there are possible $\rho$ candidates at 2150 and 
2210 MeV\cite{pdg94}. 
The predictions for decays
of 3S vectors are given in Table B3; it is notable that the simple S+S modes
have small couplings, with the exception of $\rho(1900)\to\rho\rho$.
Unfortunately the relatively obscure 2S+S modes are favored, especially for the
$\omega(1900)$. Some S+P modes have sufficiently strong couplings 
to the 3S vectors
to be attractive 
experimentally, notably $\rho(1900)\to a_2\pi$ and $\omega(1900)\to b_1\pi$.
As noted previously,
the $b_1\pi$ mode is forbidden to an $\omega$ vector hybrid
by the singlet selection rule, since this hybrid decay would have 
S$_{q\bar q}=0$ for
all states.
 
\section{2P States}

The 2P states are especially important because the expected mass of this multiplet
($\approx 1700$ MeV) 
is close to the predicted mass of the lowest hybrid multiplet
in the flux tube model,
$\approx 1.8$-$1.9$ GeV \cite{paton85,bcs}. 
Furthermore, the position of the 1P and 2P unmixed $n\bar n$ levels
and the 1P $s\bar s$ level are needed for input to quarkonium - 
glueball mixing studies\cite{cafe}
 based on the lattice expectations for glueballs in
this region\cite{glueball}. Determining the nature of the $f_J(1710)$ will
be important in this regard.
Since the quantum numbers 
$1^{++}$ and
$1^{+-}$ occur in both the hybrid and 2P multiplets, 
these states need to be identified to avoid confusion with 
hybrids. 
As we shall see, a recently discovered $1^{++}$ state, the $a_1(1700)$, appears
to be our first confirmed member of the 2P multiplet, in that it 
passes a very nontrivial \3p0 model 
amplitude test and thereby for the first time establishes the mass scale 
of the 2P multiplets.

\subsection{$1^{++}: \;  2^3$P$_1 \;  a_1(1700)$}

A recent experiment at BNL \cite{bnl1} reported 
a candidate $1^{-+}$ exotic, produced by 
$\pi \rho$ 
and decaying to $\pi f_1$. They also see a 
$1^{++}$ state in this channel 
at $\approx 1.7$ GeV, with a width of $\approx 0.4$ GeV;
the relative phase of the $1^{++}$ and $1^{-+}$ waves was used to support
the claim of a resonant $1^{-+}$. A similar $1^{++}$ signal has been reported
by VES in $\rho\pi$ \cite{ves951,ves}.

The challenge is to establish whether this $1^{++}$ $a_1(1700)$ is a 
hybrid $a_{1(H)}$ 
(perhaps a partner of the reported $1^{-+}$ exotic) or a radial
$2^3$P$_1$ $n\bar n$ state.  
The predicted total width of a $1^{++}$ $a_1(1700)$ 
hybrid in the model of Close and Page \cite{cp95} is 
$\approx 300$ MeV, comparable to the observed width.
However the total width predicted for a
$a_1(1700)$ $2^3$P$_1$ $n\bar n$ state is similar, about 250 MeV
(see Table B5).
Some differences between these 
assignments are evident when we compare partial widths
(see Table IV).
\begin{table}
\caption{
Partial widths of 2P and hybrid $a_1(1700)$ states.} 
\label{taba11700}
\begin{tabular}{lccccccccc}
 & $\rho\pi$ & $\rho\omega$  &  $\rho(1465)\pi$ & $b_1\pi$ & $f_0(1300)\pi $ 
& $f_1\pi$ 
& $f_2\pi$ 
& K$^*$K & total \\
\hline
$a_{1(2P)}(1700)$  & 57.  & 15. &   41.  & 41.  & 2.   & 18.  & 39. & 33. & 246.    \\
$a_{1(H)}(1700)$    & 30  & 0    & 110     & 0     & 6     
& 60 
& 70 
& 20 
& $\approx 300$ 
\end{tabular}
\end{table}
Clearly the 2P state couples more strongly to S+S 
modes than does the hybrid, as usual,
so an accurate determination of 
the branching fractions to $\rho\pi$ and $\rho\omega$ 
would be interesting. The other modes are
less characteristic with the exception of $b_1 \pi$, 
which should come exclusively
from the quarkonium state.
The absence of the decay $a_{1(H)}\to b_1\pi$ 
is a special case of the singlet selection rule 
cited 
previously as forbidding the transition
$\omega_H \to b_1\pi$.
We therefore urge that
experiments that observe $a_1(1700) \rightarrow \pi f_1$ also seek
a signal, or a limit, for $a_1(1700)\to \pi b_1$.

A crucial test of 2P versus H assignments for the $a_1(1700)$ arises in the
decay amplitudes to $\rho\pi$. From Appendix A, Eqs.(A53,A58,A59), the
transition $2 {}^3{\rm P}_1 \to {}^3{\rm S}_1 + {}^1{\rm S}_0 $ has both S and D
amplitudes, and the D/S ratio is (where $x \equiv |\vec{p}_f|/\beta$)

\begin{equation}
{{\rm D}\over {\rm S}}\bigg|_{2 {}^3{\rm P}_1 \to {}^3{\rm S}_1 + {}^1{\rm S}_0} =
-{ 2^{1/2} 7 \over 3^2 5} {x^2 (1-{2\over 21} x^2 ) \over 
(1 - {4\over 9}x^2 + {4\over 135} x^4 )}  \  .
\end{equation}
The inverse of this ratio is shown versus $\beta$ in Fig.4; note that the S-wave
amplitude has a zero very close to the preferred value $\beta=0.4$ GeV. This is
a striking and unusual result, 
since in most cases we find that the lower partial
waves are dominant. 
In contrast, for a hybrid one expects S-wave dominance,
$a_{1(H)} \to (\rho \pi)_S:(\rho \pi)_D \approx 20:1$.

\begin{figure}
$$\epsfxsize=5truein\epsffile{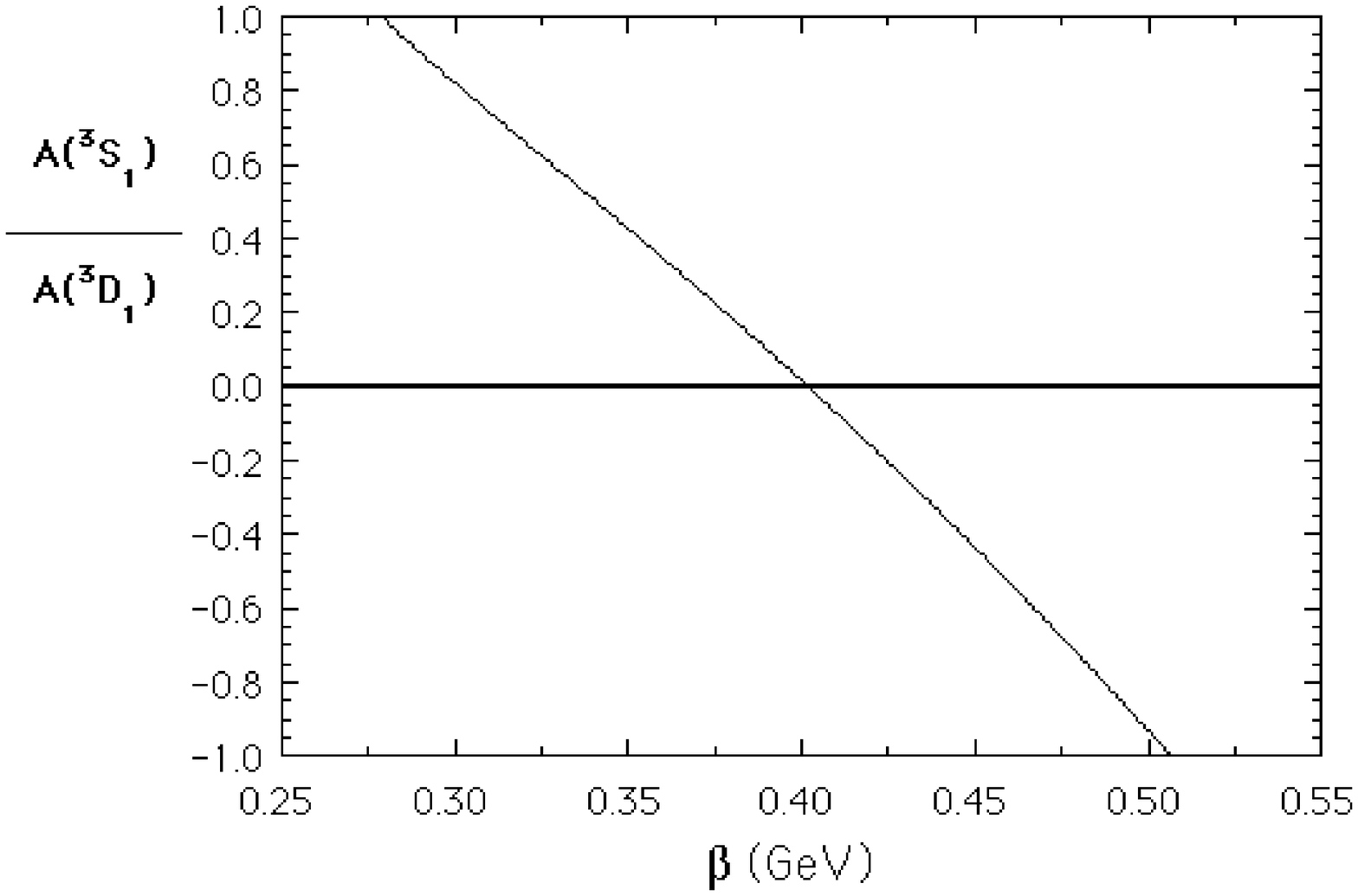}$$
{Figure~4.
The S/D amplitude ratio 
in the transition 2$^3$P$_1$ $a_1(1700)\to\rho\pi$ 
predicted by the \3p0 model.
}
\end{figure}

Experimentally, 
VES sees the $a_1(1700)$ prominently in 
the $\rho\pi$ D-wave 
(see Fig.2c of Ref.\cite{ves951}); 
the resonance near 1.7 GeV dominates the
entire 1-2 GeV region. In contrast, the $\rho\pi$ S-wave 
(Fig.2a of \cite{ves951}) 
is dominated by the $a_1(1230)$ and shows no clear evidence for the
$a_1(1700)$. E852 similarly sees this resonance clearly in
the $\rho\pi$ D-wave, with a mass and width of 
M $\approx 1.66$ GeV and
$\Gamma\approx 0.22$ GeV \cite{suchung}. 
This D-wave dominance of the $\rho\pi$ final state
appears to be dramatic confirmation
that the $a_1(1700)$ is a 2$^3$P$_1$ radial excitation.
Furthermore the successful predictions of $a_1 \to \rho \pi$ being in S wave
and $a_{1R} \to \rho \pi$ being in D wave supports the extension of the
model to radial excitations.

With the $a_1(1700)$ established as a 2P $n\bar n$ state, the multiplet partners
are expected nearby in mass (multiplet splittings due to spin-orbit and tensor
forces appear to be small even at
L$_{q\bar q}=1$) and searches for these states should be carried out.
In the next sections we will discuss the decay modes predicted for these
other 2P states.

\subsection{$0^{++},2^{++} \ 2^3$P$_0, 2^3$P$_2 \; $: 
$a_0(1700),
a_2(1700)$
}

With the $a_1(1700)$ as the 2$^3$P$_1$
``$a_{1R}$'' radial state, one may ask why the
$a_{0R}$ and $a_{2R}$ 
partners are not seen in the same experiments. A simple 
explanation 
follows from the partial widths shown in Table B5.
Since the production mechanism of the 
$a_1(1700)$ in $\pi p \to \pi f_1 p$ 
apparently involves natural parity exchange (probably $\rho$ or $f_2$ exchange),
the $0^{++}$ scalar state $a_{0R}$ cannot be produced. Although the $2^{++}$
$a_{2R}$ can be produced (note the large $\rho\pi$ coupling), it has a
weak coupling to the $\pi f_1$ final state and hence is not readily observable
in this channel. 

There is some very recent evidence for a 2$^3$P$_2$ state from the
Crystal Barrel, who report an $a_2(1650)$ in $\eta\pi^o$ 
final states in $p\bar p \to \eta \eta \pi^o$ \cite{Degener}. Although we
expect $\eta\pi$ to be a relatively minor mode, with a branching fraction
of 7\%, the mass and reported width of $\Gamma=260(15)$ MeV 
are consistent with expectations (Table B5). 
The final states $\rho\pi$ and $\rho\omega$ are predicted
to have large couplings to an $a_{2R}$ state, so we expect a large
signal in these $ 3 \pi$ and $5\pi$ 
final states.

The prediction of a large 
coupling to vector meson pairs suggests $\gamma \gamma \to 2^3$P$_{\rm J}
\to VV$ as a possible source of the $a_{0R}$ and $a_{2R}$  states.
Indeed, ARGUS has evidence that the $\rho\omega$ final state near threshold
is mainly in the partial wave J$^{PC}=2^{++}$, J$_z=2$, 
and the $\gamma\gamma\to \rho^o\omega$ cross section is at maximum near 1.7 GeV
\cite{argusa2r}.
The J$_z=2$ signal is characteristic of a $2^{++}$
resonance, as there is a selection rule\cite{abc} that
$\gamma \gamma \rightarrow ({\rm J}=2^{++}, \lambda=0) = 0$ in the 
nonrelativistic quark model; hence $\lambda=2$
dominates. A study of $\gamma \gamma
\rightarrow 5 \pi$ with improved statistics, perhaps at LEP2, 
may help to isolate these states.
Of course the interpretation of any $\gamma\gamma \rightarrow VV$ 
reaction should be regarded as tentative until the large
$\gamma\gamma\to\rho^o\rho^o$ signal \cite{mpwrev} is understood, as this 
reaction also is dominated by J$^{PC}=2^{++}$, J$_z=2$, 
but contains both I=0 and I=2 projections in s-channel and hence
cannot come from a single $q\bar q$ resonance. 
Finally, the reaction 
$\gamma \gamma \to a_{0R} \to \pi b_1$ may also lead to 
a significant signal in $5 \pi$ final states, and could be isolated if
the $\lambda = 0$ selection rule is used to suppress the $a_{2R}$ signal.

\subsection{$2^{++} \ 2^3$P$_2 \;  $: $f_2(1600 - 1800)$}

Encouraged by the likely confirmation of the radial $1^{++}$ $a_1(1700)$, 
we now turn our attention to the 2P isoscalar multiplet. First we consider the
$f_2(1700)$ 2$^3$P$_2$ $n\bar n$ radial tensor.
We predict a large $\rho\rho$ width 
for the 2$^3$P$_2$ $f_2(1700)$, and the modes
$\omega \omega$, $\pi \pi$ and perhaps $\pi a_2$ should also be
important (see Table B6).
(Note that the simple branching fraction ratio 
$\rho\rho/\omega\omega \approx 3$ follows trivially from flavor counting.)
The total width is predicted to be $\approx 400$ MeV.

Although there is no strong evidence for such a state, there are
suggestions of its presence in several processes. A large 
$2^{++}$ enhancement referred to as the $X(1600)$, with $\Gamma=400(200)$ MeV, 
is well
known in $\gamma\gamma\to\rho^o\rho^o$ \cite{pdg96,argus91}. The small 
charged to neutral $\rho\rho$ ratio however precludes the identification of
this signal with a single $f_2(1700)$ resonance. 
There are also reports of
a rather narrow $f_2(1640)$ with a width of  $\approx 60$-$120$ MeV in
$\omega \omega$\cite{pdg96,argusww,ves92,obelix92}. 
Although the predicted 2$^3$P$_2$ $f_2(1700)$ width is much larger, it would be 
reduced somewhat by threshold effects in the $\omega\omega$ channel.
Indeed, if the resonance mass is around 1700 MeV and its width is 
several hundred 
MeV, as suggested by our analysis, it may decay strongly into $\rho \rho$
(due to the large $\rho$ width leading to a favorable phase space), but 
the narrowness of the $\omega$ may cause only
the upper part of the resonance to feed the $\omega \omega$ channel.
Thus the resonance width in $\omega\omega$ may appear smaller than in
$\rho\rho$, so
both the X(1600) and the $f_2(1640)$ may be aspects of a single state.

A recent reanalysis of MarkIII data on $\psi\to\gamma\pi^+\pi^+\pi^-\pi^-$
\cite{bsz} similarly sees evidence of a $2^{++}$ state near
M $=1.64$ GeV, with
$\Gamma=0.14$ GeV, which couples strongly to $\rho\rho$. (In contrast 
they observe $0^{++}$ states dominantly in $\sigma\sigma$.) This preference
of the tensor state for $\rho\rho$ is consistent with \3p0 model
expectations for a 2$^3$P$_2$ $f_2(1700)$ state (Table B6).
 
Finally, it is possible that the 
$f_2(1520)$ or ``AX" state seen
in $p\bar{p} \to 3 \pi$ \cite{axstate} may be
the low-mass tail of the $f_2(1700)$.

\subsection{$0^{++}\ 2^3$P$_0 \; $: $f_0(1500), f_{\rm J}(1710)$}

The $0^{++}$ $f_0$ 
sector in the 1.5 GeV mass region is clearly of interest for glueball
searches. It is thus important to identify the $^3$P$_0$ 
quarkonia in this mass region.
We stress that one should not be overly naive in this endeavor since strong
recoupling effects, including couplings of quarkonia to nearby glueballs, 
are expected \cite{cafe}. 
Nonetheless for initial theoretical guidance it will be useful to
consider the predictions of the naive \3p0 model for the decays of unmixed
\3p0 $n\bar n$ quarkonia. 

The decays predicted for the 2P scalar $f_0(1700)$ state in the \3p0 model are
given in Table B6. Fortunately they are very characteristic. The dominant modes
are $\rho\pi\pi$, with approximately equal contributions from $\pi(1300)\pi$ and
$a_1(1230)\pi$. The channels 
$\rho\rho$ and $\pi\pi$ are also important, and the total width
is predicted to be $\approx 400$ MeV. The $\eta\eta$ and KK amplitudes are both close to nodes and are predicted to be quite small.

The two well known scalar resonances in this mass region which can be 
compared to these predictions are the glueball candidate $f_0(1500)$ and
the $f_{\rm J}(1710)$.
These states have PDG masses and total widths of 
M $=1503(11)$ MeV, $\Gamma = 120(19)$ MeV and 
M $=1697(4)$ MeV, $\Gamma = 175(9)$ MeV; both are rather narrow relative to 
expectations for a 2P $n\bar n$ state.
BES has recently reported\cite{besnew} a spin parity analysis of the
K$^+$K$^-$ system in $\psi$ radiative decays; they see both J=0 and J=2
states. Both have widths of $\approx 100$~MeV, much narrower than we
expect for 2P $n\bar n$ states. The presence of a significant 
$\eta\eta$ mode
for both the $f_0(1500)$ and $f_{\rm J}(1710)$
argues against a 2P $n\bar n$ assignment. The possibility 
that a node in the 2P decay amplitude is consistent with
the observed weakness of $f_{\rm J}(1710)\to\pi\pi$ is found to be 
unrealistic in practice; although there are actually two nodes, the modes
that are strongly suppressed by these in the \3p0 
model are $\eta\eta$ and KK, not
$\pi\pi$.

The disagreement of predicted decay modes of 2P $n\bar n$ 
states with experiment
for the $f_0(1500)$ and $f_{\rm J}(1710)$ 
supports the suggestions that neither
of these states is a quarkonium.
Amsler and Close \cite{cafe} 
have noted that the $f_0(1500)$ could be a glueball that is mixed with 
the nearby
$n\bar n$ and $s\bar s$ basis states, which explains the observed 
branching fractions. Conversely, Weingarten \cite{wein} 
suggests that the $f_{\rm J}(1710)$
is the scalar glueball, based on its mass and on lattice QCD evidence that 
flavor symmetry may be inaccurate in glueball decays, together with a different
pattern of $q\bar q\leftrightarrow G$ mixing. 
It may be that the glueball, $n\bar n$ and $s\bar s$ basis states are
all strongly mixed in this sector, so that an assumed separation
into glueball and quarkonium states is inaccurate \cite{glynnys}.

An alternative suggestion is that the $f_{\rm J}(1710)$ may be a vector-vector
molecule, analogous to the $f_0(980)$ and $a_0(980)$ 
K$\bar{\rm K}$ candidates. 
The two possibilities discussed in the literature are
K$^*\bar{\rm K}^*$ \cite{T} and 
K$^*\bar{\rm K}^*$+$\omega \phi$ \cite{dsb}; 
these both predict small nonstrange
modes and large couplings to KK$\pi\pi$ final states.
The weakness of the 
$\pi\pi$ mode is due to the presence of a hidden $s\bar s$ pair (just as for
$f_0(980)\to\pi\pi$), since both models assume 
that the $f_{\rm J}(1710)$
is dominantly $ns\bar n \bar s$ in flavor. 

In any case the 2P scalar 
$n\bar n$ states (or resonances with large $2^3$P$_0$ $n\bar n$
components) should appear in 
$\rho\pi\pi$ final states, so it would be useful to search for these states,
especially 
in reactions that produce the $f_0(1500)$ or $f_{\rm J}(1710)$.

Finally, we should consider the possibility that the $f_{\rm J}(1710)$
is dominantly a 2$^3$P$_2$ $n\bar n$ tensor state (see Table B6), since
the quantum numbers have not been determined definitively.
Again the quarkonium assignment 
is inconsistent with experiment; the $\eta\eta$ coupling is predicted to
be small, and $\pi\pi$ is predicted to be quite large. The largest mode,
$\rho\rho$, has not been reported for the $f_{\rm J}(1710)$. The total width
of the $n\bar n$ state is again rather larger than reported for 
the $f_{\rm J}(1710)$.
One must conclude that the $f_{\rm J}(1710)$ does not appear to be
consistent with any $n\bar n$
quarkonium assignment.

\subsection{$1^{+-}\ 2^1$P$_1 \; $: $b_1(1700), h_1(1700)$}

Predictions for the missing spin-singlet 2P states are given in Table B7. 
These are expected to be only about 250 MeV wide,
so they may be easy to detect. Reactions that produce the $h_1(1170)$ and 
$b_1(1231)$ are obviously the most promising for searches for their
 radial excitations.
The $h_1(1700)$ couples dominantly to $\rho\pi$, so it may be 
observable for example in 
$\pi^-p\to \rho\pi n$, 
in production through 
natural-parity exchange. Its partner $b_1(1700)$ can be produced 
similarly in
$\omega\pi $ final states, and less characteristically in  
$\rho\rho$. 

\section{1D States:}

\subsection{$2^{-+}\ ^1$D$_2 \; $}

Studies of the decays of hybrids in the flux tube model conclude that 
a $2^{-+}$ member of the lowest hybrid multiplet may be observably
narrow \cite{cp95}.
This hybrid multiplet is expected at $\approx 1.8$-$1.9$ GeV
\cite{paton85,bcs}, which overlaps the
Godfrey-Isgur quark model predictions of  
1.68 GeV for the  $^1$D$_2$ $n\bar n$,
1.89 GeV for $^1$D$_2 \; s\bar{s}$,
and 2.13 GeV for 2$^1$D$_2$ $n\bar n$ \cite{god}. 
Thus it may be necessary to use 
characteristic branching fractions to 
distinguish quarkonia from hybrids 
in this mass region. Of course the $\pi_2(1670)$ is presumably
$n\bar n$ because it has well established 1D multiplet partners 
such as the $\rho_3(1691)$, but distinguishing the higher-mass
$s\bar s$ and 2D 
quarkonia from hybrids may not be so straightforward.
 
\subsection{$\pi_2$}

Experimentally, 
the $\pi_2(1670)$ couples most strongly to $f_2(1275)\pi$ ($\approx 56\% $) 
and $\rho\pi$ ($\approx 31\% $), with
weaker couplings (at the 5-10$\% $ level) 
to $f_0(1300)\pi$ and K$^*$K. The 1996 
PDG total width is 258(18) MeV \cite{pdg96}.
In comparison, the \3p0 model predicts a total width of 250 MeV, with 
branching fractions of 
$f_2(1275)\pi$ ($\approx 30\% $), 
$\rho\pi$ ($\approx 47\% $) and
K$^*$K ($\approx 12\% $); these are in reasonable qualitative agreement with 
experiment.  
There is however disagreement with experiment 
in that little $f_0(1300)\pi$ is expected; we predict
a branching fraction of only $0.2\% $ to this mode, whereas the PDG value is
$8.7(3.4)\% $. The largest as yet unreported mode should be
$\rho\omega$, predicted to have a branching fraction of $11\% $.

In addition to the plausible quarkonium state $\pi_2(1670)$, 
the ACCMOR Collaboration
in 1981 noted a $2^{-+}$ structure near
1.8 GeV, coupled to $f_2\pi $ and weakly to $f_0(1300)\pi$ and $\rho\pi$
\cite{acc}. 
This is 
similar to reports of a possible $2^{-+}$ (or even $1^{-+}$) 
seen in photoproduction of $3\pi$ states
near $1.77$~GeV with a width of 100-200 MeV, 
which couples to $\rho\pi$ and $f_2\pi$
\cite{pi2phot}. 
The VES Collaboration also claims a  
peak near
1.8 GeV, which they believe however to be
non-resonant \cite{rya}. Lastly, two-photon 
experiments which see the $\pi_2(1670)$ in
$\gamma\gamma\rightarrow\pi_2\rightarrow\pi^0\pi^0\pi^0$
\cite{crystalball} and $\gamma\gamma\rightarrow\pi_2\rightarrow\pi^+\pi^-\pi^0$
\cite{cello} also see indications of a possible
contribution around 1.8 GeV. (In both cases the data appear 
skewed towards the higher masses relative to simple Breit Wigner and
PDG values.) This may be expected for $\pi_{2(D)}$ through VMD as its $\rho
\omega$ coupling is predicted to be large and thereby provide a further
probe for any 2D component in $\pi_2(1800)$ state.
It may be possible for LEP2 to clarify this situation.

If there is indeed a second $\pi_2$ state near 1.8 GeV, 
it is much too light to
be a radial excitation of the $\pi_2(1670)$, and may instead be
a hybrid. 
To test this possibility we have calculated the branching fractions of a 
$\pi_2(1800)$ hybrid in the flux tube model, and for comparison we
show the partial widths of a hypothetical 
1D quarkonium $\pi_2(1800)$.
These are given in Table V. (The partial widths to
$a_1(1230)\eta$ and K$^*_1(1273)$K are
$<1$ MeV in both models, so these modes are not displayed.)

\begin{table}
\caption{
Partial widths of 1D and hybrid $\pi_2(1800)$ states.} 
\label{tabpi2}
\begin{tabular}{lccccccccc}
 & $\rho\pi$ & $\omega\rho$  &  $\rho_R\pi$  & $b_1\pi$ & $f_0\pi$ 
& $f_1\pi$ & $ f_2\pi $ & K$^*$K & total \\
\hline
$\pi_{2(1D)}(1800)$    & 162. & 69.  & 0. & 0. & 1. & 5. & 86. & 49. & 372.   \\
$\pi_{2(H)}(1800)$     & 8    & 0    & 5  & 15 & 1  & 0  & 50  & 1   & 80 
\end{tabular}
\end{table}

Evidently there are 
very characteristic differences between hybrid
and 1D ($\pi_2$) branching fractions. 
First, note that a large $f_2(1275)\pi$ mode is
{\it not} distinctive; this is expected from both states. 
A 1D quarkonium should also couple
strongly to $\rho\pi$, $\omega\rho$ and K$^*$K, and the
total width should be about 400 MeV. In contrast, 
these S+S modes are weak for a hybrid; the second largest mode (after $f_2\pi$)
should be $b_1\pi$, which is forbidden to quarkonium by the singlet
selection rule. 
Clearly a study of $b_1\pi $ final states in processes that report a
$\pi_2(1800)$
would be very useful as a hybrid search.
Other modes are quite small, so the hybrid should be
a relatively narrow state, with a total width of only about 100 MeV. 
In summary, the characteristic signature of a $\pi_{2(H)}(1800)$ hybrid is
a strong $f_2\pi $ mode and some $b_1\pi$ but weak couplings to
$\rho\pi$, $\omega\rho$ and K$^*$K.

\subsection{$\eta_2$}

A doubling of $2^{-+}$ peaks has 
also been reported  by 
Crystal Barrel, in the isoscalar sector in
$p\bar p  \to (\eta\pi^{o}\pi^{o})\pi^{o}$ 
\cite{bugg}. Masses and widths of 
M $ = 1645(14)(15)$~MeV,
$\Gamma = 180 {+40\atop-21}(25)$~MeV
and 
M $ = 1875(20)(35)$~MeV,
$\Gamma = 200(25)(45)$~MeV
have been reported for the two $2^{-+}$ states.
This $\eta_2(1645)$ 
is seen in 
$a_2(1318)\pi$ \cite{eta2cb}, and in view of the approximate degeneracy 
with the
$\pi_2(1670)$ and other 1D candidates is probably the $^1$D$_2$ $n\bar n$
isosinglet partner of $\pi_2(1670)$.
The higher-mass state $\eta_2(1875)$ has been seen only
in $f_2(1275)\eta$ (only 
50 MeV above
threshold), and no evidence of it is found in
$a_0 (980)\pi$,  
$f_0(980)\eta$ or $f_0(1300)\eta$. 
The Crystal Ball Collaboration some time ago reported
a $2^{-+}$ (or possibly
$0^{-+}$) at 1880 MeV, with a width of 220 MeV, 
decaying equally to $a_2 (1318)\pi$
and $a_0 (980)\pi$ \cite{crystalball}. 
These data are also consistent with 
a contribution from $\eta_2(1645)$. One expects
$\gamma\gamma\rightarrow\eta_2 > \gamma\gamma\rightarrow\pi_2$, with the
magnitude of the signal 
in  $\gamma\gamma\rightarrow \eta\pi\pi$ depending on 
BR($\eta_2\rightarrow \eta\pi\pi$). Here again LEP2 may have much to contribute.

\begin{table}
\caption{
Partial widths of 1D and hybrid $\eta_2(1875)$ states.} 
\label{tabeta2}
\begin{tabular}{lcccccccc}
 & $\rho\rho$ & $\omega\omega$  &  $f_2\eta$ & $a_0(1450)\pi$ & $a_1\pi$ 
& $ a_2\pi $ & K$^*$K & total \\
\hline
$\eta_{2(1D)}(1875)$    & 147. & 46.  & 45. & 1. & 43. & 264. & 61. & 607.    \\
$\eta_{2(H)}(1875)$     & 0    & 0    & 20  & 2 &  0   & 160  & 10  
& $\approx 190$    
\end{tabular}
\end{table}

In Table VI we compare the decay modes expected for a hybrid at 1875 MeV
with \3p0 model predictions for a hypothetical 
$^1$D$_2$ $\eta_2(1875)$ quarkonium. Both assignments lead to a significant
$f_2\eta$ signal, and both predict a much larger $a_2\pi$
mode. 

The most characteristic modes are $\rho\rho$ and 
$\omega\omega$, which should be
very weak for a hybrid but large for a 1D quarkonium. Similar results
follow for ${\rm K}^*{\rm K}$ and $a_1\pi$. Clearly searches for $a_2\pi$,
$\rho\rho$ and $\omega\omega$ would be most useful. The large predicted
coupling to $\rho \rho$ for the $\eta_{2(1D)}$ encourages a search in 
$\gamma \gamma$ for this state.  

\subsection{$^3$D$_{\rm J}$ states}

Here we consider only the 
$^3$D$_3$ and
$^3$D$_2$ states since the 
$^3$D$_1$ vectors were previously discussed with the 2$^3$S$_1$ states.
The $3^{--}$ states $\rho_3(1691)$ and $\omega_3(1667)$ are well established
$^3$D$_3$ $n\bar n$ quarkonia, with masses as expected for 1D states and
widths of about 200 MeV. The $\rho_3$ (Table B7) is expected to decay
mainly to 
$\rho\rho$ 
$(41\% )$ 
and 
$\pi\pi$
$(34\% )$, 
with a somewhat weaker
$\omega\pi$ mode
$(11\% )$. 
Experimentally the decays to
$4\pi$ are about
$70\% $, of which 
$16(6)\% $ is
$\omega\pi$. The $\pi\pi$ branching fraction is observed to be
$23.6(1.6)\% $. There are also KK and ${\rm K}^*{\rm K}$ modes of a
few percent, roughly as predicted. The total width is predicted to be
174 MeV with these parameters, consistent with observation. Thus the
$\rho_3(1691)$ appears to decay approximately as predicted by the 
\3p0 model, which supports the application of the model to decays of
high-L states.

Its isoscalar partner $\omega_3(1667)$ is a more interesting case.
Since few modes are open and the couplings are rather weak, we predict
a total width of only 69 MeV. Although this appears inconsistent with
the PDG width of 168(10) MeV, this observed value is presumably broadened
by the hadronic width of the $\rho$ and $b_1$ in the two-body modes
$\rho\pi$ and $b_1\pi$. The reported modes are $\rho\pi$ and $\omega\pi\pi$;
we expect $\rho\pi$ to be dominant, with $\approx 10\% $ branches to
$b_1\pi$ (the source of $\omega\pi\pi$?) and KK. The KK mode affords an
opportunity to measure the actual width of the $\omega_3$, which may be
much smaller than it appears in $\rho\pi$ and $b_1\pi$ modes.

Our results for the $^3$D$_2$ $2^{--}$ states $\rho_2(1670)$ and
$\omega_2(1670)$ are especially interesting because these are 
``missing mesons" in the quark model. We find that these are rather broad
states, with total widths of about 300-400 MeV. 
The $\rho_2$ is predicted to have
a large branching fraction 
of $54 \% $
to $a_2\pi$, so it should be observable in this final state or in the
secondary modes $\omega\pi$ or K$^*$K. The $\omega_2$ is predicted
to have an even larger branching fraction
of $74 \% $ to $\rho\pi$. It too couples significantly to
K$^*$K, and may also be observable in $\omega\eta$.
 
\section{1F States}

The 1F states provide us with an opportunity to test the accuracy of
the \3p0 decay model predictions for higher quarkonium states, 
since the $4^{++}$ and $3^{+\pm}$ states expected
near 2.05~GeV do not have competing assignments
as glueballs or hybrids. At present only two of these states
are reasonably well established, the $f_4(2044)$ and $a_4(2037)$ \cite{pdg96}. 
There is also some evidence for an $a_3(2080)$ \cite{pdg94}. 

We do not yet have experimental branching fractions
for the I=1 1F states.
The $a_4(2037)$ is
seen in KK and $3\pi$, and the $a_3(2080)$ is reported in $3\pi$ and
$\rho_3(1691)\pi$, with $\rho_3\pi$ dominant.
The branching fractions of the $f_4(2044)$ are known with more accuracy;
$\omega\omega$ and $\pi\pi$ are important modes, 
$26(6) \% $ 
and 
$17.0(1.5) \% $. 
KK and $\eta\eta$ modes are both known, with reported 
branching fractions of
about $0.7\% $ and $0.2 \% $ respectively.

\3p0 predictions for the decays of these $^3$F$_{\rm J}$ states 
are given in Tables B11 and B12.
The $a_4(2050)$ is indeed expected to appear 
in $3\pi$ (mainly $\rho\pi$), and the dominant
mode is predicted to be $\rho\omega$. 
This state is predicted to
be rather narrower than reported. The $a_3(2080)$ is predicted to decay dominantly
to $\rho_3\pi$, as is observed. 
The $3\pi$ mode is also predicted to be large,
and to arise from both $\rho\pi$ and $f_2\pi$.
The $f_4(2044)$ \3p0 model predictions are also in qualitative agreement with
experiment, in that $\pi\pi$ and $\omega\omega$ are expected to be important
modes, as observed. 
The $f_4$ partial widths to pseudoscalar pairs are uniformly 
too large, for example 
$\Gamma_{f_4 \to \pi\pi}^{thy.} = 62.$ MeV 
but
$\Gamma_{f_4 \to \pi\pi}^{expt.} = 35(4)$ MeV. This decay however
is G-wave, so the rate has a prefactor of $|\vec p_{\pi} / \beta|^9$;
this extreme sensitivity means that a small increase of $\beta$ by
$\approx 10\%$, halves the decay rate and gives agreement with experiment.
Thus this disagreement is quite sensitive to parameters and is probably not
significant.

The predictions for branching fractions of the five 
missing I=0,1 1F states suggest that
several of them may easily be found 
by reconstructing the appropriate final states.
The total widths of all except the $^3$F$_2$ states are predicted to be
$\sim 300$ MeV, so they should be observable
experimentally. The 
$f_3(2050)$ is predicted to couple dominantly to $a_2\pi$. In the
spin-singlet $^1$F$_3$ sector, the $h_3(2050)$ should appear in $\rho\pi$ and 
$\rho_3(1691)\pi$, just as we found for the $a_3(2080)$. The 
$b_3(2050)$ should be evident in $a_2\pi$, and less strongly in
$\omega_3\pi$, $\omega\pi$ and $\rho\rho$. Modes such as
$a_2\pi$ are preferable because the two-body mesons are not excessively
broad and they are far from threshold,
so a resonance can be distinguished from a threshold effect.
In some cases the amplitude structure of these final states is also
characteristic; these can be determined from the results quoted in App.A.

The missing $^3$F$_2$ states may be more difficult to identify, as we
predict large total widths of
$\approx 600$ MeV for these states.
The $a_2(2050)$ couples most strongly to $b_1\pi$; 
$\eta_2(1645)\pi$ and ${\rm K}^*_1(1273){\rm K}$ are other important modes.
Its I=0 partner $f_2(2050)$ should be evident in $\pi_2(1670)\pi$ and will
also populate ${\rm K}^*_1(1273){\rm K}$ final states.

Identification of these 1F states and 
determination of their branching fractions
and decay amplitudes
will be a very useful contribution to the study of resonances, as it
will allow detailed tests of the usefulness of the \3p0 
model as a means for identifying
quarkonium states in this crucial 2 GeV region.

\section{Summary and Experimental Strategy}

We have established that the $a_1(1700)$ is very likely a 2P radial
excitation. This follows from the weak S-wave and strong D-wave
in $\rho \pi$. This also establishes the natural mass scale for the 2P
multiplets as $\approx 1.7$ GeV.  
We have been unable to identify radial scalars. These are predicted to
be broad, and so their non-appearance is not surprising. Conversely it
raises interest in the (relatively narrow) $f_0(1500)$ and
possible scalar $f_{\rm J}(1710)$.
We do identify some (more speculative) potential
candidates for $2^{++}$ 2P members. 
We note that $\gamma \gamma$
production may help identify these radial 2P states and also
clarify the nature of $f_0(1500)$ and $f_{\rm J}(1710)$\cite{glynnys}.

The $\pi(1300)$ and $\eta(1295)$ appear to be convincing 2S states. This 
conclusion is based on their relative widths; the large $\rho \pi$
mode of the $\pi(1300)$ has no analog for its $\eta$ counterparts. The
status of the $\eta(1440)$ remains open; the mass and width
suggest a dominantly $s\bar s$ state, 
but the $\gamma\rho$ mode argues against it.
Studies of $\psi \to \eta(1295,1440) +
(\omega, \phi)$  and $\psi \to  \gamma +
(\gamma\omega,\gamma\rho, \gamma\phi)$ may identify the flavor content
of these $\eta$ states.

The $\rho(1465)$ and $\omega(1419)$ have masses that are consistent
with radial 2S but their decays show characteristics of hybrids, as
noted previously \cite{cp95}.
We suggest that
these states 
may be 2S-hybrid mixtures analogous to the
3S-hybrid mixing suggested for the $c\bar c$\cite{pagecc}. 
This can be tested by accurate measurement of the partial widths of these
states and their vector 
partners at 1.6-1.7 GeV to $\pi\pi$, $\omega\pi$,
and especially $h_1\pi$ and $a_1\pi$.

The 3S $\pi$ is expected in the $1800$ MeV mass region as is 
a $\pi_H$
hybrid. We find that the decay patterns of these states
are very different.
A strong $f_0(1300)\pi $ from the hybrid contrasted with a large
$\rho \omega$ mode from the 3S quarkonium is the sharpest
discriminant. The VES state $\pi(1800)$ clearly exhibits
this hybrid signature. It is now necessary to establish the
presence of $0^{-+}$ in the $\rho \omega$ channel, and to see if any
resonant state is present that is 
distinct from the $\pi(1800)$ seen in 
$f_0(1300)\pi $. It is possible that there are two 
$\pi(\approx 1800)$ states, $q\bar q$ and hybrid, whose
production mechanisms and decay fractions differ sufficiently
so that they can be separated.
We suggest that the possibility of two such $\pi(\approx 1800)$ states
be allowed for in 
data analyses.

In the immediate future there are opportunities for $\gamma \gamma$
physics at LEP2 and at B~factories. Possible strategies for isolating some of
these higher quarkonia include:

$\bullet$ $\gamma \gamma \to 5\pi$
contains (i) $\rho \omega$ which may access the radial $a_{0R}$ and $a_{2R}$
near 1700 MeV and a possible $\pi_{3S}(1800)$.
(ii) $\pi b_1$ which can isolate the $a_{0R}$ if
the helicity selection rule\cite{abc} is used to suppress the $a_{2R}$.

$\bullet$ $\gamma \gamma \to 4 \pi$ may access the radial $f_{2R}$ near
1700 MeV through its decay into $\rho \rho$. The $4 \pi$ channel
may also be searched for the $f_0(1500)$ since this state is known to
have a significant branching fraction to $4\pi $ but should have
a suppressed $\gamma \gamma$ coupling if it is a glueball \cite{glynnys}.

$\bullet$ $\gamma \gamma \to 3 \pi$ may be searched for $2^{-+}$ states
in order to verify whether the established $\pi_2(1670)$ is accompanied
by a higher $\pi_2(1800)$ in $3\pi^o$ and $\pi^+ \pi^- \pi^o$. This
$3\pi$ system may also be studied for evidence of one or more
$\pi(1800)$ states.

$\bullet$ $\gamma \gamma \to \eta \pi \pi$ may access the isoscalar
partners of these $\pi_2$ states.

In the near future it will be possible to study $e^+ e^-$ annihilation
up to $\approx 2$ GeV at DAFNE. The channels $e^+ e^- \to 4\pi$ should be
measured and $\pi a_1$ and $\pi  h_1$ states separated in
order to carry out the analysis of hybrid and radial vector
components in section 3B.
The isoscalar partners of the vectors also need confirmation, and
final states with kaons
are needed to investigate possible $\omega$-$\phi$ mixing; 
a potential weakness of the present data analyses is that such flavor
mixing is assumed to be unimportant.

In the next century there will be new opportunities at the COMPASS facility
at CERN. This will enable further studies of central production and
also of diffractive excitation. For the latter one may anticipate
improved studies of the $\pi$ excitations
(such as the $\pi(1300)$ and $\pi(1800)$ states),
possibly including Primakoff excitation. Judicious studies of 
specific final states as
discussed above may help separate 3S and hybrid states. The use
of K beams will allow analogous studies of the strange 
counterparts of these states and may help to clarify the spectrum
of quarkonia, glueballs and hybrids.

Experiments with $\pi$ beams can access the following interesting channels.

$\bullet$ $\pi p \to (\pi f_1) p$, to confirm the D-wave dominance of
$a_{1R}(1700)$ and to seek its partner $a_{2R}$.

$\bullet$ $\pi p \to (\pi f_2) p$ can access both $\pi_{2(1D)}$ and $\pi_
{2(H)}$. These can be separated in $b_1\pi$;
the singlet selection rule forbids this mode
for $\pi_{2(1D)}$ but allows it for $\pi_{2(H)}$. 
$(\pi \rho) p$ can also separate $\pi_{2(1D)}$ from 
$\pi_{2(H)}$; $\pi_{2(1D)} \to \rho \pi$ is the dominant mode 
whereas $\pi_{2(H)}$
is much suppressed into S+S hadrons.

$\bullet$ $(\pi\pi), (\pi\omega)$, 
$(a_1\pi) $ and
$(h_1\pi) $ are important in the interpretation of the vectors
between 1.4 and 1.7 GeV, which may contain large hybrid components.

$\bullet$ $(f_0\pi), (f_2\pi)$ and 
$(\rho\omega) $ can all be searched for evidence of $\pi(1800)$ states.

$\bullet$ $\pi^- p \to (\pi \rho)^o n$ or $ (\pi \omega)^o n$ access
respectively $h_{1R}$ and $b_{1R}$.

Finally, many two-body channels are predicted to couple strongly
to specific 2P, 1D and 1F states, as shown in Appendix B. These include
``missing mesons" such as the $^3$F$_2$ and most 2P states, and
studies
of these two-body final states may reveal the missing resonances. The modes
$a_2\pi$, $\rho\rho$ and $b_1\pi$ are important for many of these missing 
states and merit careful investigation.

We reiterate that it is in general a good strategy to study decays into
both S+S and S+P meson modes, as the relative couplings of these modes
are usually quite distinct for hybrid versus quarkonium
assignments.

\newpage
\acknowledgements

We would like to acknowledge useful communications with 
C.Amsler, D.V.Bugg, S.U.Chung, 
G.Condo, K.Danyo, A.Dzierba, S.Godfrey, I.Kachaev, 
Y.Khokhlov, 
A.Kirk, 
D.Ryabchikov
and A.Zaitsev. This work was supported in part by
the United States Department
of Energy under contracts
DE-FG02-96ER40944 at North Carolina State University and
DE-AC05-96OR22464 managed by
Lockheed Martin Energy Systems Inc. at Oak Ridge National Laboratory.
FEC is supported in part by European Community Human Capital Mobility
Programme Eurodafne, Contract CHRX-CT92-0026.

\newpage 

\appendix{}

\section
{A Compilation of $^3$P$_0$ Model Decay Amplitudes.}

We quote results for the \3p0 model A$\to$BC meson decay amplitudes 
in terms of an
invariant amplitude 
${\cal M}_{\rm L_{BC}S_{BC}}$, 
which is 
the L$_{BC}$S$_{BC}$ projection of the \3p0 pair creation Hamiltonian
matrix element divided by a momentum conserving delta function,
\begin{equation}
{\cal M}_{L_{BC}S_{BC}}^{A\to BC} \; 
= \langle {J_A,L_{BC},S_{BC}}|BC\rangle \; \langle BC | 
H_I(^3{\rm P}_0) | A \rangle 
 / \delta(\vec A - \vec B - \vec C \, ) \ . 
\end{equation}
This amplitude and the derivation of the $^3$P$_0$ matrix elements
are discussed in detail in Appendix A of Ackleh {\it et al.} \cite{abs}.
The partial widths $\Gamma_{A\to BC}$ are related to these decay amplitudes by
\begin{equation}
\Gamma_{A\to BC}
 = 
2\pi \,  {P E_B E_C \over M_A}  \,
\sum_{\rm LS}
|{\cal M}_{\rm LS}|^2\, 
\ .
\end{equation}

The full \3p0 decay amplitude is the sum of two Feynman diagrams, called
$d_1$ and $d_2$ (Fig.A1). 

In a specified flavor channel these diagrams have 
flavor weight factors that multiply the spin-space matrix element. 
The flavor factors for all the processes considered in this paper are given
in Table A1. The ${\cal M}$ amplitudes listed 
below are for unit flavor
factors, $I_{flavor}(d_1)=+1$ and $I_{flavor}(d_2) = \pm 1$, with the phase
chosen so they add rather than cancel. 
(The cancelations are due to flavor symmetries 
such as G-parity.) Thus for a physical decay such as
$\rho^+\to\pi^+\pi^o$ one should multiply the unit-flavor amplitude
${\cal M}$ in A3 by $+1/\sqrt{2}$ before computing the decay width
using (A2). 
Some states populate several decay channels, for example
$f\to\pi^o\pi^o$ as well as $\to\pi^+\pi^-$; to sum over all channels
one should multiply the width by the flavor multiplicity factor ${\cal F}$
in the Table. 
In these flavor weights the pairs 
$(\pi,a), (\rho,b), (\omega, h)$ and $(f,\eta_{n\bar n})$ are equivalent,
up to factors due to identical particles in the final state.

\begin{figure}
$$\epsfxsize=4truein\epsffile{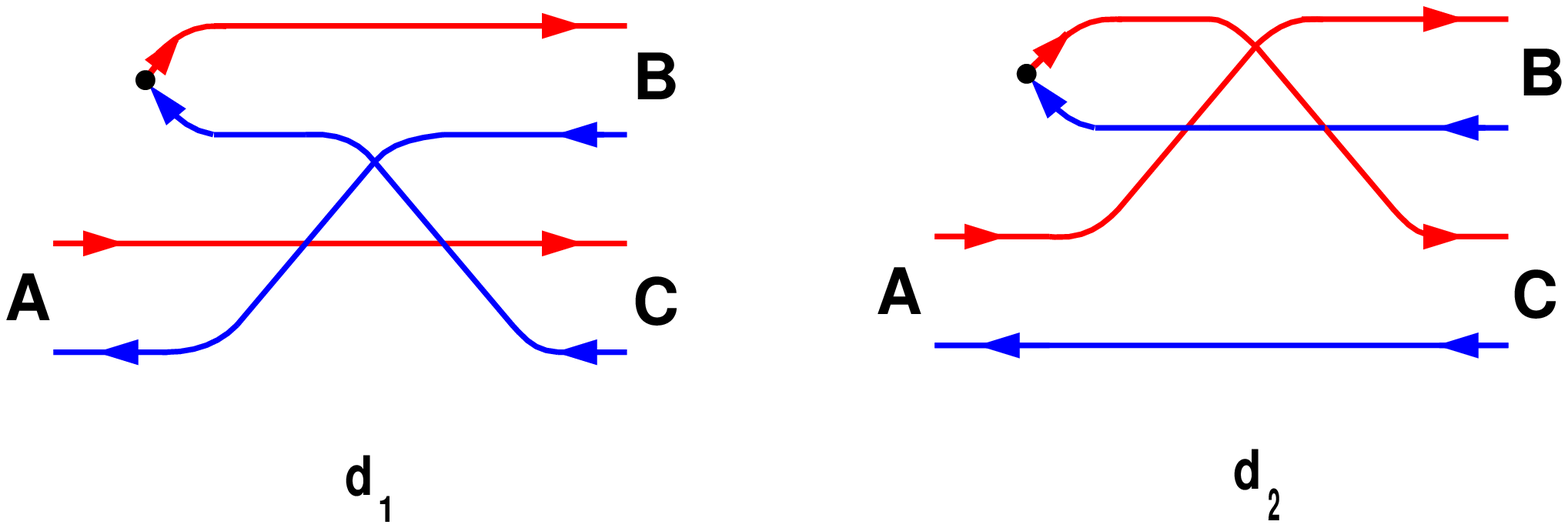}$$
\begin{center}
{Figure~A1.
$q\bar q$ meson decay diagrams in the $^3$P$_0$ decay model.
}
\end{center}
\end{figure}

\begin{center}
\begin{tabular}{||c|c|c|c|c||}  \hline
\multicolumn{5}{||c||}{Table A1. Flavor Weight Factors.} \\ \hline
$\ $ Generic Decay $\ $  & $\ $ Subprocess $\ $ & $\ I_{flavor}(d_1)\ $
& $\ I_{flavor}(d_2)\ $  &
$\ {\cal F}\ $
\\ \hline
$\rho \to \pi\pi$ & $\rho^+ \to \pi^+\pi^o$ & $+1/\sqrt{2}$ & $-1/\sqrt{2}$    &  $1$
\\  \hline
$f \to \pi\pi$ & $f \to \pi^+\pi^-$ & $-1/\sqrt{2}$ & $-1/\sqrt{2}$    &  $\  3/2\  $
\\
\hline
$f \to {\rm K}{\rm K} $ & $f \to {\rm K}^+ {\rm K}^-$
& $0$ & $-1/\sqrt{2}$  &  $2$  \\  \hline
$ a\to \rho\pi$ & $a^+ \to \rho^+\pi^o$ & $+1/\sqrt{2}$ & $-1/\sqrt{2}$    &  $2$
\\  \hline
$a \to {\rm K}{\rm K} $ & $a^+ \to {\rm K}^+ {\rm K}^o$
& $0$ & $-1$  &  $1$  \\  \hline
$b \to \omega\pi$ & $b^+ \to \omega\pi^+$ & $+1/\sqrt{2}$ & $+1/\sqrt{2}$    &
$1$
\\  \hline
$h \to \rho\pi$ & $h \to \rho^+\pi^-$ & $-1/\sqrt{2}$ & $-1/\sqrt{2}$    &  $3$
\\  \hline
${\rm K}^*\to {\rm K}\pi$ & ${\rm K}^{*+}\to {\rm K}^+\pi^o$
& $+1/\sqrt{2}$ & $0$  &  $3$  \\  \hline
$\phi \to {\rm K}{\rm K} $ & $\phi \to {\rm K}^+ {\rm K}^-$
& $+1$ & $0$  &  $2$  \\  \hline
\end{tabular}
\end{center}

\vskip 1cm

We take all spatial wavefunctions to be SHO forms with the same width
parameter $\beta$; as a result the 
${\cal M}_{\rm LS}$ decay amplitudes  
are proportional to an overall Gaussian in $x=P/\beta $ times 
a channel-dependent polynomial ${\cal P}_{\rm LS}(x)$, 
\begin{equation}
{\cal M}_{\rm LS} = {\gamma \over \pi^{1/4} \beta^{1/2} } \;
{\cal P}_{\rm LS}(x) \, e^{-x^2/12} \ ,
\end{equation}
where $\gamma$ is the \3p0 pair production coupling constant \cite{abs}.
To specify these amplitudes it suffices
to quote the polynomial ${\cal P}_{\rm LS}(x)$
for each decay channel. The complete set of \3p0 decay amplitudes for 
all $q\bar q$ resonances with ``excitation level''
${\cal N}_A=$N$_A+$L$_A \leq 4$ 
decaying into 
final states with
${\cal N}_B \leq {\cal N}_A -1$ and C $ = {}^1$S$_0$ (and
C = $ {}^3$S$_1$ in most cases)
is given below. 
For the relatively obscure transitions
3S $\to$ 1D + C, 
1F $\to$ 1P + C, 
1F $\to$ 2P + C and
1F $\to$ 1D + C we restrict C to $ {}^1$S$_0$; this does not exclude
any decays allowed by phase space.

We include a few additional amplitudes in this list.
Some of these are of interest as couplings to virtual two-body
states, although phase space 
nominally forbids the decay.

\newpage

\hrule
\vskip .1cm
\hrule
\vskip .5cm
\fbox{\bf 1S $\to$ 1S + 1S} 

\begin{equation}
f_P = -{2^5\over 3^3 } \; x
\end{equation}

\begin{eqnarray}
  & & \nonumber \\ \fbox{$^3${\bf S}$_1$} & & \nonumber \\  & & \nonumber \\ 
 & {\cal P}_{10}^{ ( ^3{\rm S}_1 \to ^1{\rm S}_0 + ^1{\rm S}_0 )} \  & =
\begin{array}{lr}
f_P 
\ &   ^1{\rm P}_1   \\
\end{array}
\\ 
& {\cal P}_{11}^{ ( ^3{\rm S}_1 \to ^3{\rm S}_1 + ^1{\rm S}_0 )} \  & =
\begin{array}{lr}
-\sqrt{2} \; f_P 
\ &   ^3{\rm P}_1   \\
\end{array}
\\ 
 & {\cal P}_{\rm LS}^{ ( ^3{\rm S}_1 \to ^3{\rm S}_1 + ^3{\rm S}_1 )} \  & =
\left\{
\begin{array}{lr}
\sqrt{1\over 3 }\; f_P \;
\ &   ^1{\rm P}_1   \\
0 \;
\ &   ^3{\rm P}_1   \\
-\sqrt{20 \over 3} \; f_P \;
\ &   ^5{\rm P}_1   \\
0 \;
\ &   ^5{\rm F}_1   \\
\end{array}
\right. 
\\
  & & \nonumber \\ \fbox{$^1${\bf S}$_0$} & & \nonumber \\  & & \nonumber \\ 
 & {\cal P}_{\rm LS}^{ ( ^1{\rm S}_0 \to ^1{\rm S}_0 + ^1{\rm S}_0 )} \  & =
 0 \;
\\
 & {\cal P}_{11}^{ ( ^1{\rm S}_0 \to ^3{\rm S}_1 + ^1{\rm S}_0 )} \  & =
\begin{array}{lr}
-\sqrt{3} \; f_P 
\ &   ^3{\rm P}_1   \\
\end{array}
\\
 & {\cal P}_{11}^{ ( ^1{\rm S}_0 \to ^3{\rm S}_1 + ^3{\rm S}_1 )} \  & =
\begin{array}{lr}
\sqrt{6} \; f_P 
\ &   ^3{\rm P}_1   \\
\end{array}
\end{eqnarray}

\newpage

\vskip .5cm
\hrule
\vskip .1cm
\hrule
\vskip .5cm

\fbox{\bf 2S $\to$ 1S + 1S} 

{(\small \it See 1S $\to$ 1S + 1S for channel coefficients.)}
\vskip .5cm

\begin{equation}
f_P = -{2^{9/2} 5 \over 3^{9/2} } \; x\; \Big( 1 - {2\over 15} x^2 \Big) \;
\end{equation}

\vskip .5cm
\hrule

\vskip .5cm
\fbox{\bf 2S $\to$ 1P + 1S} 

\begin{eqnarray}
 & f_S  = &{2^4\over 3^4} \; \Big(1 - {7\over 9} x^2 + {2\over 27} x^4 \Big) \; 
\\
 & f_D  = &{2^{9/2} (13) \over 3^6} \; x^2 \Big(1 - {2\over 39} x^2 \Big) \; 
\end{eqnarray}

\begin{eqnarray}
  & & \nonumber \\ \fbox{$2^3${\bf S}$_1$} & & \nonumber \\  & & \nonumber \\ 
 & {\cal P}_{\rm LS}^{ ({\rm 2}^3{\rm S}_1 \to ^1{\rm P}_1 + ^1{\rm S}_0 )} \  & =
\left\{
\begin{array}{cr}
f_S \; 
\ &   ^3{\rm S}_1   \\
f_D \; 
\ &   ^3{\rm D}_1   \\
\end{array}
\right. 
\\
 & {\cal P}_{\rm LS}^{ ({\rm 2}^3{\rm S}_1 \to ^3{\rm P}_1 + ^1{\rm S}_0 )} \  & =
\left\{
\begin{array}{cr}
-\sqrt{2}\; f_S \; 
\ &   ^3{\rm S}_1   \\
\sqrt{1\over 2}\; f_D \; 
\ &   ^3{\rm D}_1   \\
\end{array}
\right. 
\\
 & {\cal P}_{22}^{ (2^3{\rm S}_1 \to ^3{\rm P}_2 + ^1{\rm S}_0 )} \  & =
\begin{array}{cr}
-\sqrt{3\over 2}\; f_D \;
\ &   ^5{\rm D}_1   \\
\end{array}
\\ 
 & {\cal P}_{\rm LS}^{ ({\rm 2}^3{\rm S}_1 \to ^1{\rm P}_1 + ^3{\rm S}_1 )} \  & =
\left\{
\begin{array}{cr}
-\sqrt{1\over 2}\; f_D \; 
\ &   ^3{\rm D}_1   \\
\sqrt{3\over 2}\; f_D \; 
\ &   ^5{\rm D}_1   \\
\end{array}
\right. 
\\
 & {\cal P}_{\rm LS}^{ ({\rm 2}^3{\rm S}_1 \to ^3{\rm P}_0 + ^3{\rm S}_1 )} \  & =
\left\{
\begin{array}{cr}
-\sqrt{3}\; f_S \; 
\ &   ^3{\rm S}_1   \\
0 \; 
\ &   ^3{\rm D}_1   \\
\end{array}
\right. 
\\
 & {\cal P}_{\rm LS}^{ ({\rm 2}^3{\rm S}_1 \to ^3{\rm P}_1 + ^3{\rm S}_1 )} \  & =
\left\{
\begin{array}{cr}
- 2 \; f_S \; 
\ &   ^3{\rm S}_1   \\
-{1\over 2} \; f_D \; 
\ &   ^3{\rm D}_1   \\
\sqrt{3\over 4} \; f_D \; 
\ &   ^5{\rm D}_1   \\
\end{array}
\right. 
\\
 & {\cal P}_{\rm LS}^{ ({\rm 2}^3{\rm S}_1 \to ^3{\rm P}_2 + ^3{\rm S}_1 )} \  & =
\left\{
\begin{array}{cr}
0\; 
\ &   ^3{\rm S}_1   \\
\sqrt{3\over 20} \; f_D \; 
\ &   ^3{\rm D}_1   \\
{1\over 2} \; f_D \; 
\ &   ^5{\rm D}_1   \\
-\sqrt{28 \over 5} \; f_D \; 
\ &   ^7{\rm D}_1   \\
0\; 
\ &   ^5{\rm G}_1   \\
\end{array}
\right.  \\
  & & \nonumber \\ \fbox{$2^1${\bf S}$_0$} & & \nonumber \\  & & \nonumber \\ 
 & {\cal P}_{00}^{ ( 2^1{\rm S}_0 \to ^3{\rm P}_0 + ^1{\rm S}_0 )} \  & =
\begin{array}{cr}
- \sqrt{3} \; f_S \; 
\ &   ^1{\rm S}_0   \\
\end{array}
\\
 & {\cal P}_{22}^{ ( 2^1{\rm S}_0 \to ^3{\rm P}_2 + ^1{\rm S}_0 )} \  & =
\begin{array}{cr}
- \sqrt{3} \; f_D \; 
\ &   ^5{\rm D}_0   \\
\end{array}
\\ 
 & {\cal P}_{\rm LS}^{ ({\rm 2}^1{\rm S}_0 \to ^1{\rm P}_1 + ^3{\rm S}_1 )} \  & =
\left\{
\begin{array}{cr}
- \sqrt{3} \; f_S \; 
\ &   ^1{\rm S}_0   \\
- \sqrt{3} \; f_D \; 
\ &   ^5{\rm D}_0   \\
\end{array}
\right. 
\\
 & {\cal P}_{\rm LS}^{ ({\rm 2}^1{\rm S}_0 \to ^3{\rm P}_1 + ^3{\rm S}_1 )} \  & =
\left\{
\begin{array}{cr}
 \sqrt{6} \; f_S \; 
\ &   ^1{\rm S}_0   \\
-\sqrt{3\over 2} \; f_D \; 
\ &   ^5{\rm D}_0   \\
\end{array}
\right. 
\\
 & {\cal P}_{\rm 22}^{ ( 2^1{\rm S}_0 \to ^3{\rm P}_2 + ^3{\rm S}_1 )} \  & =
\begin{array}{cr}
\sqrt{9\over 2} \; f_D \; 
\ &   ^5{\rm D}_0   \\
\end{array}
\end{eqnarray}

\newpage

\vskip .5cm
\hrule
\vskip .1cm
\hrule
\vskip .5cm

\fbox{\bf 3S $\to$ 1S + 1S} 

{(\small \it See 1S $\to$ 1S + 1S for channel coefficients.)}
\vskip .5cm

\begin{equation}
f_P = -{2^{7/2} 5^{1/2} 7 \over 3^{11/2} } \; x\; 
\Big( 1 - {4\over 15} x^2 + {4\over 315} x^4 \Big) \;
\end{equation}

\vskip .5cm
\hrule
\vskip .5cm

\fbox{\bf 3S $\to$ 2S + 1S} 

\begin{equation}
f_P = -{2^4 5^{3/2}  \over 3^5 } \; x\; 
\Big( 1 - {1\over 4} x^2 + {1\over 75} x^4 
-{1\over 6075} x^6 \Big) \;
\end{equation}

\begin{eqnarray}
  & & \nonumber \\ \fbox{$3^3${\bf S}$_1$} & & \nonumber \\  & & \nonumber \\ 
 & {\cal P}_{10}^{ (3^3{\rm S}_1 \to 2^1{\rm S}_0 + ^1{\rm S}_0 )} \  & =
\begin{array}{cr}
f_P
\ &   ^1{\rm P}_1   \\
\end{array}
\\
 & {\cal P}_{11}^{ (3^3{\rm S}_1 \to 2^3{\rm S}_1 + ^1{\rm S}_0 )} \  & =
\begin{array}{cr}
-\sqrt{2} f_P
\ &   ^3{\rm P}_1   \\
\end{array}
\\
 & {\cal P}_{11}^{ (3^3{\rm S}_1 \to 2^1{\rm S}_0 + ^3{\rm S}_1 )} \  & =
\begin{array}{cr}
\sqrt{2} f_P
\ &   ^3{\rm P}_1   \\
\end{array}
\\
 & {\cal P}_{\rm LS}^{( 3^3{\rm S}_1 \to 2^3{\rm S}_1 + ^3{\rm S}_1 ) }\  & =
\left\{
\begin{array}{cr}
\sqrt{1\over 3}  f_P
\ &   ^1{\rm P}_1   \\
0 \;
\ &   ^3{\rm P}_1   \\
-\sqrt{20\over 3}  f_P
\ &   ^5{\rm P}_1   \\
0 \;
\ &   ^5{\rm F}_1   \\
\end{array}
\right. \\
  & & \nonumber \\ \fbox{$3^1${\bf S}$_0$} & & \nonumber \\  & & \nonumber \\ 
 & {\cal P}_{11}^{ (3^1{\rm S}_0 \to 2^3{\rm S}_1 + ^1{\rm S}_0 )} \  & =
\begin{array}{cr}
-\sqrt{3} f_P
\ &   ^3{\rm P}_0   \\
\end{array}
\\
 & {\cal P}_{11}^{ (3^1{\rm S}_0 \to 2^1{\rm S}_0 + ^3{\rm S}_1 )} \  & =
\begin{array}{cr}
-\sqrt{3} f_P
\ &   ^3{\rm P}_0   \\
\end{array}
\\
 & {\cal P}_{11}^{ (3^1{\rm S}_0 \to 2^3{\rm S}_1 + ^3{\rm S}_1 )} \  & =
\begin{array}{cr}
\sqrt{6} f_P
\ &   ^3{\rm P}_0   \\
\end{array}
\end{eqnarray}

\vskip .5cm
\hrule
\vskip .5cm

\fbox{\bf 3S $\to$ 1P + 1S} 

{(\small \it See 2S $\to$ 1P + 1S for channel coefficients.)}
\vskip .5cm

\begin{eqnarray}
 & f_S  = &{2^3 5^{3/2} \over 3^5} \; 
\Big(1 - {3\over 5} x^2 + {16\over 225} x^4 -{4\over 2025} x^6 \Big) \; 
\\
 & f_D  = &{2^{7/2} 7^2 \over 3^6 5^{1/2} } \; x^2 
\Big(1 - {20\over 147} x^2 + {4\over 1323} x^4 \Big) \; 
\end{eqnarray}

\vskip .5cm
\hrule
\vskip .5cm

\fbox{\bf 3S $\to$ 2P + 1S} 

{(\small \it See 2S $\to$ 1P + 1S for channel coefficients.)}
\vskip .5cm

\begin{eqnarray}
 & f_S  = &{2^{5/2} \over 3^4} \; 
\Big(1 - {47\over 18} x^2 + {1\over 2} x^4 -{8\over 405} x^6 + {2\over 10935}x^8 \Big) \; 
\\
 & f_D  = &{2^6 5 \over 3^6 } \; x^2 
\Big(1 - {57\over 400} x^2 + {13\over 2700} x^4 - {1\over 24300} x^6 \Big) \; 
\end{eqnarray}

\vskip .5cm
\hrule
\vskip .5cm

\fbox{\bf 3S $\to$ 1D + 1$^1$S$_0$} 

\begin{eqnarray}
 & f_P  = 
&-{2^3 \over 3^5} \; x\;
\Big(1 - {23\over 15} x^2 + {8\over 45} x^4 -{4\over 1215} x^6 \Big) \; 
\\
 & f_F  = 
&-{2^{5/2} (43) \over 3^{11/2} 5 } \; x^3 
\Big(1 - {92\over 1161} x^2 + {4\over 3483} x^4 \Big) \; 
\end{eqnarray}

\begin{eqnarray}
  & & \nonumber \\ \fbox{3$^3${\bf S}$_1$} & & \nonumber \\  & & \nonumber \\ 
 & {\cal P}_{\rm LS}^{( 3^3{\rm S}_1 \to ^1{\rm D}_2 + ^1{\rm S}_0 ) }\  & =
\left\{
\begin{array}{cr}
f_P
\ &   ^5{\rm P}_1   \\
f_F
\ &   ^5{\rm F}_1   \\
\end{array}
\right. \\
 & {\cal P}_{\rm 11}^{( 3^3{\rm S}_1 \to ^3{\rm D}_1 + ^1{\rm S}_0 ) }\  & =
\begin{array}{cr}
\sqrt{ {1\over 2} }\; f_P
\ &   ^3{\rm P}_1   \\
\end{array}
\\
 & {\cal P}_{\rm LS}^{( 3^3{\rm S}_1 \to ^3{\rm D}_2 + ^1{\rm S}_0 ) }\  & =
\left\{
\begin{array}{cr}
-\sqrt{ {3\over 2} }\; f_P
\ &   ^5{\rm P}_1   \\
\sqrt{ {2\over 3} }\; f_F
\ &   ^5{\rm F}_1   \\
\end{array}
\right. \\
 & {\cal P}_{\rm 33}^{( 3^3{\rm S}_1 \to ^3{\rm D}_3 + ^1{\rm S}_0 ) }\  & =
\begin{array}{cr}
-\sqrt{ {4\over 3} }\; f_F
\ &   ^7{\rm F}_1   \\
\end{array}
\\
  & & \nonumber \\ \fbox{3$^1${\bf S}$_0$} & & \nonumber \\  & & \nonumber \\ 
 & {\cal P}_{11}^{ ( 3^1{\rm S}_0 \to ^3{\rm D}_1 + ^1{\rm S}_0 )} \  & =
\begin{array}{cr}
-\sqrt{ 3 }\; f_P
\ &   ^3{\rm P}_0   \\
\end{array}
\\
 & {\cal P}_{33}^{ ( 3^1{\rm S}_0 \to ^3{\rm D}_3 + ^1{\rm S}_0 )} \  & =
\begin{array}{cr}
-\sqrt{ 3 }\; f_F
\ &   ^7{\rm F}_0   \\
\end{array}
\end{eqnarray}

\newpage 

\vskip .5cm
\hrule
\vskip .1cm
\hrule
\vskip .5cm

\fbox{\bf 1P $\to$ 1S + 1S}

\begin{eqnarray}
 & f_S  = &{2^5  \over 3^{5/2}  }\; \Big(1 - {2\over 9} x^2\Big) \; 
\\
 & f_D  = &{2^6  \over 3^4 5^{1/2} }\; x^2  \; 
\end{eqnarray}

\begin{eqnarray}
  & & \nonumber \\ \fbox{$^3${\bf P}$_2$} & & \nonumber \\  & & \nonumber \\ 
 & {\cal P}_{20}^{ ( ^3{\rm P}_2 \to ^1{\rm S}_0 + ^1{\rm S}_0 )} \  & =
f_D \; 
\\
 & {\cal P}_{21}^{( ^3{\rm P}_2 \to ^3{\rm S}_1 + ^1{\rm S}_0 ) }\  & =
-\sqrt{3\over 2} \; f_D \; 
\\
 & {\cal P}_{\rm LS}^{( ^3{\rm P}_2 \to ^3{\rm S}_1 + ^3{\rm S}_1 ) }\  & =
\left\{
\begin{array}{cr}
-\sqrt{2}\; f_S \;
\ &   ^5{\rm S}_2   \\
\sqrt{1\over 3} \; f_D \;
\ &   ^1{\rm D}_2  \\
-\sqrt{7\over 3} \; f_D \;
\ &   ^5{\rm D}_2  \\
\end{array}
\right. \\
  & & \nonumber \\ \fbox{$^3${\bf P}$_1$} & & \nonumber \\  & & \nonumber \\ 
 & {\cal P}_{\rm LS}^{( ^3{\rm P}_1 \to ^3{\rm S}_1 + ^1{\rm S}_0 ) }\  & =
\left\{
\begin{array}{cr}
f_S  \;
\ &   ^3{\rm S}_1   \\
-\sqrt{5\over 6} \; f_D \;
\ &   ^3{\rm D}_1   \\
\end{array}
\right. 
\\
 & {\cal P}_{\rm LS}^{( ^3{\rm P}_1 \to ^3{\rm S}_1 + ^3{\rm S}_1 ) }\  & =
\left\{
\begin{array}{cr}
0 \;
\ &   ^3{\rm S}_1   \\
0 \;
\ &   ^3{\rm D}_1   \\
-\sqrt{5} \; f_D \;
\ &   ^5{\rm D}_1   \\
\end{array}
\right. \\
  & & \nonumber \\ \fbox{$^3${\bf P}$_0$} & & \nonumber \\  & & \nonumber \\ 
 & {\cal P}_{00}^{( ^3{\rm P}_0 \to ^1{\rm S}_0 + ^1{\rm S}_0 ) }\  & =
\begin{array}{cr}
\sqrt{3\over 2} \; f_S \;
\ &   ^1{\rm S}_0   \\
\end{array}
\\
& {\cal P}_{\rm LS}^{( ^3{\rm P}_0 \to ^3{\rm S}_1 + ^3{\rm S}_1 ) }\  & =
\left\{
\begin{array}{cr}
\sqrt{1\over 2} \; f_S \;
\ &   ^1{\rm S}_0   \\
-\sqrt{20\over 3} \; f_D \;
\ &   ^5{\rm D}_0   \\
\end{array}
\right. \\
  & & \nonumber \\ \fbox{$^1${\bf P}$_1$} & & \nonumber \\  & & \nonumber \\ 
 & {\cal P}_{\rm LS}^{( ^1{\rm P}_1 \to ^3{\rm S}_1 + ^1{\rm S}_0 ) }\  & =
\left\{
\begin{array}{cr}
-\sqrt{1\over 2} \; f_S \;
\ &   ^3{\rm S}_1   \\
-\sqrt{5\over 3} \; f_D \;
\ &   ^3{\rm D}_1   \\
\end{array}
\right. 
\\
 & {\cal P}_{\rm LS}^{( ^1{\rm P}_1 \to ^3{\rm S}_1 + ^3{\rm S}_1 ) }\  & =
\left\{
\begin{array}{cr}
f_S \;
\ &   ^3{\rm S}_1   \\
\sqrt{10 \over 3} \; f_D \;
\ &   ^3{\rm D}_1   \\
0 \;
\ &   ^5{\rm D}_1   \\
\end{array}
\right. 
\end{eqnarray}

\newpage

\vskip .5cm
\hrule
\vskip .1cm
\hrule
\vskip .5cm

\fbox{\bf 2P $\to$ 1S + 1S} 

{(\small \it See 1P $\to$ 1S + 1S for channel coefficients.)}
\vskip .5cm

\begin{eqnarray}
 & f_S  = &{2^{9/2} 5^{1/2} \over 3^{7/2}} \; 
\Big(1 - {4\over 9} x^2 + {4\over 135} x^4 \Big) \; 
\\
 & f_D  = &{2^{11/2} 7 \over 3^5 5 } \; x^2 
\Big(1 - {2\over 21} x^2 \Big) \; 
\end{eqnarray}

\vskip .5cm
\hrule
\vskip .5cm

\fbox{\bf 2P $\to$ 2S + 1S} 

\begin{eqnarray}
 & f_S  = &{2^4 5^{1/2} 7 \over 3^5 } \; 
\Big(1 - {1\over 2} x^2 + {2\over 45} x^4 - {2\over 2835} x^6 \Big) \; 
\\
 & f_D  = &{2^6 (11) \over 3^{11/2} 5 } \; x^2 
\Big(1 - {13\over 132} x^2  + {1\over 594} x^4 \Big) \; 
\end{eqnarray}

\begin{eqnarray}
  & & \nonumber \\ \fbox{$2^3${\bf P}$_2$} & & \nonumber \\  & & \nonumber \\ 
 & {\cal P}_{20}^{ ( 2 ^3{\rm P}_2 \to 2 ^1{\rm S}_0 + ^1{\rm S}_0 )} \  & =
\begin{array}{cr}
f_D \; 
\ &   ^1{\rm D}_2  \\
\end{array}
\\
 & {\cal P}_{21}^{( 2 ^3{\rm P}_2 \to 2 ^3{\rm S}_1 + ^1{\rm S}_0 ) }\  & =
\begin{array}{cr}
-\sqrt{3\over 2} \; f_D \; 
\ &   ^3{\rm D}_2  \\
\end{array}
\\
 & {\cal P}_{21}^{( 2 ^3{\rm P}_2 \to 2 ^1{\rm S}_0 + ^3{\rm S}_1 ) }\  & =
\begin{array}{cr}
+\sqrt{3\over 2} \; f_D \; 
\ &   ^3{\rm D}_2  \\
\end{array}
\\
 & {\cal P}_{\rm LS}^{( 2 ^3{\rm P}_2 \to 2 ^3{\rm S}_1 + ^3{\rm S}_1 ) }\  & =
\left\{
\begin{array}{cr}
-\sqrt{2} \; f_S \;
\ &   ^5{\rm S}_2   \\
\sqrt{1\over 3} \; f_D \;
\ &   ^1{\rm D}_2  \\
0  \;
\ &   ^3{\rm D}_2  \\
-\sqrt{7\over 3} \; f_D \;
\ &   ^5{\rm D}_2  \\
0  \;
\ &   ^5{\rm G}_2  \\
\end{array}
\right.  \\
  & & \nonumber \\ \fbox{$2^3${\bf P}$_1$} & & \nonumber \\  & & \nonumber \\ 
 & {\cal P}_{\rm LS}^{( 2 ^3{\rm P}_1 \to 2 ^3{\rm S}_1 + ^1{\rm S}_0 ) }\  & =
\left\{
\begin{array}{cr}
f_S \;
\ &   ^3{\rm S}_1   \\
-\sqrt{5\over 6} \; f_D \;
\ &   ^3{\rm D}_1  \\
\end{array}
\right.
\\
 & {\cal P}_{\rm LS}^{( 2 ^3{\rm P}_1 \to 2 ^1{\rm S}_0 + ^3{\rm S}_1 ) }\  & =
\left\{
\begin{array}{cr}
-f_S \;
\ &   ^3{\rm S}_1   \\
\sqrt{5\over 6} \; f_D \;
\ &   ^3{\rm D}_1  \\
\end{array}
\right.
\\
 & {\cal P}_{22}^{ ( 2 ^3{\rm P}_1 \to 2 ^3{\rm S}_1 + ^3{\rm S}_1 )} \  & =
\begin{array}{cr}
-\sqrt{5} \; f_D 
\ &   ^5{\rm D}_1  \\
\end{array}
\\
  & & \nonumber \\ \fbox{$2^3${\bf P}$_0$} & & \nonumber \\  & & \nonumber \\ 
 & {\cal P}_{00}^{ ( 2 ^3{\rm P}_0 \to 2 ^1{\rm S}_0 + ^1{\rm S}_0 )} \  & =
\begin{array}{cr}
\sqrt{3\over 2}  \; f_S 
\ &   ^1{\rm S}_0  \\
\end{array}
\\
 & {\cal P}_{\rm LS}^{( 2 ^3{\rm P}_0 \to 2 ^3{\rm S}_1 + ^3{\rm S}_1 ) }\  & =
\left\{
\begin{array}{cr}
\sqrt{1\over 2} \; f_S 
\ &   ^1{\rm S}_0   \\
-\sqrt{20\over 3} \; f_D \;
\ &   ^5{\rm D}_0  \\
\end{array}
\right. \\
  & & \nonumber \\ \fbox{$2^1${\bf P}$_1$} & & \nonumber \\  & & \nonumber \\ 
 & {\cal P}_{\rm LS}^{( 2 ^1{\rm P}_1 \to 2 ^3{\rm S}_1 + ^1{\rm S}_0 ) }\  & =
\left\{
\begin{array}{cr}
-\sqrt{1\over 2} \; f_S \;
\ &   ^3{\rm S}_1   \\
-\sqrt{5\over 3} \; f_D \;
\ &   ^3{\rm D}_1  \\
\end{array}
\right.
\\
 & {\cal P}_{\rm LS}^{( 2 ^1{\rm P}_1 \to 2 ^1{\rm S}_0 + ^3{\rm S}_1 ) }\  & =
\left\{
\begin{array}{cr}
-\sqrt{1\over 2} \; f_S \;
\ &   ^3{\rm S}_1   \\
-\sqrt{5\over 3} \; f_D \;
\ &   ^3{\rm D}_1  \\
\end{array}
\right.
\\
 & {\cal P}_{\rm LS}^{ ( 2 ^1{\rm P}_1 \to 2 ^3{\rm S}_1 + ^3{\rm S}_1 )} \  & =
\left\{
\begin{array}{cr}
f_S 
\ &   ^3{\rm S}_1  \\
\sqrt{10\over 3} \; f_D 
\ &   ^3{\rm D}_1  \\
0 \;
\ &   ^5{\rm D}_1  \\
\end{array}
\right.
\end{eqnarray}

\vskip .5cm
\hrule
\vskip .5cm

\fbox{\bf 2P $\to$ 1P + 1S} 

\begin{eqnarray}
  & & \nonumber \\ \fbox{$2^3${\bf P}$_2$} & & \nonumber \\  & & \nonumber \\ 
 & {\cal P}_{\rm LS}^{( 2 ^3{\rm P}_2 \to  ^1{\rm P}_1 + ^1{\rm S}_0 ) }\  & =
\left\{
\begin{array}{cr}
-{2^{9/2} (13) \over 3^5  5^{1/2} } \;
x \;
\Big( 1 - {8\over 39} x^2  + {4\over 585} x^4 \Big) \;
\ &   ^3{\rm P}_2   \\
-{2^5 \over 3^{9/2}  5^{1/2} } \;
x^3 \;
\Big( 1 - {2\over 45} x^2 \Big) \;
\ &   ^3{\rm F}_2   \\
\end{array}
\right.
\\
 & {\cal P}_{\rm LS}^{( 2 ^3{\rm P}_2 \to  ^3{\rm P}_1 + ^1{\rm S}_0 ) }\  & =
\left\{
\begin{array}{cr}
{2^6  \over 3^4  5^{1/2} } \;
x \;
\Big( 1 - {1\over 4} x^2  + {1\over 90} x^4 \Big) \;
\ &   ^3{\rm P}_2   \\
-{2^{9/2} \over 3^{9/2}  5^{1/2} } \;
x^3 \;
\Big( 1 - {2\over 45} x^2 \Big) \;
\ &   ^3{\rm F}_2   \\
\end{array}
\right.
\\
 & {\cal P}_{\rm LS}^{( 2 ^3{\rm P}_2 \to  ^3{\rm P}_2 + ^1{\rm S}_0 ) }\  & =
\left\{
\begin{array}{cr}
{2^5 7 \over 3^{9/2}  5^{1/2} } \;
x \;
\Big( 1 - {1\over 6} x^2 + {1\over 315} x^4  \Big) \;
\ &   ^5{\rm P}_2   \\
{2^5 \over 3^{9/2}  5^{1/2} } \;
x^3 \;
\Big( 1 - {2\over 45} x^2 \Big) \;
\ &   ^5{\rm F}_2   \\
\end{array}
\right.
\\
 & {\cal P}_{\rm LS}^{( 2 ^3{\rm P}_2 \to  ^1{\rm P}_1 + ^3{\rm S}_1 ) }\  & =
\left\{
\begin{array}{cr}
-{2^6 \over 3^4 5^{1/2} } \;
x \;
\Big( 1 - {1\over 4} x^2 + {1\over 90} x^4 \Big) \;
\ &   ^3{\rm P}_2   \\
-{2^5 7  \over 3^{9/2} 5^{1/2} } \;
x \;
\Big( 1 - {1\over 6} x^2 + {1\over 315} x^4 \Big) \;
\ &   ^5{\rm P}_2   \\
{2^{9/2} \over 3^{9/2}  5^{1/2} } \;
x^3 \;
\Big( 1 - {2\over 45} x^2 \Big) \;
\ &   ^3{\rm F}_2   \\
-{2^5 \over 3^{9/2}  5^{1/2} } \;
x^3 \;
\Big( 1 - {2\over 45} x^2 \Big) \;
\ &   ^5{\rm F}_2   \\
\end{array}
\right.
\\
 & {\cal P}_{\rm LS}^{( 2 ^3{\rm P}_2 \to  ^3{\rm P}_0 + ^3{\rm S}_1 ) }\  & =
\left\{
\begin{array}{cr}
{2^{11/2}5^{1/2}  \over 3^{11/2} } \;
x \;
\Big( 1 - {11\over 30} x^2 + {1\over 45} x^4 \Big) \;
\ &   ^3{\rm P}_2   \\
0 \;
\ &   ^3{\rm F}_2   \\
\end{array}
\right.
\\
 & {\cal P}_{\rm LS}^{( 2 ^3{\rm P}_2 \to  ^3{\rm P}_1 + ^3{\rm S}_1 ) }\  & =
\left\{
\begin{array}{cr}
{2^{7/2} (23) \over 3^5 5^{1/2} } \;
x \;
\Big( 1 - {19\over 69} x^2 + {14\over 1035} x^4 \Big) \;
\ &   ^3{\rm P}_2   \\
{2^{7/2} \over 3^{9/2} 5^{1/2} } \;
x \;
\Big( 1 + {1\over 3} x^2 - {2\over 45} x^4 \Big) \;
\ &   ^5{\rm P}_2   \\
{2^4 \over 3^{9/2}  5^{1/2} } \;
x^3 \;
\Big( 1 - {2\over 45} x^2 \Big) \;
\ &   ^3{\rm F}_2   \\
-{2^{9/2} \over 3^{9/2}  5^{1/2} } \;
x^3 \;
\Big( 1 - {2\over 45} x^2 \Big) \;
\ &   ^5{\rm F}_2   \\
\end{array}
\right.
\\
 & {\cal P}_{\rm LS}^{( 2 ^3{\rm P}_2 \to  ^3{\rm P}_2 + ^3{\rm S}_1 ) }\  & =
\left\{
\begin{array}{cr}
-{2^{7/2} (29) \over 3^{11/2} 5 } \;
x \;
\Big( 1 - {13\over 87} x^2 + {2\over 1305} x^4 \Big) \;
\ &   ^3{\rm P}_2   \\
{2^{7/2}  \over 3^5 5^{1/2} } \;
x \;
\Big( 1 + {1\over 3} x^2 - {2\over 45} x^4 \Big) \;
\ &   ^5{\rm P}_2   \\
{2^{9/2} 7^{1/2} (41)  \over 3^5 5 } \;
x \;
\Big( 1 - {22\over 123} x^2 + {8\over 1845} x^4 \Big) \;
\ &   ^7{\rm P}_2   \\
-{2^4  \over 3^4 5 } \;
x^3 \;
\Big( 1 - {2\over 45} x^2  \Big) \;
\ &   ^3{\rm F}_2   \\
-{2^{9/2}  \over 3^5 5^{1/2} } \;
x^3 \;
\Big( 1 - {2\over 45} x^2  \Big) \;
\ &   ^5{\rm F}_2   \\
{2^{13/2}  \over 3^{9/2} 5 } \;
x^3 \;
\Big( 1 - {2\over 45} x^2  \Big) \;
\ &   ^7{\rm F}_2   \\
0\;
\ &   ^7{\rm H}_2   \\
\end{array}
\right. \\
  & & \nonumber \\ \fbox{$2^3${\bf P}$_1$} & & \nonumber \\  & & \nonumber \\ 
 & {\cal P}_{\rm 11}^{( 2 ^3{\rm P}_1 \to  ^1{\rm P}_1 + ^1{\rm S}_0 ) }\  & =
\begin{array}{cr}
{2^{9/2} 5^{1/2}  \over 3^4 } \;
x \;
\Big( 1 - {2\over 15} x^2 \Big) \;
\ &   ^3{\rm P}_1   \\
\end{array}
\\
 & {\cal P}_{\rm 10}^{( 2 ^3{\rm P}_1 \to  ^3{\rm P}_0 + ^1{\rm S}_0 ) }\  & =
\begin{array}{cr}
{2^4 5^{1/2}  \over 3^{9/2} } \;
x \;
\Big( 1 - {2\over 15} x^2 \Big) \;
\ &   ^1{\rm P}_1   \\
\end{array}
\\
 & {\cal P}_{\rm 11}^{( 2^3{\rm P}_1 \to ^3{\rm P}_1 + ^1{\rm S}_0 ) }\  & =
\begin{array}{cr}
{2^4 5^{3/2}  \over 3^5 } \;
x \;
\Big( 1 - {17\over 75} x^2 + {2\over 225} x^4 \Big) \;
\ &   ^3{\rm P}_1   \\
\end{array}
\\
 & {\cal P}_{\rm LS}^{( 2^3{\rm P}_1 \to ^3{\rm P}_2 + ^1{\rm S}_0 ) }\  & =
\left\{
\begin{array}{cr}
-{2^4  \over 3^{5/2} } \;
x \;
\Big( 1 - {5\over 27} x^2 + {2\over 405} x^4 \Big) \;
\ &   ^5{\rm P}_1   \\
{2^{9/2}   \over 3^5 } \;
x^3 \;
\Big( 1 - {2\over 45} x^2 \Big) \;
\ &   ^5{\rm F}_1   \\
\end{array}
\right. \\
 & {\cal P}_{\rm LS}^{( 2^3{\rm P}_1 \to ^1{\rm P}_1 + ^3{\rm S}_1 ) }\  & =
\left\{
\begin{array}{cr}
-{2^4 5^{1/2} \over 3^{9/2} } \;
x \;
\Big( 1 - {2\over 15} x^2 \Big) \;
\ &   ^1{\rm P}_1   \\
-{2^4 5^{3/2} \over 3^5 } \;
x \;
\Big( 1 - {17\over 75} x^2 + {2\over 225} x^4 \Big) \;
\ &   ^3{\rm P}_1   \\
{2^4 \over 3^{5/2} } \;
x \;
\Big( 1 - {5\over 27} x^2 + {2\over 405} x^4 \Big) \;
\ &   ^5{\rm P}_1   \\
-{2^{9/2}   \over 3^5 } \;
x^3 \;
\Big( 1 - {2\over 45} x^2 \Big) \;
\ &   ^5{\rm F}_1   \\
\end{array}
\right. \\
 & {\cal P}_{\rm 11}^{( 2^3{\rm P}_1 \to ^3{\rm P}_0 + ^3{\rm S}_1 ) }\  & =
\begin{array}{cr}
{2^{9/2} 5^{3/2}  \over 3^{11/2} } \;
x \;
\Big( 1 - {17\over 75} x^2 + {2\over 225} x^4 \Big) \;
\ &   ^3{\rm P}_1   \\
\end{array}
\\
 & {\cal P}_{\rm LS}^{( 2^3{\rm P}_1 \to ^3{\rm P}_1 + ^3{\rm S}_1 ) }\  & =
\left\{
\begin{array}{cr}
0 \;
\ &   ^1{\rm P}_1   \\
{2^{9/2} 5^{1/2} \over 3^5 } \;
x \;
\Big( 1 - {11\over 30} x^2 + {1\over 45} x^4 \Big) \;
\ &   ^3{\rm P}_1   \\
{2^{9/2} 7 \over 3^{9/2} } \;
x \;
\Big( 1 - {1\over 6} x^2 + {1\over 315} x^4 \Big) \;
\ &   ^5{\rm P}_1   \\
-{2^4   \over 3^5 } \;
x^3 \;
\Big( 1 - {2\over 45} x^2 \Big) \;
\ &   ^5{\rm F}_1   \\
\end{array}
\right. \\
 & {\cal P}_{\rm LS}^{( 2^3{\rm P}_1 \to ^3{\rm P}_2 + ^3{\rm S}_1 ) }\  & =
\left\{
\begin{array}{cr}
{2^{9/2} 7 \over 3^{11/2} } \;
x \;
\Big( 1 - {1\over 6} x^2 + {1\over 315} x^4 \Big) \;
\ &   ^3{\rm P}_1   \\
{2^{9/2} 7 \over 3^5 } \;
x \;
\Big( 1 - {1\over 6} x^2 + {1\over 315} x^4 \Big) \;
\ &   ^5{\rm P}_1   \\
-{2^4   \over 3^{11/2} } \;
x^3 \;
\Big( 1 - {2\over 45} x^2 \Big) \;
\ &   ^5{\rm F}_1   \\
{2^{13/2}   \over 3^{11/2} } \;
x^3 \;
\Big( 1 - {2\over 45} x^2 \Big) \;
\ &   ^7{\rm F}_1   \\
\end{array}
\right. \\
  & & \nonumber \\ \fbox{$2^3${\bf P}$_0$} & & \nonumber \\  & & \nonumber \\ 
 & {\cal P}_{\rm 11}^{( 2 ^3{\rm P}_0 \to  ^1{\rm P}_1 + ^1{\rm S}_0 ) }\  & =
\begin{array}{cr}
-{2^{7/2} 5^{1/2} (13) \over 3^5 } \;
x \;
\Big( 1 - {8\over 39} x^2 + {4\over 585} x^4 \Big) \;
\ &   ^3{\rm P}_0   \\
\end{array}
\\
 & {\cal P}_{\rm 11}^{( 2 ^3{\rm P}_0 \to  ^3{\rm P}_1 + ^1{\rm S}_0 ) }\  & =
\begin{array}{cr}
-{2^5 5^{1/2} \over 3^4 } \;
x \;
\Big( 1 - {2\over 15} x^2 \Big) \;
\ &   ^3{\rm P}_0   \\
\end{array}
\\
 & {\cal P}_{\rm 11}^{( 2 ^3{\rm P}_0 \to  ^1{\rm P}_1 + ^3{\rm S}_1 ) }\  & =
\begin{array}{cr}
{2^5 5^{1/2} \over 3^4 } \;
x \;
\Big( 1 - {2\over 15} x^2 \Big) \;
\ &   ^3{\rm P}_0   \\
\end{array}
\\
 & {\cal P}_{\rm 11}^{( 2 ^3{\rm P}_0 \to  ^3{\rm P}_0 + ^3{\rm S}_1 ) }\  & =
\begin{array}{cr}
{2^{7/2} 5^{1/2} (13) \over 3^{11/2} } \;
x \;
\Big( 1 - {8\over 39} x^2 + {4\over 585} x^4 \Big) \;
\ &   ^3{\rm P}_0   \\
\end{array}
\\
 & {\cal P}_{\rm 11}^{( 2 ^3{\rm P}_0 \to  ^3{\rm P}_1 + ^3{\rm S}_1 ) }\  & =
\begin{array}{cr}
{2^{7/2} 5^{1/2} (13) \over 3^5 } \;
x \;
\Big( 1 - {8\over 39} x^2 + {4\over 585} x^4 \Big) \;
\ &   ^3{\rm P}_0   \\
\end{array}
\\
 & {\cal P}_{\rm LS}^{( 2^3{\rm P}_0 \to ^3{\rm P}_2 + ^3{\rm S}_1 ) }\  & =
\left\{
\begin{array}{cr}
-{2^{7/2} (13) \over 3^{11/2} } \;
x \;
\Big( 1 - {8\over 39} x^2 + {4\over 585} x^4 \Big) \;
\ &   ^3{\rm P}_0   \\
{2^6   \over 3^5 } \;
x^3 \;
\Big( 1 - {2\over 45} x^2 \Big) \;
\ &   ^7{\rm F}_0   \\
\end{array}
\right. \\
  & & \nonumber \\ \fbox{$2^1${\bf P}$_1$} & & \nonumber \\  & & \nonumber \\ 
 & {\cal P}_{\rm 11}^{( 2^1{\rm P}_1 \to ^1{\rm P}_1 + ^1{\rm S}_0 ) }\  & =
\begin{array}{cr}
0\;
\ &   ^3{\rm P}_1   \\
\end{array}
\\
 & {\cal P}_{\rm 10}^{( 2^1{\rm P}_1 \to ^3{\rm P}_0 + ^1{\rm S}_0 ) }\  & =
\begin{array}{cr}
{2^{7/2} 5^{1/2} 7 \over 3^{11/2} } \;
x \;
\Big( 1 - {4\over 15} x^2 + {4\over 315} x^4 \Big) \;
\ &   ^1{\rm P}_1   \\
\end{array}
\\
 & {\cal P}_{\rm 10}^{( 2^1{\rm P}_1 \to ^3{\rm P}_1 + ^1{\rm S}_0 ) }\  & =
\begin{array}{cr}
{2^{7/2} 5^{1/2} \over 3^4 } \;
x \;
\Big( 1 - {2\over 15} x^2  \Big) \;
\ &   ^1{\rm P}_1   \\
\end{array}
\\
 & {\cal P}_{\rm LS}^{( 2^1{\rm P}_1 \to ^3{\rm P}_2 + ^1{\rm S}_0 ) }\  & =
\left\{
\begin{array}{cr}
{2^{7/2} (41) \over 3^{11/2} } \;
x \;
\Big( 1 - {22\over 123} x^2 + {8\over 1845} x^4 \Big) \;
\ &   ^5{\rm P}_1   \\
{2^5   \over 3^5 } \;
x^3 \;
\Big( 1 - {2\over 45} x^2 \Big) \;
\ &   ^5{\rm F}_1   \\
\end{array}
\right. \\
 & {\cal P}_{\rm LS}^{( 2^1{\rm P}_1 \to ^1{\rm P}_1 + ^3{\rm S}_1 ) }\  & =
\left\{
\begin{array}{cr}
{2^{7/2} 5^{1/2} 7 \over 3^{11/2} } \;
x \;
\Big( 1 - {4\over 15} x^2 + {4\over 315} x^4 \Big) \;
\ &   ^1{\rm P}_1   \\
{2^{7/2} 5^{1/2}   \over 3^4 } \;
x \;
\Big( 1 - {2\over 15} x^2 \Big) \;
\ &   ^3{\rm P}_1   \\
{2^{7/2} (41) \over 3^{11/2} } \;
x \;
\Big( 1 - {22\over 123} x^2 + {8\over 1845} x^4 \Big) \;
\ &   ^5{\rm P}_1   \\
{2^5   \over 3^5 } \;
x^3 \;
\Big( 1 - {2\over 45} x^2 \Big) \;
\ &   ^5{\rm F}_1   \\
\end{array}
\right. \\
 & {\cal P}_{\rm 11}^{( 2^1{\rm P}_1 \to ^3{\rm P}_0 + ^3{\rm S}_1 ) }\  & =
\begin{array}{cr}
{2^4 5^{1/2}  \over 3^{9/2} } \;
x \;
\Big( 1 - {2\over 15} x^2 \Big) \;
\ &   ^3{\rm P}_1   \\
\end{array}
\end{eqnarray}

\newpage

\vskip .5cm
\hrule
\vskip .1cm
\hrule
\vskip .5cm

\fbox{\bf 1D $\to$ 1S + 1S}

\begin{eqnarray}
 &f_P  &= {2^{13/2}  \over 3^4  }\; x \Big(1 - {2\over 15} x^2\Big) \; 
\\
 &f_F  &= -{2^6  \over 3^{9/2} 5^{1/2} 7^{1/2} }\; x^3  \; 
\end{eqnarray}

\begin{eqnarray}
  & & \nonumber \\ \fbox{$^3${\bf D}$_3$} & & \nonumber \\  & & \nonumber \\ 
 & {\cal P}_{30}^{ ( ^3{\rm D}_3 \to ^1{\rm S}_0 + ^1{\rm S}_0 )} \  & =
\begin{array}{cr}
f_F \; 
\ &   ^1{\rm F}_3  \\
\end{array}
\\
 & {\cal P}_{31}^{( ^3{\rm D}_3 \to ^3{\rm S}_1 + ^1{\rm S}_0 ) }\  & =
\begin{array}{cr}
-\sqrt{4\over 3} \; f_F \; 
\ &   ^3{\rm F}_3  \\
\end{array}
\\
 & {\cal P}_{\rm LS}^{( ^3{\rm D}_3 \to ^3{\rm S}_1 + ^3{\rm S}_1 ) }\  & =
\left\{
\begin{array}{cr}
 f_P \;
\ &   ^5{\rm P}_3   \\
\sqrt{1\over 3} \; f_F \;
\ &   ^1{\rm F}_3  \\
0 \;
\ &   ^3{\rm F}_3  \\
-\sqrt{8\over 5} \; f_F \;
\ &   ^5{\rm F}_3  \\
0 \;
\ &   ^5{\rm H}_3  \\
\end{array}
\right. \\
  & & \nonumber \\ \fbox{$^3${\bf D}$_2$} & & \nonumber \\  & & \nonumber \\ 
 & {\cal P}_{\rm LS}^{( ^3{\rm D}_2 \to ^3{\rm S}_1 + ^1{\rm S}_0 ) }\  & =
\left\{
\begin{array}{cr}
-\sqrt{3\over 8} \; f_P  \;
\ &   ^3{\rm P}_2   \\
-\sqrt{14\over 15} \; f_F \;
\ &   ^3{\rm F}_2   \\
\end{array}
\right.  \\
 & {\cal P}_{\rm LS}^{( ^3{\rm D}_2 \to ^3{\rm S}_1 + ^3{\rm S}_1 ) }\  & =
\left\{
\begin{array}{cr}
{1\over 2} \; f_P  \;
\ &   ^5{\rm P}_2   \\
0\;
\ &   ^3{\rm F}_2   \\
-\sqrt{56\over 15} \; f_F \;
\ &   ^5{\rm F}_2   \\
\end{array}
\right.  \\
  & & \nonumber \\ \fbox{$^3${\bf D}$_1$} & & \nonumber \\  & & \nonumber \\ 
 & {\cal P}_{10}^{( ^3{\rm D}_1 \to ^1{\rm S}_0 + ^1{\rm S}_0 ) }\  & =
\begin{array}{cr}
-\sqrt{5\over 12} \; f_P \;
\ &   ^1{\rm P}_1   \\
\end{array}
\\
 & {\cal P}_{11}^{( ^3{\rm D}_1 \to ^3{\rm S}_1 + ^1{\rm S}_0 ) }\  & =
\begin{array}{cr}
-\sqrt{5\over 24} \; f_P \;
\ &   ^3{\rm P}_1   \\
\end{array}
\\
& {\cal P}_{\rm LS}^{( ^3{\rm D}_1 \to ^3{\rm S}_1 + ^3{\rm S}_1 ) }\  & =
\left\{
\begin{array}{cr}
-{\sqrt{5}\over 6} \; f_P \;
\ &   ^1{\rm P}_1   \\
0\;
\ &   ^3{\rm P}_1   \\
{1\over 6} \; f_P \;
\ &   ^5{\rm P}_1   \\
-\sqrt{28\over 5} \; f_F \;
\ &   ^5{\rm F}_1   \\
\end{array}
\right. \\
  & & \nonumber \\ \fbox{$^1${\bf D}$_2$} & & \nonumber \\  & & \nonumber \\ 
 & {\cal P}_{\rm LS}^{( ^1{\rm D}_2 \to ^3{\rm S}_1 + ^1{\rm S}_0 ) }\  & =
\left\{
\begin{array}{cr}
{1\over 2} \; f_P \;
\ &   ^3{\rm P}_2   \\
-\sqrt{7\over 5} \; f_F \;
\ &   ^3{\rm F}_2   \\
\end{array}
\right.  \\
 & {\cal P}_{\rm LS}^{( ^1{\rm D}_2 \to ^3{\rm S}_1 + ^3{\rm S}_1 ) }\  & =
\left\{
\begin{array}{cr}
-\sqrt{1\over 2} \; f_P \;
\ &   ^3{\rm P}_2   \\
0\;
\ &   ^5{\rm P}_2   \\
\sqrt{14\over 5} \; f_F \;
\ &   ^3{\rm F}_2   \\
0\;
\ &   ^5{\rm F}_2   \\
\end{array}
\right. 
\end{eqnarray}

\vskip .5cm
\hrule
\vskip .5cm

\fbox{\bf 1D $\to$ 2S + 1S}

\begin{eqnarray}
 &f_P  &= {2^6  \over 3^{11/2}  }\; x \Big(1 - {29\over 30} x^2 +
{1\over 45} x^4 \Big) \; 
\\
 &f_F  &= -{2^{13/2} \over 3^5 5^{1/2} 7^{1/2} }\; x^3 \Big(1-
{1\over 36}x^2 \Big)   \; 
\end{eqnarray}

\begin{eqnarray}
  & & \nonumber \\ \fbox{$^3${\bf D}$_3$} & & \nonumber \\  & & \nonumber \\ 
 & {\cal P}_{30}^{ ( ^3{\rm D}_3 \to 2^1{\rm S}_0 + ^1{\rm S}_0 )} \  & =
\begin{array}{cr}
f_F \; 
\ &   ^1{\rm F}_3   \\
\end{array}
\\
 & {\cal P}_{31}^{( ^3{\rm D}_3 \to 2^3{\rm S}_1 + ^1{\rm S}_0 ) }\  & =
\begin{array}{cr}
-\sqrt{4\over 3} \; f_F \; 
\end{array}
\\
 & {\cal P}_{31}^{( ^3{\rm D}_3 \to 2^1{\rm S}_0 + ^3{\rm S}_1 ) }\  & =
\begin{array}{cr}
\sqrt{4\over 3} \; f_F \; 
\end{array}
\\
 & {\cal P}_{\rm LS}^{( ^3{\rm D}_3 \to 2^3{\rm S}_1 + ^3{\rm S}_1 ) }\  & =
\left\{
\begin{array}{cr}
 f_P \;
\ &   ^5{\rm P}_3   \\
\sqrt{1\over 3} \; f_F \;
\ &   ^1{\rm F}_3  \\
0\;
\ &   ^3{\rm F}_3  \\
-\sqrt{8\over 5} \; f_F \;
\ &   ^5{\rm F}_3  \\
0\;
\ &   ^5{\rm H}_3  \\
\end{array}
\right. \\
  & & \nonumber \\ \fbox{$^3${\bf D}$_2$} & & \nonumber \\  & & \nonumber \\ 
 & {\cal P}_{\rm LS}^{( ^3{\rm D}_2 \to 2^3{\rm S}_1 + ^1{\rm S}_0 ) }\  & =
\left\{
\begin{array}{cr}
-\sqrt{3\over 8} \; f_P  \;
\ &   ^3{\rm P}_2   \\
-\sqrt{14\over 15} \; f_F \;
\ &   ^3{\rm F}_2   \\
\end{array}
\right.  \\
 & {\cal P}_{\rm LS}^{( ^3{\rm D}_2 \to 2^1{\rm S}_0 + ^3{\rm S}_1 ) }\  & =
\left\{
\begin{array}{cr}
\sqrt{3\over 8} \; f_P  \;
\ &   ^3{\rm P}_2   \\
\sqrt{14\over 15} \; f_F \;
\ &   ^3{\rm F}_2   \\
\end{array}
\right.  \\
 & {\cal P}_{\rm LS}^{( ^3{\rm D}_2 \to 2^3{\rm S}_1 + ^3{\rm S}_1 ) }\  & =
\left\{
\begin{array}{cr}
0\;
\ &   ^3{\rm P}_2  \\
{1\over 2} \; f_P  \;
\ &   ^5{\rm P}_2   \\
0\;
\ &   ^3{\rm F}_2  \\
-\sqrt{56 \over 15} \; f_F \;
\ &   ^5{\rm F}_2   \\
\end{array}
\right.  \\
  & & \nonumber \\ \fbox{$^3${\bf D}$_1$} & & \nonumber \\  & & \nonumber \\ 
 & {\cal P}_{10}^{( ^3{\rm D}_1 \to 2^1{\rm S}_0 + ^1{\rm S}_0 ) }\  & =
\begin{array}{cr}
-\sqrt{5\over 12} \; f_P \;
\ &   ^1{\rm P}_1  \\
\end{array}
\\
 & {\cal P}_{11}^{( ^3{\rm D}_1 \to 2^3{\rm S}_1 + ^1{\rm S}_0 ) }\  & =
\begin{array}{cr}
-\sqrt{5\over 24} \; f_P \;
\ &   ^1{\rm P}_1  \\
\end{array}
\\
 & {\cal P}_{11}^{( ^3{\rm D}_1 \to 2^1{\rm S}_0 + ^3{\rm S}_1 ) }\  & =
\begin{array}{cr}
\sqrt{5\over 24} \; f_P \;
\ &   ^1{\rm P}_1  \\
\end{array}
\\
& {\cal P}_{\rm LS}^{( ^3{\rm D}_1 \to 2^3{\rm S}_1 + ^3{\rm S}_1 ) }\  & =
\left\{
\begin{array}{cr}
-\sqrt{5\over 36} \; f_P \;
\ &   ^1{\rm P}_1   \\
0\;
\ &   ^3{\rm P}_1   \\
{1\over 6} \; f_P \;
\ &   ^5{\rm P}_1   \\
-\sqrt{28\over 5} \; f_F \;
\ &   ^5{\rm F}_1   \\
\end{array}
\right. \\
  & & \nonumber \\ \fbox{$^1${\bf D}$_2$} & & \nonumber \\  & & \nonumber \\ 
 & {\cal P}_{\rm LS}^{( ^1{\rm D}_2 \to 2^3{\rm S}_1 + ^1{\rm S}_0 ) }\  & =
\left\{
\begin{array}{cr}
{1\over 2} \; f_P \;
\ &   ^3{\rm P}_2   \\
-\sqrt{7\over 5} \; f_F \;
\ &   ^3{\rm F}_2   \\
\end{array}
\right. 
\end{eqnarray}
\begin{eqnarray}
 & {\cal P}_{\rm LS}^{( ^1{\rm D}_2 \to 2^1{\rm S}_0 + ^3{\rm S}_1 ) }\  & =
\left\{
\begin{array}{cr}
{1\over 2} \; f_P \;
\ &   ^3{\rm P}_2   \\
-\sqrt{7\over 5} \; f_F \;
\ &   ^3{\rm F}_2   \\
\end{array}
\right.  \\
 & {\cal P}_{\rm LS}^{( ^1{\rm D}_2 \to 2^3{\rm S}_1 + ^3{\rm S}_1 ) }\  & =
\left\{
\begin{array}{cr}
-\sqrt{1\over 2} \; f_P \;
\ &   ^3{\rm P}_2   \\
0\;
\ &   ^5{\rm P}_2   \\
\sqrt{14\over 5} \; f_F \;
\ &   ^3{\rm F}_2   \\
0\;
\ &   ^5{\rm F}_2   \\
\end{array}
\right. 
\end{eqnarray}

\vskip .5cm
\hrule
\vskip .5cm

\fbox{\bf 1D $\to$ 1P + 1S} 

\begin{eqnarray}
  & & \nonumber \\ \fbox{$^3${\bf D}$_3$} & & \nonumber \\  & & \nonumber \\ 
 & {\cal P}_{\rm LS}^{( ^3{\rm D}_3 \to ^1{\rm P}_1 + ^1{\rm S}_0 ) }\  & =
\left\{
\begin{array}{cr}
{2^{11/2} \over 3^4 5^{1/2} } 
\; 
x^2 
\Big( 1 - {1\over 21} x^2 \Big) 
\;
\ &   ^3{\rm D}_3   
\\
{2^{13/2} \over 3^{11/2} 5^{1/2} 7 } 
\; 
x^4 
\;
\ &   ^3{\rm G}_3   
\\
\end{array}
\right.  \\
 & {\cal P}_{\rm LS}^{( ^3{\rm D}_3 \to ^3{\rm P}_1 + ^1{\rm S}_0 ) }\  & =
\left\{
\begin{array}{cr}
-{2^6 \over 3^5 5^{1/2} } 
\; 
x^2 
\Big( 1 - {2\over 21} x^2 \Big) 
\;
\ &   ^3{\rm D}_3   
\\
{2^6 \over 3^{11/2} 5^{1/2} 7 } 
\; 
x^4 
\;
\ &   ^3{\rm G}_3   
\\
\end{array}
\right.  \\
 & {\cal P}_{\rm LS}^{( ^3{\rm D}_3 \to ^3{\rm P}_2 + ^1{\rm S}_0 ) }\  & =
\left\{
\begin{array}{cr}
-{2^{15/2}  \over 3^5 5^{1/2} } 
\; 
x^2 
\Big( 1 - {1\over 42} x^2 \Big) 
\;
\ &   ^5{\rm D}_3   
\\
-{2^6 \over 3^6 7 } 
\; 
x^4 
\;
\ &   ^5{\rm G}_3   
\\
\end{array}
\right.  
\\
 & {\cal P}_{\rm LS}^{( ^3{\rm D}_3 \to ^1{\rm P}_1 + ^3{\rm S}_1 ) }\  & =
\left\{
\begin{array}{cr}
{2^6 \over 3^5 5^{1/2} } 
\; 
x^2 
\Big( 1 - {2\over 21} x^2 \Big) 
\;
\ &   ^3{\rm D}_3   
\\
{2^{15/2} \over 3^5 5^{1/2} } 
\; 
x^2 
\Big( 1 - {1\over 42} x^2 \Big) 
\;
\ &   ^5{\rm D}_3   
\\
-{2^6 \over 3^{11/2} 5^{1/2}  7 } 
\; 
x^4 
\;
\ &   ^3{\rm G}_3   
\\
{2^6 \over 3^6 7 } 
\; 
x^4 
\;
\ &   ^5{\rm G}_3   
\\
\end{array}
\right.  \\
 & {\cal P}_{\rm 21}^{( ^3{\rm D}_3 \to ^3{\rm P}_0 + ^3{\rm S}_1 ) }\  & =
\left\{
\begin{array}{cr}
{2^{11/2} \over 3^{11/2} 5^{1/2} } 
\; 
x^2 
\Big( 1 + {1\over 3} x^2 \Big) \;
\ &   ^3{\rm D}_3   
\\
0 \;
\ &   ^3{\rm G}_3   
\\
\end{array}
\right.  \\
 & {\cal P}_{\rm LS}^{( ^3{\rm D}_3 \to ^3{\rm P}_1 + ^3{\rm S}_1 ) }\  & =
\left\{
\begin{array}{cr}
-{2^{11/2} \over 3^5 5^{1/2} } 
\; 
x^2 
\Big( 1 - {5\over 21} x^2 \Big) 
\;
\ &   ^3{\rm D}_3   
\\
-{2^6 \over 3^5 5^{1/2} } 
\; 
x^2 
\Big( 1 + {1\over 21} x^2 \Big) 
\;
\ &   ^5{\rm D}_3   
\\
-{2^{11/2} \over 3^{11/2} 5^{1/2}  7 } 
\; 
x^4 
\;
\ &   ^3{\rm G}_3   
\\
{2^{11/2} \over 3^6 7 } 
\; 
x^4 
\;
\ &   ^5{\rm G}_3   
\\
\end{array}
\right.  \\
 & {\cal P}_{\rm LS}^{( ^3{\rm D}_3 \to ^3{\rm P}_2 + ^3{\rm S}_1 ) }\  & =
\left\{
\begin{array}{cr}
-{2^7 \over 3^{7/2} } 
\; 
\Big( 1 - {5\over 18} x^2 + {1\over 135} x^4 \Big) 
\;
\ &   ^7{\rm S}_3   
\\
{2^{11/2} \over 3^{11/2} } 
\; 
x^2 
\Big( 1 - {1\over 105} x^2 \Big) 
\;
\ &   ^3{\rm D}_3   
\\
-{2^6 \over 3^{11/2} 5^{1/2} } 
\; 
x^2 
\Big( 1 + {1\over 21} x^2 \Big) 
\;
\ &   ^5{\rm D}_3   
\\
-{2^7 \over 3^5 } 
\; 
x^2 
\Big( 1 - {4\over 105} x^2 \Big) 
\;
\ &   ^7{\rm D}_3   
\\
{2^{11/2} \over 3^5 5\;  7 } 
\; 
x^4 
\;
\ &   ^3{\rm G}_3   
\\
{2^{11/2} \over 3^{13/2} 7 } 
\; 
x^4 
\;
\ &   ^5{\rm G}_3   
\\
-{2^{15/2} (11)^{1/2} \over 3^{13/2} 5\; 7 } 
\; 
x^4 
\;
\ &   ^7{\rm G}_3   
\\
\end{array}
\right.  \\
  & & \nonumber \\ \fbox{$^3${\bf D}$_2$} & & \nonumber \\  & & \nonumber \\ 
 & {\cal P}_{\rm 21}^{( ^3{\rm D}_2 \to ^1{\rm P}_1 + ^1{\rm S}_0 ) }\  & =
\begin{array}{cr}
-{2^{11/2} \over 3^4 5^{1/2} } 
\; 
x^2 
\ &   ^3{\rm D}_2   \\
\end{array}
\\
 & {\cal P}_{\rm 20}^{( ^3{\rm D}_2 \to ^3{\rm P}_0 + ^1{\rm S}_0 ) }\  & =
\begin{array}{cr}
-{2^5 \over 3^4 5^{1/2} } 
\; 
x^2 
\ &   ^1{\rm D}_2   \\
\end{array}
\\
 & {\cal P}_{\rm 21}^{( ^3{\rm D}_2 \to ^3{\rm P}_1 + ^1{\rm S}_0 ) }\  & =
\begin{array}{cr}
-{2^4 7 \over 3^5 5^{1/2} } 
\; 
x^2 
\;
\Big( 1 - {2\over 21} x^2 \Big)
\ &   ^3{\rm D}_2   \\
\end{array}
\\
 & {\cal P}_{\rm LS}^{( ^3{\rm D}_2 \to ^3{\rm P}_2 + ^1{\rm S}_0 ) }\  & =
\left\{
\begin{array}{cr}
{2^{11/2} \over 3^3 } \;
\Big( 1 - {5\over 18} x^2 + {1\over 135} x^4 \Big) \;
\ &   ^5{\rm S}_2   \\
{2^4 7^{1/2} \over 3^5 5^{1/2} } \;
x^2 \;
\Big( 1 - {2\over 21} x^2 \Big) \;
\ &   ^5{\rm D}_2   \\
-{2^7 \over 3^6 5\; 7^{1/2}  } \;
x^4 \;
\ &   ^5{\rm G}_2   \\
\end{array}
\right.
\\
 & {\cal P}_{\rm LS}^{( ^3{\rm D}_2 \to ^1{\rm P}_1 + ^3{\rm S}_1 ) }\  & =
\left\{
\begin{array}{cr}
-{2^{11/2} \over 3^3 } \;
\Big( 1 - {5\over 18} x^2 + {1\over 135} x^4 \Big) \;
\ &   ^5{\rm S}_2   \\
{2^5 \over 3^4 5^{1/2} } \;
x^2 \;
\ &   ^1{\rm D}_2   \\
{2^4 7 \over 3^5 5^{1/2} } \;
x^2 \;
\Big( 1 - {2\over 21} x^2 \Big) \;
\ &   ^3{\rm D}_2   \\
-{2^4 7^{1/2} \over 3^5 5^{1/2} } \;
x^2 \;
\Big( 1 - {2\over 21} x^2 \Big) \;
\ &   ^5{\rm D}_2   \\
{2^7 \over 3^6 5\; 7^{1/2}  } \;
x^4 \;
\ &   ^5{\rm G}_2   \\
\end{array}
\right.
\\ 
& {\cal P}_{\rm 21}^{( ^3{\rm D}_2 \to ^3{\rm P}_0 + ^3{\rm S}_1 ) }\  & =
\begin{array}{cr}
-{2^{9/2} 7 \over 3^{11/2} 5^{1/2} } 
\; 
x^2 
\;
\Big( 1 - {2\over 21} x^2 \Big)
\ &   ^3{\rm D}_2   \\
\end{array}
\\
 & {\cal P}_{\rm LS}^{( ^3{\rm D}_2 \to ^3{\rm P}_1 + ^3{\rm S}_1 ) }\  & =
\left\{
\begin{array}{cr}
-{2^5 \over 3^3 } \;
\Big( 1 - {5\over 18} x^2 + {1\over 135} x^4 \Big) \;
\ &   ^5{\rm S}_2   \\
-{2^{7/2} \over 3^5 5^{1/2} } \;
x^2 
\Big( 1 - {2\over 3} x^2 \Big) \;
\ &   ^3{\rm D}_2   \\
-{2^{7/2} 7^{3/2} \over 3^5 5^{1/2} } \;
x^2 \;
\Big( 1 - {2\over 147} x^2 \Big) \;
\ &   ^5{\rm D}_2   \\
{2^{13/2} \over 3^6 5\; 7^{1/2}  } \;
x^4 \;
\ &   ^5{\rm G}_2   \\
\end{array}
\right.
\\
 & {\cal P}_{\rm LS}^{( ^3{\rm D}_2 \to ^3{\rm P}_2 + ^3{\rm S}_1 ) }\  & =
\left\{
\begin{array}{cr}
-{2^5 \over 3^{7/2} } \;
\Big( 1 - {5\over 18} x^2 + {1\over 135} x^4  \Big) \;
\ &   ^5{\rm S}_2   \\
-{2^{7/2} 5 \over 3^{11/2} } \;
x^2 \;
\Big( 1 - {2\over 75} x^2  \Big) \;
\ &   ^3{\rm D}_2   \\
-{2^{7/2} 7^{3/2} \over 3^{11/2} 5^{1/2} } \;
x^2 \;
\Big( 1 - {2\over 147} x^2  \Big) \;
\ &   ^5{\rm D}_2   \\
-{2^6 7^{1/2} \over 3^{11/2} } \;
x^2 \;
\Big( 1 - {4\over 105} x^2  \Big) \;
\ &   ^7{\rm D}_2   \\
{2^{13/2}  \over 3^{13/2} 5\; 7^{1/2}} \;
x^4 \;
\ &   ^5{\rm G}_2   \\
-{2^{15/2}  \over 3^{13/2} 5^{1/2} 7^{1/2}} \;
x^4 \;
\ &   ^7{\rm G}_2   \\
\end{array}
\right. \\
  & & \nonumber \\ \fbox{$^3${\bf D}$_1$} & & \nonumber \\  & & \nonumber \\ 
 & {\cal P}_{\rm LS}^{( ^3{\rm D}_1 \to ^1{\rm P}_1 + ^1{\rm S}_0 ) }\  & =
\left\{
\begin{array}{cr}
{2^6 5^{1/2} \over 3^4 } \;
\Big( 1 - {5\over 18} x^2 + {1\over 135} x^4 \Big) \;
\ &   ^3{\rm S}_1   \\
{2^{15/2} \over 3^6 5^{1/2} } \;
x^2 \;
\Big( 1 - {1\over 6} x^2 \Big) \;
\ &   ^3{\rm D}_1   \\
\end{array}
\right.
\\
 & {\cal P}_{\rm LS}^{( ^3{\rm D}_1 \to ^3{\rm P}_1 + ^1{\rm S}_0 ) }\  & =
\left\{
\begin{array}{cr}
{2^{11/2} 5^{1/2} \over 3^4 } \;
\Big( 1 - {5\over 18} x^2 + {1\over 135} x^4 \Big) \;
\ &   ^3{\rm S}_1   \\
{2^4 (23) \over 3^6 5^{1/2} } \;
x^2 
\Big( 1 + {2\over 69} x^2 \Big) \;
\ &   ^3{\rm D}_1   \\
\end{array}
\right.
\\
& {\cal P}_{\rm 22}^{( ^3{\rm D}_1 \to ^3{\rm P}_2 + ^1{\rm S}_0 ) }\  & =
\begin{array}{cr}
{2^4 (13) \over 3^{11/2}  5^{1/2} } 
\; 
x^2 
\Big( 1 - {2\over 39} x^2 \Big)
\ &   ^5{\rm D}_1   \\
\end{array}
\\
 & {\cal P}_{\rm LS}^{( ^3{\rm D}_1 \to ^1{\rm P}_1 + ^3{\rm S}_1 ) }\  & =
\left\{
\begin{array}{cr}
-{2^{11/2} 5^{1/2} \over 3^4 } \;
\Big( 1 - {5\over 18} x^2 + {1\over 135} x^4 \Big) \;
\ &   ^3{\rm S}_1   \\
-{2^4 (23) \over 3^6 5^{1/2} } \;
x^2 
\Big( 1 + {2\over 69} x^2 \Big) \;
\ &   ^3{\rm D}_1   \\
-{2^4 (13) \over 3^{11/2}  5^{1/2} } 
\; 
x^2 
\Big( 1 - {2\over 39} x^2 \Big)
\ &   ^5{\rm D}_1   \\
\end{array}
\right.
\\
& {\cal P}_{\rm LS}^{( ^3{\rm D}_1 \to ^3{\rm P}_0 + ^3{\rm S}_1 ) }\  & =
\left\{
\begin{array}{cr}
0 \;
\ &   ^3{\rm S}_1   \\
-{2^{9/2} (13) \over 3^{11/2} 5^{1/2} } \;
x^2 \;
\Big( 1 - {2\over 39} x^2 \Big) \;
\ &   ^3{\rm D}_1   \\
\end{array}
\right.
\\
 & {\cal P}_{\rm LS}^{( ^3{\rm D}_1 \to ^3{\rm P}_1 + ^3{\rm S}_1 ) }\  & =
\left\{
\begin{array}{cr}
-{2^5 5^{1/2} \over 3^4 } \;
\Big( 1 - {5\over 18} x^2 + {1\over 135} x^4 \Big) \;
\ &   ^3{\rm S}_1   \\
-{2^{7/2} (47) \over 3^6 5^{1/2} } \;
x^2 
\Big( 1 - {10\over 141} x^2 \Big) \;
\ &   ^3{\rm D}_1   \\
-{2^{7/2} (31) \over 3^{11/2}  5^{1/2} } 
\; 
x^2 
\Big( 1 - {2\over 93} x^2 \Big)
\ &   ^5{\rm D}_1   \\
\end{array}
\right.
\\
 & {\cal P}_{\rm LS}^{( ^3{\rm D}_1 \to ^3{\rm P}_2 + ^3{\rm S}_1 ) }\  & =
\left\{
\begin{array}{cr}
{2^5  \over 3^{7/2} } \;
\Big( 1 - {5\over 18} x^2 + {1\over 135} x^4 \Big) \;
\ &   ^3{\rm S}_1   \\
-{2^{7/2} \over 3^{11/2}  } 
\; 
x^2 
\Big( 1 + {2\over 15} x^2 \Big)
\ &   ^3{\rm D}_1   \\
-{2^{7/2} (31) \over 3^6  5^{1/2} } 
\; 
x^2 
\Big( 1 - {2\over 93} x^2 \Big)
\ &   ^5{\rm D}_1   \\
-{2^{11/2} 7^{1/2} \over 3^6 } 
\; 
x^2 
\Big( 1 - {4\over 105} x^2 \Big)
\ &   ^7{\rm D}_1   \\
-{2^{15/2} \over 3^{11/2}  5 \; 7^{1/2}  } 
\; 
x^4 
\ &   ^7{\rm G}_1   \\
\end{array}
\right.
\\
  & & \nonumber \\ \fbox{$^1${\bf D}$_2$} & & \nonumber \\  & & \nonumber \\ 
 & {\cal P}_{\rm 21}^{( ^1{\rm D}_2 \to ^1{\rm P}_1 + ^1{\rm S}_0 ) }\  & =
\begin{array}{cr}
0 \;
\ &   ^3{\rm D}_2   \\
\end{array}
\\
 & {\cal P}_{\rm 20}^{( ^1{\rm D}_2 \to ^3{\rm P}_0 + ^1{\rm S}_0 ) }\  & =
\begin{array}{cr}
-{ 2^{13/2} \over 3^{11/2} 5^{1/2}} \; x^2 \;
\Big( 1 - {1\over 6} x^2 \Big) \;
\    &^1{\rm D}_2   \\
\end{array}
\\
 & {\cal P}_{\rm 21}^{( ^1{\rm D}_2 \to ^3{\rm P}_1 + ^1{\rm S}_0 ) }\  & =
\begin{array}{cr}
-{ 2^{9/2} \over 3^{7/2} 5^{1/2}} \; x^2 \;
\    &^3{\rm D}_2   \\
\end{array}
\\
 & {\cal P}_{\rm LS}^{( ^1{\rm D}_2 \to ^3{\rm P}_2 + ^1{\rm S}_0 ) }\  & =
\left\{
\begin{array}{cr}
-{2^6 \over 3^{7/2} } \;
\Big( 1 - {5\over 18} x^2 + {1\over 135} x^4 \Big) \;
\ &   ^5{\rm S}_2   \\
-{2^{9/2} 5^{1/2} 7^{1/2} \over 3^{11/2} } \;
x^2 \;
\Big( 1 - {4\over 105} x^2 \Big) \;
\ &   ^5{\rm D}_2   \\
-{2^{13/2} \over 3^{11/2}5\; 7^{1/2} } \;
x^4 \;
\ &   ^5{\rm G}_2   \\
\end{array}
\right.
\\
 & {\cal P}_{\rm LS}^{( ^1{\rm D}_2 \to ^1{\rm P}_1 + ^3{\rm S}_1 ) }\  & =
\left\{
\begin{array}{cr}
-{2^6 \over 3^{7/2} } \;
\Big( 1 - {5\over 18} x^2 + {1\over 135} x^4 \Big) \;
\ &   ^5{\rm S}_2   \\
-{2^{13/2}  \over 3^{11/2} 5^{1/2}} \;
x^2 \;
\Big( 1 - {1\over 6} x^2 \Big) \;
\ &   ^1{\rm D}_2   \\
-{2^{9/2}   \over 3^{7/2} 5^{1/2}} \;
x^2 \;
\ &   ^3{\rm D}_2   \\
-{2^{9/2} 5^{1/2} 7^{1/2} \over 3^{11/2} } \;
x^2 \;
\Big( 1 - {4\over 105} x^2 \Big) \;
\ &   ^5{\rm D}_2   \\
-{2^{13/2} \over 3^{11/2}5\; 7^{1/2} } \;
x^4 \;
\ &   ^5{\rm G}_2   \\
\end{array}
\right.
\\
 & {\cal P}_{\rm 21}^{( ^1{\rm D}_2 \to ^3{\rm P}_0 + ^3{\rm S}_1 ) }\  & =
\begin{array}{cr}
-{2^5  \over 3^4 5^{1/2}} \;
x^2 \;
\    &^3{\rm D}_2   \\
\end{array}
\\
 & {\cal P}_{\rm LS}^{( ^1{\rm D}_2 \to ^3{\rm P}_1 + ^3{\rm S}_1 ) }\  & =
\left\{
\begin{array}{cr}
-{2^{11/2} \over 3^{7/2} } \;
\Big( 1 - {5\over 18} x^2 + {1\over 135} x^4 \Big) \;
\    &^5{\rm S}_2   \\
{2^7  \over 3^{11/2} 5^{1/2}} \;
x^2 \;
\Big( 1 - {1\over 6} x^2 \Big) \;
\    &^1{\rm D}_2   \\
{2^4 \over 3^{7/2} 5^{1/2}} \;
x^2 \;
\    &^3{\rm D}_2   \\
-{2^4 5^{1/2} 7^{1/2} \over 3^{11/2} } \;
x^2 \;
\Big( 1 - {4\over 105} x^2 \Big) \;
\    &^5{\rm D}_2   \\
-{2^6 \over 3^{11/2}5\; 7^{1/2} } \;
x^4 \;
\    &^5{\rm G}_2   \\
\end{array}
\right.
\\
 & {\cal P}_{\rm LS}^{( ^1{\rm D}_2 \to ^3{\rm P}_2 + ^3{\rm S}_1 ) }\  & =
\left\{
\begin{array}{cr}
{2^{11/2} \over 3^3 } \;
\Big( 1 - {5\over 18} x^2 + {1\over 135} x^4 \Big) \;
\ &   ^5{\rm S}_2   \\
{2^4  \over 3^4 } \;
x^2 \;
\    &^3{\rm D}_2   \\
{2^4 5^{1/2} 7^{1/2} \over 3^5 } \;
x^2 \;
\Big( 1 - {4\over 105} x^2 \Big) \;
\    &^5{\rm D}_2   \\
0 \;
\    &^7{\rm D}_2   \\
{2^6 \over 3^5 5\; 7^{1/2} } \;
x^4 \;
\    &^5{\rm G}_2   \\
0 \;
\    &^7{\rm G}_2   \\
\end{array}
\right.
\end{eqnarray}

\newpage

\vskip .5cm
\hrule
\vskip .1cm
\hrule
\vskip .5cm

\fbox{\bf 1F $\to$ 1S + 1S}

\begin{eqnarray}
 &f_D  &=  -{2^{13/2}  \over 3^{9/2} 5^{1/2}  }\; x^2 \Big(1 - {2\over 21} x^2\Big) \; 
\\
 &f_G  &=   {2^{15/2}  \over 3^7 5^{1/2} 7^{1/2} }\; x^4  \; 
\end{eqnarray}

\begin{eqnarray}
  & & \nonumber \\ \fbox{$^3${\bf F}$_4$} & & \nonumber \\  & & \nonumber \\ 
 & {\cal P}_{40}^{ ( ^3{\rm F}_4 \to ^1{\rm S}_0 + ^1{\rm S}_0 )} \  & =
\begin{array}{cr}
f_G \; 
\    &^1{\rm G}_4   \\
\end{array}
\\
 & {\cal P}_{41}^{( ^3{\rm F}_4 \to ^3{\rm S}_1 + ^1{\rm S}_0 ) }\  & =
\begin{array}{cr}
-\sqrt{5\over 4} \; f_G \; 
\    &^3{\rm G}_4   \\
\end{array}
\\
 & {\cal P}_{\rm LS}^{( ^3{\rm F}_4 \to ^3{\rm S}_1 + ^3{\rm S}_1 ) }\  & =
\left\{
\begin{array}{cr}
 f_D \;
\ &   ^5{\rm D}_4   \\
\sqrt{1\over 3} \; f_G \;
\ &   ^1{\rm G}_4  \\
0 \;
\ &   ^3{\rm G}_4   \\
-\sqrt{55\over 42} \; f_G \;
\ &   ^5{\rm G}_4   \\
0 \;
\ &   ^5{\rm I}_4   \\
\end{array}
\right.  \\
  & & \nonumber \\ \fbox{$^3${\bf F}$_3$} & & \nonumber \\  & & \nonumber \\ 
 & {\cal P}_{\rm LS}^{( ^3{\rm F}_3 \to ^3{\rm S}_1 + ^1{\rm S}_0 ) }\  & =
\left\{
\begin{array}{cr}
-\sqrt{1\over 3} \; f_D \; 
\ &   ^3{\rm D}_3   \\
-\sqrt{27\over 28} \; f_G \;
\ &   ^3{\rm G}_3  \\
\end{array}
\right.
\\
 & {\cal P}_{\rm LS}^{( ^3{\rm F}_3 \to ^3{\rm S}_1 + ^3{\rm S}_1 ) }\  & =
\left\{
\begin{array}{cr}
0 \;
\ &   ^3{\rm D}_3   \\
\sqrt{1\over 3} \;  f_D \;
\ &   ^5{\rm D}_3   \\
0 \;
\ &   ^3{\rm G}_3   \\
-\sqrt{45\over 14} \; f_G \;
\ &   ^5{\rm G}_3  \\
\end{array}
\right. \\
  & & \nonumber \\ \fbox{$^3${\bf F}$_2$} & & \nonumber \\  & & \nonumber \\ 
 & {\cal P}_{20}^{ ( ^3{\rm F}_2 \to ^1{\rm S}_0 + ^1{\rm S}_0 )} \  & =
\begin{array}{cr}
- \sqrt{7\over 20} \; f_D \; 
\ &   ^1{\rm D}_2  \\
\end{array}
\\
 & {\cal P}_{21}^{( ^3{\rm F}_2 \to ^3{\rm S}_1 + ^1{\rm S}_0 ) }\  & =
\begin{array}{cr}
-\sqrt{7\over 30} \; f_D \; 
\ &   ^3{\rm D}_2  \\
\end{array}
\\
 & {\cal P}_{\rm LS}^{( ^3{\rm F}_2 \to ^3{\rm S}_1 + ^3{\rm S}_1 ) }\  & =
\left\{
\begin{array}{cr}
0 \;
\ &   ^5{\rm S}_2   \\
-\sqrt{7\over 60} \;  f_D \;
\ &   ^1{\rm D}_2   \\
0 \;
\ &   ^3{\rm D}_2   \\
\sqrt{1\over 15} \;  f_D \;
\ &   ^5{\rm D}_2   \\
-\sqrt{36\over 7} \; f_G \;
\ &   ^5{\rm G}_2  \\
\end{array}
\right.  \\
  & & \nonumber \\ \fbox{$^1${\bf F}$_3$} & & \nonumber \\  & & \nonumber \\ 
 & {\cal P}_{\rm LS}^{( ^1{\rm F}_3 \to ^3{\rm S}_1 + ^1{\rm S}_0 ) }\  & =
\left\{
\begin{array}{cr}
{1\over 2} \; f_D \; 
\ &   ^3{\rm D}_3   \\
-\sqrt{9\over 7} \; f_G \;
\ &   ^3{\rm G}_3  \\
\end{array}
\right.
\\
 & {\cal P}_{\rm LS}^{( ^1{\rm F}_3 \to ^3{\rm S}_1 + ^3{\rm S}_1 ) }\  & =
\left\{
\begin{array}{cr}
-\sqrt{1\over 2} \;  f_D \;
\ &   ^3{\rm D}_3   \\
0 \;
\ &   ^5{\rm D}_3   \\
\sqrt{18\over 7} \; f_G \;
\ &   ^3{\rm G}_3  \\
0 \;
\ &   ^5{\rm G}_3   \\
\end{array}
\right.
\end{eqnarray}

\vskip .5cm
\hrule
\vskip .5cm

\fbox{\bf 1F $\to$ 2S + 1S}

\begin{eqnarray}
 &f_D  & =-{2^6  \over 3^5 5^{1/2}  }\; x^2 \Big(1 - {13\over 42} x^2
+ {1\over 189} x^4 \Big) \; 
\\
 &f_G  & ={2^{10}  \over 3^{17/2} 5^{1/2} 7^{1/2} }\; x^4 
\Big(1 - {1\over 48} x^2\Big) \;
\end{eqnarray}

\begin{eqnarray}
  & & \nonumber \\ \fbox{$^3${\bf F}$_4$} & & \nonumber \\  & & \nonumber \\ 
 & {\cal P}_{40}^{ ( ^3{\rm F}_4 \to 2^1{\rm S}_0 + ^1{\rm S}_0 )} \  & =
\begin{array}{cr}
f_G \; 
\ &   ^1{\rm G}_4   \\
\end{array}
\\
 & {\cal P}_{41}^{( ^3{\rm F}_4 \to 2^3{\rm S}_1 + ^1{\rm S}_0 ) }\  & =
\begin{array}{cr}
-\sqrt{5\over 4} \; f_G \; 
\ &   ^3{\rm G}_4   \\
\end{array}
\\
 & {\cal P}_{41}^{( ^3{\rm F}_4 \to 2^1{\rm S}_0 + ^3{\rm S}_1 ) }\  & =
\begin{array}{cr}
\sqrt{5\over 4} \; f_G \; 
\ &   ^3{\rm G}_4   \\
\end{array}
\\
 & {\cal P}_{\rm LS}^{( ^3{\rm F}_4 \to 2^3{\rm S}_1 + ^3{\rm S}_1 ) }\  & =
\left\{
\begin{array}{cr}
 f_D \;
\ &   ^5{\rm D}_4   \\
\sqrt{1\over 3} \; f_G \;
\ &   ^1{\rm G}_4  \\
0 \;
\ &   ^3{\rm G}_4  \\
-\sqrt{55\over 42} \; f_G \;
\ &   ^5{\rm G}_4  \\
0 \;
\ &   ^5{\rm I}_4  \\
\end{array}
\right. \\
  & & \nonumber \\ \fbox{$^3${\bf F}$_3$} & & \nonumber \\  & & \nonumber \\ 
 & {\cal P}_{\rm LS}^{( ^3{\rm F}_3 \to 2^3{\rm S}_1 + ^1{\rm S}_0 ) }\  & =
\left\{
\begin{array}{cr}
-\sqrt{1\over 3} \; f_D \; 
\ &   ^3{\rm D}_3   \\
-\sqrt{27\over 28} \; f_G \;
\ &   ^3{\rm G}_3  \\
\end{array}
\right.
\\
 & {\cal P}_{\rm LS}^{( ^3{\rm F}_3 \to 2^1{\rm S}_0 + ^3{\rm S}_1 ) }\  & =
\left\{
\begin{array}{cr}
\sqrt{1\over 3} \; f_D \; 
\ &   ^3{\rm D}_3   \\
\sqrt{27\over 28} \; f_G \;
\ &   ^3{\rm G}_3  \\
\end{array}
\right.
\\
 & {\cal P}_{\rm LS}^{( ^3{\rm F}_3 \to 2^3{\rm S}_1 + ^3{\rm S}_1 ) }\  & =
\left\{
\begin{array}{cr}
0 \;
\ &   ^3{\rm D}_3   \\
\sqrt{1\over 3} \;  f_D \;
\ &   ^5{\rm D}_3   \\
0 \;
\ &   ^3{\rm G}_3   \\
-\sqrt{45\over 14} \; f_G \;
\ &   ^5{\rm G}_3  \\
\end{array}
\right. \\
  & & \nonumber \\ \fbox{$^3${\bf F}$_2$} & & \nonumber \\  & & \nonumber \\ 
 & {\cal P}_{20}^{ ( ^3{\rm F}_2 \to 2^1{\rm S}_0 + ^1{\rm S}_0 )} \  & =
\begin{array}{cr}
- \sqrt{7\over 20} \; f_D \; 
\ &   ^1{\rm D}_2  \\
\end{array}
\\
 & {\cal P}_{21}^{( ^3{\rm F}_2 \to 2^3{\rm S}_1 + ^1{\rm S}_0 ) }\  & =
\begin{array}{cr}
-\sqrt{7\over 30} \; f_D \; 
\ &   ^3{\rm D}_2  \\
\end{array}
\\
 & {\cal P}_{21}^{( ^3{\rm F}_2 \to 2^1{\rm S}_0 + ^3{\rm S}_1 ) }\  & =
\begin{array}{cr}
\sqrt{7\over 30} \; f_D \; 
\ &   ^3{\rm D}_2  \\
\end{array}
\\
 & {\cal P}_{\rm LS}^{( ^3{\rm F}_2 \to 2^3{\rm S}_1 + ^3{\rm S}_1 ) }\  & =
\left\{
\begin{array}{cr}
0\;
\ &   ^5{\rm S}_2   \\
-\sqrt{7\over 60} \;  f_D \;
\ &   ^1{\rm D}_2   \\
0\;
\ &   ^3{\rm D}_2   \\
\sqrt{1\over 15} \;  f_D \;
\ &   ^5{\rm D}_2   \\
-\sqrt{36\over 7} \; f_G \;
\ &   ^5{\rm G}_2  \\
\end{array}
\right. \\
  & & \nonumber \\ \fbox{$^1${\bf F}$_3$} & & \nonumber \\  & & \nonumber \\ 
 & {\cal P}_{\rm LS}^{( ^1{\rm F}_3 \to 2^3{\rm S}_1 + ^1{\rm S}_0 ) }\  & =
\left\{
\begin{array}{cr}
{1\over 2} \; f_D \; 
\ &   ^3{\rm D}_3   \\
-\sqrt{9\over 7} \; f_G \;
\ &   ^3{\rm G}_3  \\
\end{array}
\right.
\\
 & {\cal P}_{\rm LS}^{( ^1{\rm F}_3 \to 2^1{\rm S}_0 + ^3{\rm S}_1 ) }\  & =
\left\{
\begin{array}{cr}
{1\over 2} \; f_D \; 
\ &   ^3{\rm D}_3   \\
-\sqrt{9\over 7} \; f_G \;
\ &   ^3{\rm G}_3  \\
\end{array}
\right.
\\
 & {\cal P}_{\rm LS}^{( ^1{\rm F}_3 \to 2^3{\rm S}_1 + ^3{\rm S}_1 ) }\  & =
\left\{
\begin{array}{cr}
-\sqrt{1\over 2} \;  f_D \;
\ &   ^3{\rm D}_3   \\
0\;
\ &   ^5{\rm D}_3   \\
\sqrt{18\over 7} \; f_G \;
\ &   ^3{\rm G}_3  \\
0\;
\ &   ^5{\rm G}_3   \\
\end{array}
\right.
\end{eqnarray}

\vskip .5cm
\hrule
\vskip .5cm

\fbox{\bf 1F $\to$ 3S + 1S} 

{(\small \it See 1F $\to$ 2S + 1S for channel coefficients.)}
\vskip .5cm

\begin{eqnarray}
 & f_D  =&{2^5  \over 3^5 5 }\; x^2 \Big(1 + x^2 
- {29\over 756} x^4
+ {1\over 3402} x^6 \Big) \; 
\\
 & f_G  =&{2^6 7^{1/2} (11)  \over 3^{19/2} 5 }\; x^4 
\Big(1 - {10\over 231} x^2 + {1\over 2772} x^4 \Big) \;
\end{eqnarray}

\vskip .5cm
\hrule
\vskip .5cm

\fbox{\bf 1F $\to$ 1P + 1$^1$S$_0$} 

\begin{eqnarray}
  & & \nonumber \\ \fbox{$^3${\bf F}$_4$} & & \nonumber \\  & & \nonumber \\ 
 & {\cal P}_{\rm LS}^{( ^3{\rm F}_4 \to ^1{\rm P}_1 + ^1{\rm S}_0 ) }\  & =
\left\{
\begin{array}{cr}
-{2^6 5^{1/2} \over 3^6 7^{1/2} } \;
x^3 \;
\Big( 1 - {4\over 135} x^2 \Big) \;
\ &   ^3{\rm F}_4   \\
-{2^7 \over 3^9  7^{1/2}  } \;
x^5 \;
\ &   ^3{\rm H}_4   \\
\end{array}
\right. 
\\
 & {\cal P}_{\rm LS}^{( ^3{\rm F}_4 \to ^3{\rm P}_1 + ^1{\rm S}_0 ) }\  & =
\left\{
\begin{array}{cr}
{2^{15/2}  \over 3^6 5^{1/2}  7^{1/2}} \;
x^3 \;
\Big( 1 - {5\over 108} x^2  \Big) \;
\ &   ^3{\rm F}_4   \\
-{2^{13/2} \over 3^9 7^{1/2}  } \;
x^5 \;
\ &   ^3{\rm H}_4   \\
\end{array}
\right. 
\\
 & {\cal P}_{\rm LS}^{( ^3{\rm F}_4 \to ^3{\rm P}_2 + ^1{\rm S}_0 ) }\  & =
\left\{
\begin{array}{cr}
{2^{13/2}  \over 3^{11/2} 7^{1/2} } \;
x^3 \;
\Big( 1 - {1\over 54} x^2  \Big) \;
\ &   ^5{\rm F}_4   \\
{2^6 \over 3^{17/2} 7^{1/2}  } \;
x^5 \;
\ &   ^5{\rm H}_4   \\
\end{array}
\right. 
\\
  & & \nonumber \\ \fbox{$^3${\bf F}$_3$} & & \nonumber \\  & & \nonumber \\ 
 & {\cal P}_{31}^{ ( ^3{\rm F}_3 \to ^1{\rm P}_1 + ^1{\rm S}_0 )} \  & =
\begin{array}{cr}
{2^6 \over 3^5 5^{1/2} 7^{1/2} } \;
x^3 \;
\ &   ^3{\rm F}_3   \\
\end{array}
\\
 & {\cal P}_{30}^{ ( ^3{\rm F}_3 \to ^3{\rm P}_0 + ^1{\rm S}_0 )} \  & =
\begin{array}{cr}
{2^6 \over 3^5 5^{1/2} 7^{1/2} } \;
x^3 \;
\ &   ^1{\rm F}_3   \\
\end{array}
\\
 & {\cal P}_{31}^{ ( ^3{\rm F}_3 \to ^3{\rm P}_1 + ^1{\rm S}_0 )} \  & =
\begin{array}{cr}
{2^{13/2} \over 3^5 5^{1/2} 7^{1/2} } \;
x^3 \;
\Big( 1 - {1\over 18} x^2 \Big) \;
\ &   ^3{\rm F}_3   \\
\end{array}
\\
 & {\cal P}_{\rm LS}^{( ^3{\rm F}_3 \to ^3{\rm P}_2 + ^1{\rm S}_0 ) }\  & =
\left\{
\begin{array}{cr}
-{2^{15/2} \over 3^5 } \;
x \;
\Big( 1 - {1\over 6} x^2 + {1\over 315} x^4 \Big) \;
\ &   ^5{\rm P}_3   \\
{2^{11/2} \over 3^{15/2} 5 \; 7^{1/2}  } \;
x^5 \;
\ &   ^5{\rm F}_3   \\
{2^6 \over 3^{15/2}  7  } \;
x^5 \;
\ &   ^5{\rm H}_3   \\
\end{array}
\right. 
\\
  & & \nonumber \\ \fbox{$^3${\bf F}$_2$} & & \nonumber \\  & & \nonumber \\ 
 & {\cal P}_{\rm LS}^{( ^3{\rm F}_2 \to ^1{\rm P}_1 + ^1{\rm S}_0 ) }\  & =
\left\{
\begin{array}{cr}
-{2^{13/2} 7^{1/2} \over 3^{9/2} 5^{1/2} } \;
x \;
\Big( 1 - {1\over 6} x^2 + {1\over 315} x^4 \Big) \;
\ &   ^3{\rm P}_2   \\
-{2^5 \over 3^5 5^{1/2} 7^{1/2}  } \;
x^3 \;
\Big( 1 - {2\over 15} x^2  \Big) \;
\ &   ^3{\rm F}_2   \\
\end{array}
\right. 
\\
 & {\cal P}_{\rm LS}^{( ^3{\rm F}_2 \to ^3{\rm P}_1 + ^1{\rm S}_0 ) }\  & =
\left\{
\begin{array}{cr}
-{2^6 7^{1/2} \over 3^{9/2} 5^{1/2} } \;
x \;
\Big( 1 - {1\over 6} x^2 + {1\over 315} x^4 \Big) \;
\ &   ^3{\rm P}_2   \\
-{2^{11/2} \over 3^5 5^{1/2} 7^{1/2}  } \;
x^3 \;
\Big( 1 + {2\over 45} x^2  \Big) \;
\ &   ^3{\rm F}_2   \\
\end{array}
\right. 
\\
 & {\cal P}_{\rm LS}^{( ^3{\rm F}_2 \to ^3{\rm P}_2 + ^1{\rm S}_0 ) }\  & =
\left\{
\begin{array}{cr}
-{2^6 7^{1/2} \over 3^5 5^{1/2} } \;
x \;
\Big( 1 - {1\over 6} x^2 + {1\over 315} x^4 \Big) \;
\ &   ^5{\rm P}_2   \\
-{2^6 \over 3^5 5^{1/2} 7^{1/2}  } \;
x^3 \;
\Big( 1 - {2\over 45} x^2  \Big) \;
\ &   ^5{\rm F}_2   \\
\end{array}
\right. 
\\
  & & \nonumber \\ \fbox{$^1${\bf F}$_3$} & & \nonumber \\  & & \nonumber \\ 
 & {\cal P}_{31}^{ ( ^1{\rm F}_3 \to ^1{\rm P}_1 + ^1{\rm S}_0 )} \  & =
\begin{array}{cr}
0 \;
\ &   ^3{\rm F}_3   \\
\end{array}
\\
 & {\cal P}_{30}^{ ( ^1{\rm F}_3 \to ^3{\rm P}_0 + ^1{\rm S}_0 )} \  & =
\begin{array}{cr}
{2^5 \over 3^{9/2} 5^{1/2} 7^{1/2} } \;
x^3 \;
\Big( 1 - {2\over 27} x^2  \Big) \;
\ &   ^1{\rm F}_3   \\
\end{array}
\\
 & {\cal P}_{31}^{ ( ^1{\rm F}_3 \to ^3{\rm P}_1 + ^1{\rm S}_0 )} \  & =
\begin{array}{cr}
{2^{11/2} \over 3^{9/2} 5^{1/2} 7^{1/2} } \;
x^3 \;
\ &   ^3{\rm F}_3   \\
\end{array}
\\
 & {\cal P}_{\rm LS}^{( ^1{\rm F}_3 \to ^3{\rm P}_2 + ^1{\rm S}_0 ) }\  & =
\left\{
\begin{array}{cr}
{2^{13/2} \over 3^{9/2} } \;
x \;
\Big( 1 - {1\over 6} x^2 + {1\over 315} x^4 \Big) \;
\ &   ^5{\rm P}_3   \\
{2^{11/2} \over 3^5  7^{1/2}  } \;
x^3 \;
\Big( 1 -  {4\over 135} x^2 \Big) \;
\ &   ^5{\rm F}_3   \\
{2^7 \over 3^8  7  } \;
x^5 \;
\ &   ^5{\rm H}_3   \\
\end{array}
\right. 
\end{eqnarray}

\vskip .5cm
\hrule
\vskip .5cm

\fbox{\bf 1F $\to$ 2P + 1$^1$S$_0$} 

\begin{eqnarray}
  & & \nonumber \\ \fbox{$^3${\bf F}$_4$} & & \nonumber \\  & & \nonumber \\ 
 & {\cal P}_{\rm LS}^{( ^3{\rm F}_4 \to 2^1{\rm P}_1 + ^1{\rm S}_0 ) }\  & =
\left\{
\begin{array}{cr}
-{2^{11/2} (37) \over 3^7 5 \; 7^{1/2} }\;
x^3 \;
\Big( 1 - {125\over 1998} x^2 + {2\over 2997} x^4 \Big) 
\ &   ^3{\rm F}_4   \\
-{2^{15/2} 5^{1/2} \over 3^{10}  7^{1/2} }\;
x^5 \;
\Big( 1 - {1\over 60} x^2  \Big) 
\ &   ^3{\rm H}_4   \\
\end{array}
\right. 
\\
 & {\cal P}_{\rm LS}^{( ^3{\rm F}_4 \to 2^3{\rm P}_1 + ^1{\rm S}_0 ) }\  & =
\left\{
\begin{array}{cr}
{2^6 (13) \over 3^7 5 \; 7^{1/2} }\;
x^3 \;
\Big( 1 - {34\over 351} x^2 + {5\over 4212} x^4 \Big) 
\ &   ^3{\rm F}_4   \\
-{2^7 5^{1/2} \over 3^{10}  7^{1/2} }\;
x^5 \;
\Big( 1 - {1\over 60} x^2  \Big) 
\ &   ^3{\rm H}_4   \\
\end{array}
\right. 
\\
 & {\cal P}_{\rm LS}^{( ^3{\rm F}_4 \to 2^3{\rm P}_2 + ^1{\rm S}_0 ) }\  & =
\left\{
\begin{array}{cr}
{2^9  \over 3^{13/2} 5^{1/2}  7^{1/2} }\;
x^3 \;
\Big( 1 - {19\over 432} x^2 + {1\over 2592} x^4 \Big) 
\ &   ^5{\rm F}_4   \\
{2^{13/2} 5^{1/2} \over 3^{19/2}  7^{1/2} }\;
x^5 \;
\Big( 1 - {1\over 60} x^2  \Big) 
\ &   ^5{\rm H}_4   \\
\end{array}
\right. 
\\
  & & \nonumber \\ \fbox{$^3${\bf F}$_3$} & & \nonumber \\  & & \nonumber \\ 
 & {\cal P}_{31}^{( ^3{\rm F}_3 \to 2^1{\rm P}_1 + ^1{\rm S}_0 ) }\  & =
\begin{array}{cr}
{2^{11/2}  \over 3^4 5 \; 7^{1/2} }\;
x^3 \;
\Big( 1 - {1\over 54} x^2  \Big) 
\ &   ^3{\rm F}_3   \\
\end{array}
\\
 & {\cal P}_{30}^{( ^3{\rm F}_3 \to 2^3{\rm P}_0 + ^1{\rm S}_0 ) }\  & =
\begin{array}{cr}
{2^{11/2}  \over 3^4 5 \; 7^{1/2} }\;
x^3 \;
\Big( 1 - {1\over 54} x^2  \Big) 
\ &   ^1{\rm F}_3   \\
\end{array}
\\
 & {\cal P}_{31}^{( ^3{\rm F}_3 \to 2^3{\rm P}_1 + ^1{\rm S}_0 ) }\  & =
\begin{array}{cr}
{2^7  \over 3^5 5 \; 7^{1/2} }\;
x^3 \;
\Big( 1 - {13\over 108} x^2 + {1\over 648} x^4 \Big) 
\ &   ^3{\rm F}_3   \\
\end{array}
\\
 & {\cal P}_{\rm LS}^{( ^3{\rm F}_3 \to 2^3{\rm P}_2 + ^1{\rm S}_0 ) }\  & =
\left\{
\begin{array}{cr}
-{2^7  \over 3^5 5^{1/2} }\;
x \;
\Big( 1 - {7\over 15} x^2 + {5\over 252} x^4 - {1\over 5670} x^6 \Big) 
\ &   ^5{\rm P}_3   \\
{2^6  \over 3^{11/2} 5^{3/2} 7^{1/2} }\;
x^3 \;
\Big( 1 + {5\over 27} x^2 - {1\over 324} x^4  \Big) 
\ &   ^5{\rm F}_3   \\
{2^{13/2} 5^{1/2}  \over 3^{17/2}  7 }\;
x^5 \;
\Big( 1 - {1\over 60} x^2  \Big) 
\ &   ^5{\rm H}_3   \\
\end{array}
\right. 
\\
  & & \nonumber \\ \fbox{$^3${\bf F}$_2$} & & \nonumber \\  & & \nonumber \\ 
 & {\cal P}_{\rm LS}^{( ^3{\rm F}_2 \to 2^1{\rm P}_1 + ^1{\rm S}_0 ) }\  & =
\left\{
\begin{array}{cr}
-{2^6 7^{1/2} \over 3^{9/2} 5   }\;
x \;
\Big( 1 - {7\over 15} x^2 + {5\over 252} x^4  - {1\over 5670} x^6 \Big) 
\ &   ^3{\rm P}_2   \\
-{2^{9/2}  \over 3^4 5^2  7^{1/2} }\;
x^3 \;
\Big( 1 - {5\over 6} x^2 + {1 \over 81} x^4  \Big) 
\ &   ^3{\rm F}_2   \\
\end{array}
\right. 
\\
 & {\cal P}_{\rm LS}^{( ^3{\rm F}_2 \to 2^3{\rm P}_1 + ^1{\rm S}_0 ) }\  & =
\left\{
\begin{array}{cr}
-{2^{11/2} 7^{1/2} \over 3^{9/2} 5   }\;
x \;
\Big( 1 - {7\over 15} x^2 + {5\over 252} x^4  - {1\over 5670} x^6 \Big) 
\ &   ^3{\rm P}_2   \\
-{2^5 (19)  \over 3^5 5^2 7^{1/2}  }\;
x^3 \;
\Big( 1 + {25\over 1026} x^2 - {1 \over 1539} x^4  \Big) 
\ &   ^3{\rm F}_2   \\
\end{array}
\right. 
\\
 & {\cal P}_{\rm LS}^{( ^3{\rm F}_2 \to 2^3{\rm P}_2 + ^1{\rm S}_0 ) }\  & =
\left\{
\begin{array}{cr}
-{2^{11/2} 7^{1/2} \over 3^5 5  }\;
x \;
\Big( 1 - {7\over 15} x^2 + {5\over 252} x^4  - {1\over 5670} x^6 \Big) 
\ &   ^5{\rm P}_2   \\
-{2^{11/2} (11)  \over 3^5 5^2  7^{1/2} }\;
x^3 \;
\Big( 1 - {5\over 54} x^2 + {1 \over 891} x^4  \Big) 
\ &   ^5{\rm F}_2   \\
\end{array}
\right. 
\\
  & & \nonumber \\ \fbox{$^1${\bf F}$_3$} & & \nonumber \\  & & \nonumber \\ 
 & {\cal P}_{31}^{ ( ^1{\rm F}_3 \to 2^1{\rm P}_1 + ^1{\rm S}_0 )} \  & =
\begin{array}{cr}
0 \;
\ &   ^3{\rm F}_3   \\
\end{array}
\\
 & {\cal P}_{30}^{( ^1{\rm F}_3 \to 2^3{\rm P}_0 + ^1{\rm S}_0 ) }\  & =
\begin{array}{cr}
{2^{9/2} \over 3^{11/2} 7^{1/2} } \;
x^3 \;
\Big( 1 - {49\over 270} x^2 + {1\over 405} x^4  \Big) \;
\ &   ^1{\rm F}_3   \\
\end{array}
\\
 & {\cal P}_{31}^{( ^1{\rm F}_3 \to 2^3{\rm P}_1 + ^1{\rm S}_0 ) }\  & =
\begin{array}{cr}
{2^5 \over 3^{7/2} 5\;  7^{1/2} } \;
x^3 \;
\Big( 1 - {1\over 54} x^2  \Big) \;
\ &   ^3{\rm F}_3   \\
\end{array}
\\
 & {\cal P}_{\rm LS}^{( ^1{\rm F}_3 \to 2^3{\rm P}_2 + ^1{\rm S}_0 ) }\  & =
\left\{
\begin{array}{cr}
{2^6 \over 3^{9/2} 5^{1/2} } \;
x \;
\Big( 1 - {7\over 15} x^2 + {5\over 252} x^4  - {1\over 5670} x^6 \Big) \;
\ &   ^5{\rm P}_3   \\
{2^5 (37) \over 3^6 5^{3/2} 7^{1/2} } \;
x^3 \;
\Big( 1 -  {125\over 1998} x^2 + {2\over 2997} x^4 \Big) \;
\ &   ^5{\rm F}_3   \\
{2^{15/2} 5^{1/2} \over 3^9  7  } \;
x^5 \;
\Big( 1 -  {1\over 60} x^2 \Big) \;
\ &   ^5{\rm H}_3   \\
\end{array}
\right. 
\end{eqnarray}

\vskip .5cm
\hrule
\vskip .5cm

\fbox{\bf 1F $\to$ 1D + 1$^1$S$_0$} 

\begin{eqnarray}
  & & \nonumber \\ \fbox{$^3${\bf F}$_4$} & & \nonumber \\  & & \nonumber \\ 
 & {\cal P}_{\rm LS}^{( ^3{\rm F}_4 \to ^1{\rm D}_2 + ^1{\rm S}_0 ) }\  & =
\left\{
\begin{array}{cr}
{2^7  \over 3^{11/2} 5^{1/2} } \;
x^2 \;
\Big( 1 - {5\over 42} x^2 + {1\over 567}x^4 \Big) \;
\ &   ^5{\rm D}_4   \\
{2^{11/2} 5\; (11)^{1/2} \over 3^{17/2}  7  } \;
x^4 \;
\Big( 1 - {4\over 165} x^2  \Big) \;
\ &   ^5{\rm G}_4   \\
{2^{13/2}   \over 3^{19/2}  7^{1/2}  (11)^{1/2} } \;
x^6 \;
\ &   ^5{\rm I}_4   \\
\end{array}
\right. 
\\
 & {\cal P}_{41}^{( ^3{\rm F}_4 \to ^3{\rm D}_1 + ^1{\rm S}_0 ) }\  & =
\begin{array}{cr}
{2^5 7^{1/2} \over 3^{17/2}  5^{1/2}  } \;
x^4 \;
\Big( 1 - {1\over 21} x^2  \Big) \;
\ &   ^3{\rm G}_4   \\
\end{array}
\\
 & {\cal P}_{\rm LS}^{( ^3{\rm F}_4 \to ^3{\rm D}_2 + ^1{\rm S}_0 ) }\  & =
\left\{
\begin{array}{cr}
-{2^{13/2}  \over 3^6 5^{1/2} } \;
x^2 \;
\Big( 1 - {11\over 42} x^2 + {5\over 1134}x^4 \Big) \;
\ &   ^5{\rm D}_4   \\
{2^5 (11)^{1/2} \over 3^8  7  } \;
x^4 \;
\Big( 1 + {1\over 33} x^2  \Big) \;
\ &   ^5{\rm G}_4   \\
{2^7   \over 3^{10}  7^{1/2}  (11)^{1/2} } \;
x^6 \;
\ &   ^5{\rm I}_4   \\
\end{array}
\right. 
\\
 & {\cal P}_{\rm LS}^{( ^3{\rm F}_4 \to ^3{\rm D}_3 + ^1{\rm S}_0 ) }\  & =
\left\{
\begin{array}{cr}
-{2^{15/2}  \over 3^6  } \;
x^2 \;
\Big( 1 - {1\over 21} x^2 + {1\over 2268}x^4 \Big) \;
\ &   ^7{\rm D}_4   \\
-{2^7 (11)^{1/2} \over 3^7 5^{1/2} 7  } \;
x^4 \;
\Big( 1 - {1\over 66} x^2  \Big) \;
\ &   ^7{\rm G}_4   \\
-{2^{11/2}   \over 3^{10} (11)^{1/2} } \;
x^6 \;
\ &   ^7{\rm I}_4   \\
\end{array}
\right. 
\\
  & & \nonumber \\ \fbox{$^3${\bf F}$_3$} & & \nonumber \\  & & \nonumber \\ 
 & {\cal P}_{\rm LS}^{( ^3{\rm F}_3 \to ^1{\rm D}_2 + ^1{\rm S}_0 ) }\  & =
\left\{
\begin{array}{cr}
-{2^7  \over 3^5 5^{1/2} } \;
x^2 \;
\Big( 1 - {1\over 42} x^2  \Big) \;
\ &   ^5{\rm D}_3   \\
-{2^{11/2}  \over 3^6  7  } \;
x^4 \;
\ &   ^5{\rm G}_3   \\
\end{array}
\right. 
\\
 & {\cal P}_{\rm LS}^{( ^3{\rm F}_3 \to ^3{\rm D}_1 + ^1{\rm S}_0 ) }\  & =
\left\{
\begin{array}{cr}
-{2^8  \over 3^{11/2} 5 } \;
x^2 \;
\Big( 1 - {2\over 21} x^2 + {1\over 756} x^4  \Big) \;
\ &   ^3{\rm D}_3   \\
-{2^5 (11) \over 3^7 5 \;  7  } \;
x^4 \;
\Big( 1 + {1\over 33} x^2  \Big) \;
\ &   ^3{\rm G}_3   \\
\end{array}
\right. 
\\
 & {\cal P}_{\rm LS}^{( ^3{\rm F}_3 \to ^3{\rm D}_2 + ^1{\rm S}_0 ) }\  & =
\left\{
\begin{array}{cr}
-{2^{13/2}  \over 3^{11/2} 5^{1/2} } \;
x^2 \;
\Big( 1 - {1\over 6} x^2 + {1\over 378} x^4  \Big) \;
\ &   ^5{\rm D}_3   \\
-{2^5  \over 3^{15/2}  } \;
x^4 \;
\Big( 1 - {1\over 21} x^2  \Big) \;
\ &   ^5{\rm G}_3   \\
\end{array}
\right. 
\\
 & {\cal P}_{\rm LS}^{( ^3{\rm F}_3 \to ^3{\rm D}_3 + ^1{\rm S}_0 ) }\  & =
\left\{
\begin{array}{cr}
{2^{15/2}  \over 3^{9/2} } \;
\Big( 1 - {1\over 3} x^2 + {1\over 60} x^4 -{1\over 5670} x^6 \Big) \;
\ &   ^7{\rm S}_3   \\
{2^{15/2}  \over 3^5 5  } \;
x^2 \;
\Big( 1 - {2\over 21} x^2 + {1\over 756} x^4  \Big) \;
\ &   ^7{\rm D}_3   \\
-{2^6 (11)^{1/2} \over 3^{15/2} 5\; 7  } \;
x^4 \;
\Big( 1 + {1\over 33} x^2 \Big) \;
\ &   ^7{\rm G}_3   \\
-{2^{11/2} \over 3^8  7 \; (11)^{1/2} } \;
x^6 \;
\ &   ^7{\rm I}_3   \\
\end{array}
\right. 
\\
  & & \nonumber \\ \fbox{$^3${\bf F}$_2$} & & \nonumber \\  & & \nonumber \\ 
 & {\cal P}_{\rm LS}^{( ^3{\rm F}_2 \to ^1{\rm D}_2 + ^1{\rm S}_0 ) }\  & =
\left\{
\begin{array}{cr}
{2^{13/2} 7^{1/2}  \over 3^4 5^{1/2}   } \;
\Big( 1 - {1\over 3} x^2 + {1\over 60} x^4 - {1\over 5670} x^6 \Big) \;
\ &   ^5{\rm S}_2   \\
{2^6  \over 3^4 5  } \;
x^2 \;
\Big( 1 - {5\over 42} x^2 + {1\over 567} x^4  \Big) \;
\ &   ^5{\rm D}_2   \\
{2^5 \over 3^5 5^{3/2} 7 } \;
x^4 \;
\Big( 1 - {2\over 9} x^2  \Big) \;
\ &   ^5{\rm G}_2   \\
\end{array}
\right. 
\\
 & {\cal P}_{\rm 21}^{( ^3{\rm F}_2 \to ^3{\rm D}_1 + ^1{\rm S}_0 ) }\  & =
\begin{array}{cr}
{2^{11/2} 7^{1/2}  \over 3^{11/2} 5^{1/2} } \;
x^2 \;
\Big( 1 + {1\over 30} x^2 - {1\over 945} x^4 \Big) \;
\ &   ^3{\rm D}_2   \\
\end{array}
\\
 & {\cal P}_{\rm LS}^{( ^3{\rm F}_2 \to ^3{\rm D}_2 + ^1{\rm S}_0 ) }\  & =
\left\{
\begin{array}{cr}
{2^7 7^{1/2} \over 3^{9/2} 5^{1/2} } \;
\Big( 1 - {1\over 3} x^2 + {1\over 60} x^4 - {1\over 5670} x^6 \Big) \;
\ &   ^5{\rm S}_2   \\
{2^{11/2} (11)  \over 3^{11/2} 5  } \;
x^2 \;
\Big( 1 - {23\over 462} x^2 + {1\over 2079} x^4  \Big) \;
\ &   ^5{\rm D}_2   \\
{2^{15/2} (11)   \over 3^{15/2} 5^{3/2} 7 } \;
x^4 \;
\Big( 1 + {1\over 33} x^2  \Big) \;
\ &   ^5{\rm G}_2   \\
\end{array}
\right. 
\\
 & {\cal P}_{\rm LS}^{( ^3{\rm F}_2 \to ^3{\rm D}_3 + ^1{\rm S}_0 ) }\  & =
\left\{
\begin{array}{cr}
{2^7  \over 3^{11/2} 5^{1/2} } \;
x^2 \;
\Big( 1 - {17\over 210} x^2 + {1\over 945} x^4  \Big) \;
\ &   ^7{\rm D}_2   \\
{2^{11/2} (23)  \over 3^{15/2} 5 \; 7 } \;
x^4 \;
\Big( 1 - {2\over 69} x^2  \Big) \;
\ &   ^7{\rm G}_2   \\
\end{array}
\right. 
\\
  & & \nonumber \\ \fbox{$^1${\bf F}$_3$} & & \nonumber \\  & & \nonumber \\ 
 & {\cal P}_{\rm LS}^{( ^1{\rm F}_3 \to ^1{\rm D}_2 + ^1{\rm S}_0 ) }\  & =
\left\{
\begin{array}{cr}
0 \; 
\ &   ^5{\rm D}_3   \\
0 \; 
\ &   ^5{\rm G}_3   \\
\end{array}
\right. 
\\
 & {\cal P}_{\rm LS}^{( ^1{\rm F}_3 \to ^3{\rm D}_1 + ^1{\rm S}_0 ) }\  & =
\left\{
\begin{array}{cr}
-{2^6  \over 3^5 5 } \;
x^2 \;
\Big( 1 - {13\over 42} x^2 + {1\over 189} x^4 \Big)  \;
\ &   ^3{\rm D}_3   \\
-{2^5  \over 3^{15/2}  5 \; 7 } \;
x^4 \;
\Big( 1 - {4\over 3} x^2  \Big) \;
\ &   ^3{\rm G}_3   \\
\end{array}
\right.
\\
 & {\cal P}_{\rm LS}^{( ^1{\rm F}_3 \to ^3{\rm D}_2 + ^1{\rm S}_0 ) }\  & =
\left\{
\begin{array}{cr}
-{2^{13/2}  \over 3^5 5^{1/2} } \;
x^2 \;
\Big( 1 - {1\over 42} x^2 \Big)  \;
\ &   ^5{\rm D}_3   \\
-{2^5  \over 3^6  7 } \;
x^4 \;
\ &   ^5{\rm G}_3   \\
\end{array}
\right.
\\
 & {\cal P}_{\rm LS}^{( ^1{\rm F}_3 \to ^3{\rm D}_3 + ^1{\rm S}_0 ) }\  & =
\left\{
\begin{array}{cr}
-{2^{13/2}  \over 3^4  } \;
\Big( 1 - {1\over 3} x^2 + {1\over 60} x^4 - {1\over 5670} x^6 \Big)  \;
\ &   ^7{\rm S}_3   \\
-{2^{15/2}  \over 3^{9/2}  5 } \;
x^2 \;
\Big( 1 - {1\over 14} x^2 + {1\over 1134} x^4 \Big)  \;
\ &   ^7{\rm D}_3   \\
-{2^5 (11)^{1/2} \over 3^5  5 \; 7 } \;
x^4 \;
\Big( 1 - {2\over 99} x^2 \Big)  \;
\ &   ^7{\rm G}_3   \\
-{2^{13/2} \over 3^{17/2}   7 \; (11)^{1/2} } \;
x^6 \;
\ &   ^7{\rm I}_3   \\
\end{array}
\right.
\end{eqnarray}

\newpage

\section
{Numerical Decay Rates.}

In this appendix we quote numerical values for partial widths predicted
by the \3p0 model. The masses used are experimental values of well 
established candidates, usually taken from the 1996 PDG,
otherwise we used an approximate
multiplet mass.
These are 1700 MeV (2P), 1670 MeV (1D), 2050
MeV (1F), and 1900 MeV and 1800 MeV respectively for the 
3${}^3$S$_1$ and
3${}^1$S$_0$. 
The lighter meson masses assumed are 
$m_\pi=138$~MeV,
$m_K=496$ MeV,
$m_\rho=770$ MeV,
$m_\omega=782$ MeV and
$m_{K^*}=894$ MeV. For other states we used the 1996 PDG masses except 
for the broad $f_0$, which we left at 1300 MeV.

Although we found optimum parameters near
$\gamma = 0.5$ and $\beta = 0.4$ GeV in a fit to light 1S and 1P 
decays, these parameters
lead to moderate overestimates of the widths of the well established
higher-L states $\pi_2(1670)$ and $f_4(2044)$; with this $\beta$ a value
closer to $\gamma = 0.4$ is preferred. 
Consequently we quote widths for all 
these higher quarkonia with the parameters
\begin{equation}
(\gamma, \beta )  = (0.4, 0.4 \ {\rm GeV} )  \ .
\end{equation}

The tables are largely self explanatory. Except in a few cases the states
are specified uniquely by their labels. The exceptions include the 
$|\eta(547)\rangle $ and
$|\eta'(958)\rangle $, which we take to be the usual $1/\sqrt{2}$ combinations
of 
$|n\bar n \rangle $  and
$|s\bar s \rangle $  basis states. We assume that the 
$|\eta(1295)\rangle $ and
$|\eta_2(1645)\rangle $ are pure
$|n\bar n \rangle $ states. The strange mesons 
K$_1(1273)$ and
K$_1(1402)$ are taken to be the linear combinations

\begin{equation}
|{\rm K}_1(1273)\rangle = 
\sqrt{ {2\over 3} } \; | {}^1{\rm P}_1 \rangle 
+ 
\sqrt{ {1\over 3} } \; | {}^3{\rm P}_1 \rangle 
\end{equation}
and
\begin{equation}
|{\rm K}_1(1402)\rangle = 
-\sqrt{ {1\over 3} } \; | {}^1{\rm P}_1 \rangle 
+ 
\sqrt{ {2\over 3} } \; | {}^3{\rm P}_1 \rangle  \ .
\end{equation}
This gives a zero S-wave 
${\rm K}_1(1273)\to {\rm K}^*\pi $ 
coupling; 
experimentally D/S $= 1.0(0.7)$, and the small partial 
width implies a small S-wave amplitude.
The orthogonal state ${\rm K}_1(1402)$ (B3) is predicted to have
a D/S ratio of $+0.049$ in
${\rm K}^*\pi $, quite close to the  
experimental D/S $= +0.04(1)$.
The large 
${\rm K}_1(1273)\to {\rm K}\rho $ mode is not predicted and is
possibly due to a virtual intermediate state such as 
K$^*_0(1429)\pi$ followed by a final-state interaction.

The tables give partial widths 
for all
nonstrange 2S, 3S, 2P, 1D and 1F quarkonia to all
two-body modes allowed by phase space, rounded to the nearest MeV.
The 
predictions of the dominant modes of the ``missing states''
in the quark model, such as the $2^{--}$ states and most of the 1F states,
are especially interesting.
If the \3p0 model has even moderate accuracy these tables should be very useful 
in searches for these states.

\newpage

\begin{center}
\begin{tabular}{||c|c||c|c||}  \hline
\multicolumn{4}{||c||}{Table B1. Partial widths 
of 2$^3$S$_1$ states (MeV).} \\ 
\hline
\hline
\hskip 1cm Mode \hskip 1cm \
& \hskip .5cm $\rho(1465)$ \hskip .5cm \ 
& \hskip 1cm Mode \hskip 1cm \
& \hskip .5cm $\omega(1419)$ \hskip .5cm \ 
\\ 
\hline
\hline
\multicolumn{4}{||c||}{\hskip .5cm \bf{(1S)$^2$ }}  \\
\hline
$\pi \, \pi$ 
& 74. 
&  
&    
\\ 
\hline
$\omega \, \pi  $ 
& 122. 
&  
$\rho \, \pi  $ 
& 328.   
\\ 
\hline
$\rho \, \eta  $ 
& 25. 
&  
$\omega \, \eta  $ 
& 12.   
\\ 
\hline
\hline
\multicolumn{4}{||c||}{\hskip .5cm\bf{(2S)(1S)}}  \\
\hline
$\pi(1300)\, \pi$ 
& 0. 
&  
&   
\\ 
\hline
\hline
\multicolumn{4}{||c||}{\hskip .5cm \bf{(1P)(1S)}}  \\
\hline
$h_1(1170)\, \pi$ 
& 1. 
&  
$b_1(1231)\, \pi$ 
&  1. 
\\ 
\hline
$a_1(1230)\, \pi$ 
& 3. 
&  
&   
\\ 
\hline
$a_2(1318)\, \pi$ 
& 0. 
&  
&   
\\ 
\hline
\hline
\multicolumn{4}{||c||}{\hskip .5cm \bf{(1S)$^2$ strange}}  \\
\hline
${\rm K}\, {\rm K}$ 
& 35. &  & 31.  
\\ \hline
${\rm K}^*\, {\rm K}$ 
& 19. &  & 5.   
\\ 
\hline
\hline
\multicolumn{4}{||c||}{\hskip .5cm \bf{total}}  \\
\hline
$\sum_i \Gamma_i $
& 279. &  & 378.   
\\ 
\hline
$\Gamma_{expt}$
& 310(60) &   & 174(59)   
\\ 
\hline
\end{tabular}
\end{center}

\vskip 2cm

\begin{center}
\begin{tabular}{||c|c||c|c||}  \hline
\multicolumn{4}{||c||}{Table B2. Partial widths 
of 2$^1$S$_0$ states (MeV).} \\ 
\hline
\hline
\hskip 1cm Mode \hskip 1cm \
& \hskip .5cm $\pi(1300)$ \hskip .5cm \ 
& \hskip 1cm Mode \hskip 1cm \
& \hskip .5cm $\eta(1295)$ \hskip .5cm \ 
\\ 
\hline
\hline
\multicolumn{4}{||c||}{\hskip .5cm \bf{(1S)$^2$ }}  \\
\hline
$\pi \, \rho$ 
& 209.  
& {\rm none open} 
&    
\\ 
\hline
\hline
\multicolumn{4}{||c||}{\hskip .5cm \bf{total}}  \\
\hline
$\sum_i \Gamma_i $
& 209.  &  &  0. 
\\ 
\hline
$\Gamma_{expt}$
& 200 - 600 &   & 53(6)    
\\ 
\hline
\end{tabular}
\end{center}

\begin{center}
\begin{tabular}{||c|c||c|c||}  \hline
\multicolumn{4}{||c||}{Table B3. Partial widths 
of 3$^3$S$_1$ states (MeV).} \\ 
\hline
\hline
\hskip 1cm Mode \hskip 1cm \
& \hskip .5cm $\rho(1900)$ \hskip .5cm \ 
& \hskip 1cm Mode \hskip 1cm \
& \hskip .5cm $\omega(1900)$ \hskip .5cm \ 
\\ 
\hline
\hline
\multicolumn{4}{||c||}{\hskip .5cm \bf{(1S)$^2$ }}  \\
\hline
$\pi \, \pi$ 
& 1. 
&  
&    
\\ 
\hline
$\omega \, \pi  $ 
& 5. 
&  
$\rho \, \pi  $ 
& 14.   
\\ 
\hline
$\rho \, \eta  $ 
& 8. 
&  
$\omega \, \eta  $ 
& 8.   
\\ 
\hline
$\rho \, \eta'  $ 
& 11. 
&  
$\omega \, \eta'  $ 
& 10.   
\\ 
\hline
$\rho \, \rho  $ 
& 92. 
&  
&    
\\ 
\hline
\hline
\multicolumn{4}{||c||}{\hskip .5cm\bf{(2S)(1S)}}  \\
\hline
$\pi(1300)\, \pi$ 
& 70. 
&  
&   
\\ 
\hline
$\omega(1419)\, \pi$ 
& 50. 
&  
$\rho(1465)\, \pi$ 
&  121. 
\\ 
\hline
\hline
\multicolumn{4}{||c||}{\hskip .5cm \bf{(1P)(1S)}}  \\
\hline
$h_1(1170)\, \pi$ 
& 32. 
&  
$b_1(1231)\, \pi$ 
&  75. 
\\ 
\hline
$b_1(1231)\, \eta$ 
& 4. 
&  
$h_1(1170)\, \eta$ 
&  6. 
\\ 
\hline
$a_1(1230)\, \pi$ 
& 26. 
&  
&   
\\ 
\hline
$a_2(1318)\, \pi$ 
& 46. 
&  
&   
\\ 
\hline
\hline
\multicolumn{4}{||c||}{\hskip .5cm \bf{(2P)(1S)}}  \\
\hline
$h_1(1700)\, \pi$ 
& 0. 
&  
$b_1(1700)\, \pi$ 
&  0. 
\\ 
\hline
$a_1(1700)\, \pi$ 
& 0. 
&  
&   
\\ 
\hline
$a_2(1700)\, \pi$ 
& 0. 
&  
&   
\\ 
\hline
\hline
\multicolumn{4}{||c||}{\hskip .5cm \bf{(1D)(1S)}}  \\
\hline
$\pi_2(1670)\, \pi$ 
& 0. 
&  
&   
\\ 
\hline
$\omega_1(1649) \, \pi $ 
&  0. 
&  
$\rho_1(1700)\, \pi$ 
&  0. 
\\ 
\hline
$\omega_2(1670) \, \pi $ 
&  0. 
&  
$\rho_2(1670)\, \pi$ 
&  0. 
\\ 
\hline
$\omega_3(1667) \, \pi $ 
&  0. 
&  
$\rho_3(1691)\, \pi$ 
&  0. 
\\ 
\hline
\end{tabular}
\end{center}

\begin{center}
\begin{tabular}{||c|c||c|c||}  \hline
\multicolumn{4}{||c||}{Table B3 (cont.) Partial widths 
of 3$^3$S$_1$ states (MeV).} \\ 
\hline
\hline
\hskip 1cm Mode \hskip 1cm \
& \hskip .5cm $\rho(1900)$ \hskip .5cm \ 
& \hskip 1cm Mode \hskip 1cm \
& \hskip .5cm $\omega(1900)$ \hskip .5cm \ 
\\ 
\hline
\hline
\multicolumn{4}{||c||}{\hskip .5cm \bf{(1S)$^2$ strange}}  \\
\hline
${\rm K}\, {\rm K}$ 
& 1. &  & 1.  
\\ \hline
${\rm K}^*\, {\rm K}$ 
& 21. &  & 21.   
\\ 
\hline
${\rm K}^*\, {\rm K}^*$ 
& 27. &  & 27.   
\\ 
\hline
\hline
\multicolumn{4}{||c||}{\hskip .5cm \bf{(1P)(1S) strange}}  \\
\hline  
${\rm K}_1^*(1273)\, {\rm K}$ 
& 5. &  & 5.   
\\ 
\hline
${\rm K}_1^*(1402)\, {\rm K}$ 
& 4. &   & 4.   
\\ 
\hline
\hline
\multicolumn{4}{||c||}{\hskip .5cm \bf{total}}  \\
\hline
$\sum_i \Gamma_i $
& 403. &  & 292.   
\\ 
\hline
\end{tabular}
\end{center}

\newpage

\begin{center}
\begin{tabular}{||c|c||c|c||}  \hline
\multicolumn{4}{||c||}{Table B4. Partial widths 
of 3$^1$S$_0$ states (MeV).} \\ 
\hline
\hline
\hskip 1cm Mode \hskip 1cm \
& \hskip .5cm $\pi(1800)$ \hskip .5cm \ 
& \hskip 1cm Mode \hskip 1cm \
& \hskip .5cm $\eta(1800)$ \hskip .5cm \ 
\\ 
\hline
\hline
\multicolumn{4}{||c||}{\hskip .5cm \bf{(1S)$^2$ }}  \\
\hline
$\pi \, \rho$ 
&  31. 
& 
&    
\\ 
\hline
$\rho \, \omega  $ 
&  73. 
& 
$\rho \, \rho $ 
&  112.  
\\ 
\hline
&   
& 
$\omega \, \omega $ 
&  36.  
\\ 
\hline
\hline
\multicolumn{4}{||c||}{\hskip .5cm\bf{(2S)(1S)}}  \\
\hline
$\rho(1465)\, \pi$ 
& 53. 
&  
&  
\\
\hline
\hline
\multicolumn{4}{||c||}{\hskip .5cm\bf{(1P)(1S)}}  \\
\hline
$f_0(1300) \, \pi$ 
& 7. 
&  
$a_0(1450) \, \pi$ 
& 30. 
\\
\hline
$f_2(1275) \, \pi$ 
& 28. 
&  
$a_2(1318) \, \pi$ 
& 61. 
\\
\hline
\hline
\multicolumn{4}{||c||}{\hskip .5cm \bf{(1S)$^2$ strange}}  \\
\hline
${\rm K}^*\, {\rm K}$ 
& 36. &  & 36.   
\\ 
\hline
\hline
\multicolumn{4}{||c||}{\hskip .5cm \bf{total}}  \\
\hline
$\sum_i \Gamma_i $
& 228.  &  & 275.  
\\ 
\hline
$\Gamma_{expt}$
& 212(37)  &   &     
\\ 
\hline
\end{tabular}
\end{center}
\newpage 
\begin{center}
\begin{tabular}{||c|c|c|c||}  \hline
\multicolumn{4}{||c||}{Table B5. Partial widths 
of  $2\, {}^3$P$_{\rm J}$ $a_\J$ states (MeV).} \\ 
\hline
\hline
\hskip 1cm Mode \hskip 1cm \
& \hskip .5cm $a_2(1700)$ \hskip .5cm \ 
& \hskip .5cm $a_1(1700)$ \hskip .5cm \
& \hskip .5cm $a_0(1700)$ \hskip .5cm \  
\\ 
\hline
\hline
\multicolumn{4}{||c||}{\hskip .5cm \bf{(1S)$^2$ }}  \\
\hline
$\eta \, \pi$ 
& 23. &  & 5.   
\\ 
\hline
$\eta' \, \pi$ 
& 10. &  & 5.   
\\ 
\hline
$\rho \, \pi$ 
& 104. & 58. &    
\\ 
\hline
$\omega \, \rho $ 
& 109. & 15. & 46.   
\\ 
\hline
\hline
\multicolumn{4}{||c||}{\hskip .5cm\bf{(2S)(1S)}}  \\
\hline
$\eta(1295)\, \pi$ 
& 3. &  & 43.   
\\ 
\hline
$\rho(1465)\, \pi$ 
& 0. & 41. &    
\\ 
\hline
\hline
\multicolumn{4}{||c||}{\hskip .5cm\bf{(1P)(1S)}}  \\
\hline
$b_1(1231)\, \pi$ 
& 28. & 41. & 165.   
\\ 
\hline
$f_0(1300)\, \pi$ 
&  & 2. &    
\\ 
\hline
$f_1(1282)\, \pi$ 
& 4. & 18. & 30.   
\\ 
\hline
$f_2(1275)\, \pi$ 
& 20. & 39. &    
\\ 
\hline
\hline
\multicolumn{4}{||c||}{\hskip .5cm \bf{(1S)$^2$ strange}}  \\
\hline
${\rm K}\, {\rm K}$ 
& 20. &  & 0.   
\\ 
\hline
${\rm K}^*\, {\rm K}$ 
& 17. & 33. &    
\\ 
\hline
\hline
\multicolumn{4}{||c||}{\hskip .5cm \bf{total}}  \\
\hline
$\sum_i \Gamma_i $
& 336. & 246. & 293.   
\\ 
\hline
\end{tabular}
\end{center}

\begin{center}
\begin{tabular}{||c|c|c|c||}  \hline
\multicolumn{4}{||c||}{Table B6. Partial widths 
of $2\, {}^3$P$_\J$ $f_\J$ states (MeV).} \\ 
\hline
\hline
\hskip 1cm Mode \hskip 1cm \
& \hskip .5cm $f_2(1700)$ \hskip .5cm \ 
& \hskip .5cm $f_1(1700)$ \hskip .5cm \
& \hskip .5cm $f_0(1700)$ \hskip .5cm \  
\\ 
\hline
\hline
\multicolumn{4}{||c||}{\hskip .5cm \bf{(1S)$^2$ }}  \\
\hline
$\pi \, \pi$ 
& 81. &  & 47.   
\\ 
\hline
$\eta\, \eta$ 
& 4. &  & 0.   
\\ 
\hline
$\eta\, \eta'$ 
& 1. &  & 16.   
\\ 
\hline
$\rho\, \rho$ 
& 159. & 27. & 72.   
\\ 
\hline
$\omega\, \omega$ 
& 56. & 6. & 22.   
\\ 
\hline
\hline
\multicolumn{4}{||c||}{\hskip .5cm\bf{(2S)(1S)}}  \\
\hline
$\pi(1300)\, \pi$ 
& 8. &  & 130.   
\\ 
\hline
\hline
\multicolumn{4}{||c||}{\hskip .5cm \bf{(1P)(1S)}}  \\
\hline
$a_0(1450)\, \pi$ 
&  & 1. &    
\\ 
\hline
$a_1(1230)\, \pi$ 
& 16. & 70. & 122.   
\\ 
\hline
$a_2(1318)\, \pi$
& 43. & 86. &    
\\ 
\hline
\hline
\multicolumn{4}{||c||}{\hskip .5cm \bf{(1S)$^2$ strange}}  \\
\hline
${\rm K}\, {\rm K}$ 
& 20. &  & 0.   
\\ 
\hline
${\rm K}^*\, {\rm K}$ 
& 17. & 33. &    
\\ 
\hline
\hline
\multicolumn{4}{||c||}{\hskip .5cm \bf{total}}  \\
\hline
$\sum_i \Gamma_i $
& 405. & 224. & 409.   
\\ 
\hline
\end{tabular}
\end{center}

\newpage

\begin{center}
\begin{tabular}{||c|c||c|c||}  \hline
\multicolumn{4}{||c||}{Table B7. Partial widths 
of 2$^1$P$_1$ $b_1$ and $h_1$ states (MeV).} \\ 
\hline
\hline
\hskip 1cm Mode \hskip 1cm \
& \hskip .5cm $b_1(1700)$ \hskip .5cm \ 
& \hskip 1cm Mode \hskip 1cm \
& \hskip .5cm $h_1(1700)$ \hskip .5cm \ 
\\ 
\hline
\hline
\multicolumn{4}{||c||}{\hskip .5cm \bf{(1S)$^2$ }}  \\
\hline
$\omega \, \pi$ 
& 56. 
&  
$\rho \, \pi$ 
& 173.   
\\ 
\hline
$\rho \, \eta  $ 
& 18. 
&  
$\omega \, \eta  $ 
& 17.   
\\ 
\hline
$\rho \, \rho  $ 
& 60. 
&  
&
\\ 
\hline
\hline
\multicolumn{4}{||c||}{\hskip .5cm\bf{(2S)(1S)}}  \\
\hline
$\omega(1419)\, \pi$ 
& 13. 
&  
$\rho(1465)\, \pi$ 
&  31. 
\\ 
\hline
\hline
\multicolumn{4}{||c||}{\hskip .5cm \bf{(1P)(1S)}}  \\
\hline
$h_1(1170)\, \pi$ 
& 0. 
&  
$b_1(1231)\, \pi$ 
& 0.
\\
\hline
$a_0(1450)\, \pi$ 
& 2. 
&  
& 
\\ 
\hline
$a_1(1230)\, \pi$ 
& 10. 
&  
& 
\\ 
\hline
$a_2(1318)\, \pi$ 
& 67. 
&  
& 
\\ 
\hline
\hline
\multicolumn{4}{||c||}{\hskip .5cm \bf{(1S)$^2$ strange}}  \\
\hline
${\rm K}^*\, {\rm K}$ 
& 30. &  & 30.   
\\ 
\hline
\hline
\multicolumn{4}{||c||}{\hskip .5cm \bf{total}}  \\
\hline
$\sum_i \Gamma_i $
& 257. &  & 252.   
\\ 
\hline
\end{tabular}
\end{center}

\begin{center}
\begin{tabular}{||c|c|c|c||}  \hline
\multicolumn{4}{||c||}{Table B8. Partial widths 
of $^3$D$_\J$ $\rho_\J$ states (MeV).} \\ 
\hline
\hline
\hskip 1cm Mode \hskip 1cm \
& \hskip .5cm $\rho_3(1691)$ \hskip .5cm \ 
& \hskip .5cm $\rho_2(1670)$ \hskip .5cm \ 
& \hskip .5cm $\rho_1(1700)$ \hskip .5cm \ 
\\ 
\hline
\hline
\multicolumn{4}{||c||}{\hskip .5cm \bf{(1S)$^2$ }}  \\
\hline
$\pi \, \pi$ 
& 59. &  & 48.   
\\ 
\hline
$\omega\, \pi$ 
& 19. & 73. & 35.   
\\ 
\hline
$\rho\, \eta$ 
& 2. & 28. & 16.   
\\ 
\hline
$\rho\, \rho$ 
& 71. & 15. & 14.   
\\ 
\hline
\hline
\multicolumn{4}{||c||}{\hskip .5cm\bf{(2S)(1S)}}  \\
\hline
$\pi(1300)\, \pi$ 
& 0. &  & 0.   
\\ 
\hline
$\omega(1419)\, \pi$ 
& 0. & 0. & 0.   
\\ 
\hline
\hline
\multicolumn{4}{||c||}{\hskip .5cm \bf{(1P)(1S)}}  \\
\hline
$h_1(1170)\, \pi$ 
& 6. & 5. & 124.   
\\ 
\hline
$a_0(1450)\, \pi$ 
&  & 0. &    
\\ 
\hline
$a_1(1230)\, \pi$ 
& 1. & 3. & 134.   
\\ 
\hline
$a_2(1318)\, \pi$ 
& 4. & 201. &  2.  
\\ 
\hline
\hline
\multicolumn{4}{||c||}{\hskip .5cm \bf{(1S)$^2$ strange}}  \\
\hline
${\rm K}\, {\rm K}$ 
& 9. &  & 36.   
\\ 
\hline
${\rm K}^*\, {\rm K}$ 
& 2. & 44. & 26.   
\\ 
\hline
\hline
\multicolumn{4}{||c||}{\hskip .5cm \bf{total}}  \\
\hline
$\sum_i \Gamma_i $
& 174. & 369. & 435.   
\\ 
\hline
$\Gamma_{expt}$
& 215(20) &  & 235(50)   
\\ 
\hline
\end{tabular}
\end{center}

\begin{center}
\begin{tabular}{||c|c|c|c||}  \hline
\multicolumn{4}{||c||}{Table B9. Partial widths 
of $^3$D$_\J$ $\omega_\J$ states (MeV).} \\ 
\hline
\hline
\hskip 1cm Mode \hskip 1cm \
& \hskip .5cm $\omega_3(1667)$ \hskip .5cm \ 
& \hskip .5cm $\omega_2(1670)$ \hskip .5cm \ 
& \hskip .5cm $\omega_1(1649)$ \hskip .5cm \ 
\\ 
\hline
\hline
\multicolumn{4}{||c||}{\hskip .5cm \bf{(1S)$^2$ }}  \\
\hline
$\rho \, \pi$ 
& 50. & 221. & 101.   
\\ 
\hline
$\omega\, \eta$ 
& 2. & 27. & 13.   
\\ 
\hline
\hline
\multicolumn{4}{||c||}{\hskip .5cm\bf{(2S)(1S)}}  \\
\hline
$\rho(1465)\, \pi$ 
& 0. & 0. & 0.   
\\ 
\hline
\hline
\multicolumn{4}{||c||}{\hskip .5cm \bf{(1P)(1S)}}  \\
\hline
$b_1(1231)\, \pi$ 
& 7. & 8. & 371.   
\\ 
\hline
\hline
\multicolumn{4}{||c||}{\hskip .5cm \bf{(1S)$^2$ strange}}  \\
\hline
${\rm K}\, {\rm K}$ 
& 8. &  & 35.   
\\ 
\hline
${\rm K}^*\, {\rm K}$ 
& 2. & 44. & 21.   
\\ 
\hline
\hline
\multicolumn{4}{||c||}{\hskip .5cm \bf{total}}  \\
\hline
$\sum_i \Gamma_i $
& 69. & 300. & 542.   
\\ 
\hline
$\Gamma_{expt}$
& 168(10) &  & 220(35)   
\\ 
\hline
\end{tabular}
\end{center}

\begin{center}
\begin{tabular}{||c|c||c|c||}  \hline
\multicolumn{4}{||c||}{Table B10. Partial widths 
of $^1$D$_2$ $\pi_2$ and $\eta_2$ states (MeV).} \\ 
\hline
\hline
\hskip 1cm Mode \hskip 1cm \
& \hskip .5cm $\pi_2(1670)$ \hskip .5cm \ 
& \hskip 1cm Mode \hskip 1cm \
& \hskip .5cm $\eta_2(1645)$ \hskip .5cm \ 
\\ 
\hline
\hline
\multicolumn{4}{||c||}{\hskip .5cm \bf{(1S)$^2$ }}  \\
\hline
$\rho \, \pi$ 
& 118. 
&  
$\rho \, \rho$ 
& 33.   
\\ 
\hline
$\omega \, \rho $ 
& 41. 
&  
$\omega \, \omega $ 
& 8.   
\\ 
\hline
\hline
\multicolumn{4}{||c||}{\hskip .5cm\bf{(2S)(1S)}}  \\
\hline
$\rho(1465)\, \pi$ 
& 0. &  &   
\\ 
\hline
\hline
\multicolumn{4}{||c||}{\hskip .5cm \bf{(1P)(1S)}}  \\
\hline
$b_1(1231)\, \pi$ 
& 0. 
&  
& 
\\
\hline
$f_0(1300)\, \pi$ 
& 0. 
&  
$a_0(1450)\, \pi$ 
& 0.
\\ 
\hline
$f_1(1282)\, \pi$ 
& 1. 
&  
$a_1(1230)\, \pi$ 
& 5.
\\ 
\hline
$f_2(1275)\, \pi$ 
& 75. 
&  
$a_2(1318)\, \pi$ 
& 189.
\\ 
\hline
\hline
\multicolumn{4}{||c||}{\hskip .5cm \bf{(1S)$^2$ strange}}  \\
\hline
${\rm K}^*\, {\rm K}$ 
& 30. &  & 26.   
\\ 
\hline
\hline
\multicolumn{4}{||c||}{\hskip .5cm \bf{total}}  \\
\hline
$\sum_i \Gamma_i $
& 250. &  & 261.   
\\ 
\hline
$\Gamma_{expt}$
& 258(18) &  &  180 ${+40\atop -21}$(25)   
\\ 
\hline
\end{tabular}%
\end{center}


\newpage

\begin{references}

\bibitem{ikp} N.Isgur, R.Kokoski and J.Paton, Phys. Rev.
Lett. 54, 869 (1985).

\bibitem{cp95} F.E.Close and P.R.Page, Nucl. Phys. B443, 233 (1995);
Phys. Rev. D52, 1706 (1995).

\bibitem{paton85} N.Isgur and J.Paton, Phys. Rev. D31,
2910 (1985).

\bibitem{bcs} 
T.Barnes, F.E.Close and E.S.Swanson, Phys. Rev. D52,
5242 (1995); see also C.Michael \cite{glueball}.

\bibitem{glueball}
G.Bali {\it et al.} (UKQCD Collaboration),
Phys. Lett.  B309, 378 (1993);
D.Weingarten, Nucl. Phys. B (Proc. Suppl.) 34, 29 (1994);
C.Michael, Liverpool report LTH 370, hep-ph/9605243 (May 1996);
F.E.Close and M.J.Teper, 
``On the lightest Scalar Glueball",
RAL-96-040 / OUTP-96-35P (July 1996).

\bibitem{wein} 
J.Sexton, A.Vaccarino and D.Weingarten, Phys. Rev. Lett. 75, 4563 (1995). 

\bibitem{ves951} D.V.Amelin {\it et al.} (VES Collaboration), 
Phys. Lett. B356 (1995) 595.

\bibitem{suchung} S.U.Chung, private communication. 

\bibitem{fec94} F.E.Close, p.1395 in Proc. 
XXVII International Conf. on High Energy Physics,
Glasgow, (Institute of Physics, U.K., 1994, P.Bussey and I.Knowles eds.).

\bibitem{ves} A.M.Zaitsev (VES Collaboration), p.1409 in Proc.
XXVII International Conf. on High Energy Physics,
Glasgow, (Institute of Physics, U.K., 1994, P.Bussey and I.Knowles eds.).

\bibitem{abs}
E.S.Ackleh, T.Barnes and E.S.Swanson, ``On the Mechanism of Open-Flavor
Strong Decays'', ORNL-CTP-96-03, hep-ph-9604355  (April 1996), Phys. Rev. D
(to appear).

\bibitem{early3p0} A.LeYaouanc, L.Oliver, O.P\`ene and J.Raynal, 
Phys. Rev. D8, 2223 (1973);
see also
{\it ibid.},
D9, 1415 (1974);
D11, 1272 (1975);
L.Micu, Nucl. Phys. B10, 521 (1969).

\bibitem{3p0} G.Busetto and L.Oliver, Z.Phys. C20, 247 (1983);
R.Kokoski and N.Isgur, Phys. Rev. D35, 907 (1987);
P.Geiger and E.S.Swanson, Phys. Rev. D50, 6855 (1994);
H.G.Blundell and S.Godfrey, Phys. Rev. D53, 3700 (1996).

\bibitem{pdg96} Particle Data Group, Phys. Rev. D54, 1 (1996).

\bibitem{bellini}  G.Bellini et al., Phys. Rev. Lett. 48, 1697 (1982).

\bibitem{pdg94} Particle Data Group, Phys. Rev. D50, 1173 (1994).

\bibitem{yup} Yu. Prokoshkin (GAMS Collaboration), 
in Proc. of LEAP96,
Dinkelsb\"uhl, Germany, 27-31 August 1996.

\bibitem{cd} A.B.Clegg and A.Donnachie, Z. Phys. C62, 455 (1994).

\bibitem{god} S.Godfrey and N.Isgur, Phys. Rev. D32, 189 (1985).

\bibitem{kalash} A.Donnachie and Yu.S.Kalashnikova, Z.Phys. C59, 621 (1993).

\bibitem{cbrhor} A.Abele {\it et al.} (Crystal Barrel Collaboration),
``High-mass $\rho$-meson states from $\bar p d$-annihilation at
rest into $\pi^-\pi^o\pi^o p_{\rm spectator}$", Phys. Lett. B (to appear).

\bibitem{veseta} VES Collaboration,
``Diffractive reaction $\pi^- A \rightarrow \eta\eta\pi^- A$ 
study at 37 GeV/c". (unpublished)

\bibitem{bs} T.Barnes and E.S.Swanson, in preparation.

\bibitem{khok} Y.Khokhlov, private communication

\bibitem{MIIIeta} 
R.M.Baltrusaitis {\it et al.}
(MarkIII Collaboration)
Phys. Rev. Lett. 55, 1723 (1985);
Phys. Rev. D33, 1222 (1986).

\bibitem{DM2eta} 
D.Bisello {\it et al.}. 
(DM2 Collaboration) 
Phys. Lett. B192, 239 (1987);
Phys. Rev. D39, 701 (1989).

\bibitem{bsz} D.V.Bugg {\it et al.}, Phys. Lett. B353, 378 (1995).

\bibitem{cafe} 
C.Amsler and F.E.Close, 
Phys. Lett.  B353, 385 (1995);
Phys. Rev.  D53, 295 (1996).

\bibitem{bnl1} J.H.Lee {\it et al.}, Phys. Lett. B323,
227 (1994).

\bibitem{Degener} T.Degener (Crystal Barrel Collaboration), 
in Proc. of LEAP96,
Dinkelsb\"uhl, Germany, 27-31 August 1996.

\bibitem{argusa2r} 
G.Kernel (ARGUS), Proc. of PHOTON95 (World Scientific, 1995, eds. D.J.Miller,
S.L.Cartwright and V.Khoze), pp.226-231, esp. Fig.2; 
E.Kri{\v z}ni{\v c}, 
Doctoral thesis, University of Ljubljana (1993); 
E.Kri{\v z}ni{\v c}, 
Proc. XXVII International Conf. on High Energy 
Physics, Glasgow 1994, (Institute of Physics, U.K., 1994, 
P.Bussey and I.Knowles eds.), p.1413.

\bibitem{abc} Z.P.Li, F.E.Close and T.Barnes, Phys. Rev. D43, 2161 (1991);
E.S.Ackleh, T.Barnes and F.E.Close, Phys. Rev. D46, 2257 (1992); T.Barnes,
in Proc. IXth International Workshop on Photon-Photon Collisions,
La Jolla, CA 22-26 March 1992
(World Scientific, 1992, D.O.Caldwell and H.P.Paar eds.).

\bibitem{mpwrev} See 
D.Morgan, M.R.Pennington and M.R.Whalley, J. Phys. G20, A1 (1994) 
for a review of $\gamma\gamma\to VV$ data.

\bibitem{argus91} H.Albrecht {\it et al.}, Zeit. Phys. C50, 1 (1991).

\bibitem{argusww} H.Albrecht {\it et al.} (ARGUS Collaboration), Phys. Lett. 
B374, 265 (1996).

\bibitem{ves92}  G. M. Beladidze {\it et al.}, Zeit. Phys. C54, 367 (1992).

\bibitem{obelix92}  A. Adamo  {\it et al.}, Phys. Lett. B287, 368 (1992).

\bibitem{axstate} 
B.May {\it et al.} (ASTERIX Collaboration), Phys. Lett. B225, 450 (1989);
E.Aker {\it et al.} (Crystal Barrel Collaboration), Phys. Lett. B260,
249 (1991).

\bibitem{besnew} J.Bai et al (BES Collaboration) ``The Structure Analysis 
of the $f_{\rm J}(1710)$ in the Radiative Decay 
$\psi \to \gamma {\rm K}^+ {\rm K}^-$", IHEP
Report (July 1996).

\bibitem{glynnys} F.E.Close, G.Farrar and Z.P.Li, ``Determining the
Gluonic Content of Isoscalar Mesons", RAL-96-052.

\bibitem{T} N.A.T\"ornqvist, Phys. Rev. Lett. 67, 556 (1991).

\bibitem{dsb} K.Dooley, E.S.Swanson and T.Barnes, Phys. Lett. B275, 478 (1992).

\bibitem{acc} C.Daum {\it et al.},  Nucl. Phys. B182, 269 (1981).

\bibitem{pi2phot} 
G.Condo {\it et al.}, Phys. Rev. D43, 2787 (1991).
See also
Y.Eisenberg {\it et al.}, Phys. Rev. Lett. 23, 
1322 (1969);
D.Aston {\it et al.}, Nucl. Phys. B189, 15 (1981).

\bibitem{rya} D.I.Ryabchikov (VES Collaboration), 
Proc. of Hadron95, Manchester, U.K., Aug. 1995.

\bibitem{crystalball} 
D.Antreasyan {\it et al.} (Crystal Ball Collaboration), 
Z.Phys. C48, 561 (1990).

\bibitem{cello} H.J.Behrend {\it et al.} 
(Cello Collaboration), Z.Phys. C46, 583 (1990).

\bibitem{bugg} D.V.Bugg et al., Z.Phys. C (to appear),
``Study of $p\bar{p}\rightarrow\eta\pi^0\pi^0\pi^0$ at
1200 MeV/c". 

\bibitem{eta2cb} C.Amsler {\it et al.} (Crystal Barrel Collaboration),
Z. Phys. C71, 227 (1996).

\bibitem{pagecc} F.E.Close and P.R.Page, Physics Letters B366, 323 (1996). 

\end{references}
\end{document}